\documentclass{siamart1116}

\usepackage[sort]{cite}
\usepackage{graphicx,setspace}
\usepackage{url}
\usepackage{amsmath}
\usepackage{algorithmic,algorithm}
\usepackage{array,subfigure,multirow,setspace}
\usepackage{amssymb,microtype} 
\usepackage{relsize}
\usepackage{multirow}
\newcommand{\mb}[1]{\ensuremath{\mathbf{#1}}}
\newcommand{\etal}{\emph{{\it et al}. }}
\newcommand{\y}{\ensuremath{\mathbf{y}}}
\newcommand{\x}{\ensuremath{\mathbf{x}}}
\newcommand{\h}{\ensuremath{\mathbf{h}}}
\newcommand{\e}{\ensuremath{\mathbf{e}}}

\newcommand{\htil}{\tilde{\mathbf{h}}}
\newcommand{\hs}{\mb{{h}}^*}
\newcommand{\Htil}{\tilde{\mathbf{H}}}
\newcommand{\Hs}{\mb{{H}}^*}
\newcommand{\Hst}{\mb{{H}}^{* \text{\sc T}}}

\newcommand{\dH}{\Delta \mathbf{H}}
\newcommand{\dsh}{\Delta \mathbf{h}}
\newcommand{\es}{{\mathbf{e}^*}}

\newcommand{\ebls}{{\hat{\mathbf{e}}_{\textnormal{BLS}}}}
\newcommand{\ebrls}{{\hat{\mathbf{e}}_{\textnormal{BRLS}}}}
\newcommand{\dls}{\Delta\mathbf{\mb{e}}_{\textnormal{LS}}}

\newcommand{\dbrls}{\Delta \mathbf{\mb{e}}_{\textnormal{BRLS}}}
\newcommand{\w}{\ensuremath{\mathbf{w}}}

\newcommand{\argmax}[1]{\mbox{arg } \underset{#1}{\mbox{max }}}
\newcommand{\argmin}[1]{\mbox{arg } \underset{#1}{\mbox{min }}}
\newcommand{\real}[1]{\ensuremath{\mathbb{R}^{#1}}}

\newcommand{\loot}{$\ell_1/\ell_2$~}

\newcommand{\exptn}[1]{\ensuremath{\mathcal{E}\left\{#1\right\}}}

\newcommand{\tr}{\text{\sc T}}

\numberwithin{thm}{section}
\newtheorem{lem}{Lemma}
\numberwithin{lem}{section}
\newtheorem{propos}{Proposition}
\numberwithin{propos}{section}

\numberwithin{defn}{section}

\def\endproof{$\hfill\blacksquare$}

\numberwithin{theorem}{section}

\newcommand{\TheTitle}{A Non-Convex Optimization Technique for Sparse Blind Deconvolution -- Initialization Aspects and Error Reduction Properties}
\newcommand{\PageTitle}{ALPA for Sparse Blind Deconvolution}
\newcommand{\TheAuthors}{Aniruddha Adiga and Chandra Sekhar Seelamantula}

\headers{\PageTitle}{\TheAuthors}

\title{{\TheTitle}}

\author{
  Aniruddha Adiga$^*$\and
  Chandra Sekhar Seelamantula\thanks{Department of Electrical Engineering, Indian Institute of Science, Bangalore - 560 012, India (\email{{aaniruddha, chandrasekhar}@iisc.ac.in}).}}

\usepackage{amsopn}

\ifpdf
\fi

\begin{document}

\maketitle

\begin{abstract}
Sparse blind deconvolution is the problem of estimating the blur kernel and sparse excitation, both of which are unknown. Considering a linear convolution model, as opposed to the standard circular convolution model, we derive a sufficient condition for stable deconvolution. The columns of the linear convolution matrix form a Riesz basis with the tightness of the Riesz bounds determined by the autocorrelation of the blur kernel. Employing a Bayesian framework results in a non-convex, non-smooth cost function consisting of an $\ell_2$ data-fidelity term and a sparsity promoting $\ell_p$-norm ($0 \le p \le 1$) regularizer. Since the $\ell_p$-norm is not differentiable at the origin, we employ an $\epsilon$-regularized $\ell_p$-norm as a surrogate. The data term is also non-convex in both the blur kernel and excitation. An iterative scheme termed alternating minimization (Alt. Min.) $\ell_p-\ell_2$ projections algorithm (ALPA) is developed for optimization of the $\epsilon$-regularized cost function. Further, we demonstrate that, in every iteration, the $\epsilon$-regularized cost function is non-increasing and more importantly, bounds the original $\ell_p$-norm-based cost. Due to non-convexity of the cost, the accuracy of estimation is largely influenced by the initialization. Considering regularized least-squares estimate as the initialization, we analyze how the initialization errors are concentrated, first in Gaussian noise, and then in bounded noise, the latter case resulting in tighter bounds. Comparisons with state-of-the-art blind deconvolution algorithms show that the deconvolution accuracy is higher in case of ALPA. In the context of natural speech signals, ALPA results in accurate deconvolution of a voiced speech segment into a sparse excitation and smooth vocal tract response.
\end{abstract}



\section{Introduction}
Consider the measurement model
\begin{equation}
\y = \h \ast \e + \w,
\label{linmod}
\end{equation}  
where $\ast$ denotes linear convolution, the observation $\y \in \real{N}$, the blur kernel $\h \in \real{L}$, the excitation $\e \in \real{M}$, the acquisition noise $\w \in \real{N}$, and $N = L+M-1$. Such linear shift-invariant (LSI) models are frequently encountered in geophysics \cite{Seismic_takahata}, speech processing \cite{rabiner,Makhoul}, image processing \cite{PantinEtAl07,kundur,understanding_bd_algo_levin},
etc. In microscopy \cite{fluorophore_small}, astronomy \cite{PantinEtAl07}, and photography applications, etc., the vector $\h$ denotes the point-spread function (the blur kernel) of the imaging system, $\e$ is the underlying sharp image, and $\y$ is the blurred captured image. The blur kernel accounts for finite aperture of the imaging system, possible camera motion, defocus, atmospheric disturbances, etc. In a speech processing context, for voiced sounds, $\h$ models the vocal-tract filter response, $\e$ is the quasi-periodic glottal excitation, and $\y$ denotes the sampled speech signal \cite{Makhoul}.\\
\indent The goal in blind deconvolution is to estimate $\h$ and $\e$ given $\y$ and the statistics of $\w$. The problem is inherently ill-posed as there exist infinitely many combinations of $\h$ and $\e$ that give rise to the same $\y$. Taking into account available priors on $\h$ and $\e$ would constraint the solution space. We consider the widely applicable case of a sparse excitation and a relatively smooth and localized blur kernel. In several applications such as debluring of star-field images \cite{Jeffs,PantinEtAl07}, fluorophore localization in super-resolution microscopy \cite{fluorophore_small}, source-filter modeling of voiced speech signals \cite{rabiner,Makhoul}, separation of the reflectivity function from the source signature in seismic signals \cite{Seismic_takahata}, etc., the excitation is innately sparse, and the blur kernel is a lowpass function. While in some cases, sparsity manifests directly, in others, sparsity becomes apparent only after performing a suitable transformation, for instance, the wavelet transform.
\subsection{A Maximum a Posteriori (MAP) Formulation for Sparse Blind Deconvolution}
Within a Bayesian setting, the {\it maximum a posteriori} (MAP) estimates of the filter $\h$ and excitation $\e$ are given by the joint optimization:
\begin{eqnarray}
(\h_{\text{opt}},\e_{\text{opt}}) &=& \argmax{(\h,\e)} g(\y/\e;\h)f(\e),
\label{MAPgenfor}
\end{eqnarray}
where $g$ is the likelihood of the observations and $f$ denotes the prior on $\e$. Let $\e$ have i.i.d. entries following a \emph{generalized p-Gaussian} (gpG) distribution \cite{Jeffs}, which results in the prior
 \begin{equation}
 f(\e) = \left(\frac{p}{2\Gamma(1/p)\gamma\sigma_e}\right)^M \exp\left(-\sum_{i=0}^{M-1}\left(\frac{|e_i|}{\gamma\sigma_e}\right)^p\right),
 \label{exp_prior}
 \end{equation}
where $\gamma = \left(\frac{\Gamma(1/p)}{\Gamma(3/p)}\right)^{1/2}$, and $e_i$ denotes the $i^{\text{th}}$ entry of $\mb{e}$. For $0 \le p \le 1$, $f$ is a heavy-tailed distribution that yields sparse sequences \cite{Jeffs}, which is the scenario of interest in this paper. As $p \rightarrow 0$, the \emph{kurtosis}/\emph{peakedness} of the distribution increases, and the tail becomes heavier. With the gpG prior \eqref{exp_prior}, the MAP formulation is equivalent to
 \begin{eqnarray}
 (\h_{\text{opt}},\e_{\text{opt}}) 
 &=& \argmin{(\h, \e)} \|\y - \h \ast \e\|_2^2 + \frac{2\sigma_w^2}{(\gamma\sigma_e)^p} \|\e\|_p^p,
 \end{eqnarray}
 with $\ell_p$-norm regularization coming up naturally. In a practical setting, if the distribution parameters are not known, a viable alternative is to solve
\begin{eqnarray}
(\h_\text{opt},\e_\text{opt}) &=& \argmin{\h, \e} \underbrace{\|\y - \h \ast \e \|_2^2 + \delta \|\e\|_p^p}_{F(\h,\e)},
\label{gen_MAP}
\end{eqnarray}
where $\delta$ is the regularization parameter that controls the trade-off between data fidelity and sparsity and could be fixed experimentally or using cross-validation \cite{wahba}.\\
\indent $\ell_p$-quasi-norms, $0 \le p < 1$, and $\ell_1$-norm are popular sparsity promoting priors and have been employed extensively in the general class of linear inverse problems \cite{FISTA,Jeffs,focuss,ming-jun2011}, image deconvolution \cite{Jeffs}, speech coding \cite{gia} and sparse recovery problems \cite{focuss,chart}. The $\ell_1$-norm is used in the least absolute shrinkage and selection operator (LASSO) \cite{tibs}, basis-pursuit denoising problems \cite{atomic_chen}, and as a convex proxy for the $\ell_0$-quasi-norm in compressed sensing (CS) problems \cite{candes_romberg_tao2006}. However, if one were to use the $\ell_p$-norm ($0 \le p < 1$), fewer random projections would be required as compared with the $\ell_1$-norm \cite{chart2007}.\\
\indent As the $\ell_p$-norms suffer from local non-differentiability, optimization is carried out using gradient-based iterative solvers \cite{FISTA, fig_grad_proj}, majorization-minimization (MM) approaches \cite{MM_hunter} such as the iteratively reweighted least-squares (IRLS) \cite{daubechies_IRLS, chart}, iteratively reweighted $\ell_1$-norm \cite{candes} techniques, etc. Early work on deblurring of star-field images using $\ell_p$-norm priors proposed in \cite{Jeffs} employed simplex search to optimize the non-convex cost.

\subsection{Related Literature}
\indent In the computer vision and image processing communities, blind deconvolution is almost synonymous with image deblurring. A vast amount of literature exists on this topic, which makes it impossible to summarize every contribution. We refer the reader to \cite{kundur} and \cite{campisi2007blind} for a comprehensive review of blind deconvolution algorithms, most of which are set up within a Bayesian framework. Apart from the MAP formulation, there exist other approaches based on the expectation-maximization (EM) algorithm \cite{ML_EM_katsagellos,Fig_EM}, the
variational approach \cite{understanding_bd_algo_levin,variational_likas}, quasi-maximum-likelihood approach \cite{bronstein2005blind}, ADMM \cite{almeida}, etc. Specific to image deconvolution, Levin et al. showed that the naive MAP approach may lead to degeneracy (resulting in the blurred image itself as the blur kernel, and a Kronecker impulse as the excitation) and developed strategies to overcome it \cite{understanding_bd_algo_levin}. The gradients in natural images were shown to be heavy-tailed, which were parametrically modeled using Gaussian mixtures \cite{fergus2006removing}, second-order polynomials \cite{shan2008high}, non-identical but independent Gaussians \cite{wipf_revisit}, etc.\\
\indent Non-parametric regularizers include the $\ell_p$-norm \cite{levin2007image, krishnan2009fast, Joshi_lp}, the $\ell_1/\ell_2$ \cite{l1l2_krishnan, Euclid_taxicab_repetti}, the isotropic total-variation (TV) regularizer \cite{Sroubek_MC, clearTV}, and the $p^{\text{th}}$ power TV norm \cite{kotera_heavy_prior}. The $\ell_p$-norm-based cost function is optimized using IRLS \cite{levin2007image, Joshi_lp} or using separable regularizers \cite{krishnan2009fast}. The $\ell_1$-norm is influenced by the amplitudes of the estimates and may not always yield a sparse solution. In order to circumvent this problem, the scale-invariant $\ell_1/\ell_2$ function was considered in \cite{l1l2_krishnan}, and a LASSO solver is used to optimize the cost by rescaling with the $\ell_2$-norm from the previous iteration. Repetti \etal \cite{Euclid_taxicab_repetti} developed an approach consisting of a combination of MM and proximal methods to optimize a smoothed \loot regularized cost function for blind deconvolution of seismic signals.\\
\indent Wipf and Zhang \cite{wipf_revisit} solved the image deblurring problem using a variational Bayesian strategy and concave sparsity priors with the degree of concavity adapted to the noise and energy of the blur kernel. Zhang \etal \cite{zhang_multi_observation} considered the multiple measurement counterpart of this problem and showed that a single, low-noise, less blurred observation dominates the multi-observation regularizer and makes it concave. A relatively new class of approaches reformulate the blind deconvolution problem as a low-rank matrix recovery problem from linear measurements, with constraints on the subspace dimension and sparsity to ensure uniqueness, and employ convex programming techniques for recovery \cite{Ali_convex,Yanjun_TIT, Urbashi_ISIT, Chi_spike_bd}.\\
\indent Since we are dealing with a non-convex problem in general, which calls for iterative techniques, the issue of initialization becomes important, mainly from the viewpoint of avoiding local minima. In this paper, we consider the regularized least-squares (reg. LS) estimate as the initialization and analyze concentration of the error from the ground truth.

\subsection{Our Contributions}
\indent We provide a sufficiency condition for stable deconvolution for the specific case of linear convolution (Proposition~\ref{Proposition2}, Section~\ref{sec:sufficient_cond}). 
Since the cost function is non-convex and non-smooth, we develop an alternating $\ell_p$-$\ell_2$ projections algorithm (ALPA) (Section~\ref{ALPA}) considering a smoothed version of the cost. Further, we show that the iterative algorithm ensures that the smoothed cost is non-increasing in every iteration and upper bounds the original cost function (Section~\ref{conv}, Proposition~\ref{proposition1}). We then consider reg. LS estimate of the excitation as the initialization for ALPA and analyze the concentration of the mean-absolute error of this estimate from the ground truth (Section~\ref{prob_guarantees}). The error bounds depend on the condition number of the linear system, the regularization parameter, error in filter estimation, and the noise variance (Proposition~\ref{BRLS_prop}). Further, if the noise is bounded, the tail bounds can be made tighter thanks to the Hoeffding inequality (Proposition~\ref{BRLS_prop_hoeff}). On the application front, we demonstrate successful blind deconvolution of voiced speech signals into a sparse excitation and vocal-tract filter and compare the results with state-of-the-art techniques (Section~\ref{voiced_speech}).

\section{A Sufficient Condition for Stable Deconvolution}
\label{sec:sufficient_cond}
\indent Consider the matrix form of the linear measurement model in \eqref{linmod}:
\begin{eqnarray}
\y=\mathbf{H} \e + \mb{w}= \mathbf{E} \h + \mb{w},
\label{conv_mat}
\end{eqnarray}
where $\mathbf{E} \in \real{N \times L}$ and $\mathbf{H} \in \real{N \times M}$ are linear convolution matrices (cf. \eqref{convmatref} in Appendix~\ref{appendix_proposition1}) constructed from $\mb{e} \in \mathbb{R}^M$ and $\mb{h} \in \mathbb{R}^L$, respectively, and  the noise vector $\mb{w} \sim \mathcal{N}({\bf 0},\sigma_w^2\mathbf{I})$. The vector $\mb{h}$ is assumed to be deterministic. The linear convolution model \eqref{conv_mat} is a more realistic representation of practical linear, shift-invariant systems than the commonly assumed circular convolution model. Further, it does not lead to degenerate solutions (such as $\h_n = \y_n$ and $\e_n = \delta[n]$, where $n$ denotes the element index, and $\delta[n]$ is the Kronecker impulse, which is the global optimum \cite{understanding_bd_algo_levin, gribonval_pitfall}) that may be encountered in a circular convolution model, because the filter, excitation, and measurement vectors reside in different dimensional spaces.\\
\begin{lemma}
\label{Proposition1}
Let $\mb{H} \in \mathbb{R}^{N \times M}$ be a linear convolution matrix with columns $\h_s$, $0\le s \le M-1,$ obtained as shifted versions of the filter $\h$. Then the set $\{\h_s\}_{s=0}^{L-1}$ forms a Riesz basis with Riesz bounds given as
\begin{equation}
0 < \sigma^2_{\textnormal{min}} (\mb{H})\le \frac{\|\mb{Hx}\|_2^2}{\|\mb{x}\|_2^2} \le \sigma^2_{\textnormal{max}}(\mb{H}), \text{\quad} \forall \mb{x} \in \mathbb{R}^M-\{\mb{0}\},
\label{Riesz_bounds}
\end{equation}
where $\sigma_{\textnormal{min}}(\mb{H})$ and $\sigma_{\textnormal{max}}(\mb{H})$ denote the minimum and maximum singular values of $\mb{H}$, respectively.
\end{lemma}

The proof is given in Appendix~\ref{appendix_proposition1}. \\
\indent It is easy to verify that the columns of $\mb{H}$ are linearly independent and hence, the problem of determining $\e$ from $
\y$ given $\h$ is \emph{well-posed}, that is, if a solution exists, it would be unique, and a continuous function of the measurements \cite{hadamard}. The conditioning of the system is determined by the tightness of the Riesz bounds.

\begin{propos}
\label{Proposition2}
Consider a filter $\mb{h}$ with $\|\mb{h}\|_2=1$. Let $\mb{H}$ and $r_\mb{hh}$ be the corresponding linear convolution matrix and Gram sequence, respectively. The Riesz bases constituted by the columns of $\mb{H}$ will have a lower Riesz bound $\sigma_{\textnormal{min}}^2(\mb H) \ge \eta$ and an upper Riesz bound $\sigma_{\textnormal{max}}^2(\mb H) \le 2-\eta$, where $\eta \in (0,1]$, if 
\begin{equation}
0 \le \sum_{\ell = 1}^{(M-1)/2} \left| r_\mb{hh}(\ell) \right| \le \frac{1-\eta}{2}.
\end{equation}
\end{propos}
The proof is given in Appendix~\ref{appendix_theorem1}. \\
\indent The condition number of $\mb{H}^\text{\sc T}\mb{H}$ varies as $1/\eta$. As $\eta \rightarrow 1$, the Riesz bounds become tighter, and the columns of $\mb{H}^\text{\sc T} \mb{H}$ will tend to become orthonormal. A similar sufficiency condition can be derived on the excitation sequence considering the model: $\y= \mathbf{E} \h + \mb{w}.$ In particular, for periodic and sparse excitation sequences, the Riesz bounds could be tight as illustrated next. For instance, consider the unit-norm periodic sparse excitation $e(n) = \sum_{k = 0}^{K-1} a_k \delta (n-kT)$. A filter $\mb{h}$ of length $T$ yields a convolution matrix  $\mb{E}$ with orthonormal columns and hence the autocorrelation of the excitation satisfies: $\sum_{\ell = 1}^{(L-1)/2} \left| r_\mb{ee}(\ell) \right| = 0$. In this instance, since $\mb{E}^{ \text{\sc T}} \mb{E} = \mb{I}_{T \times T}$, we have that $\sigma_{\text{min}}^2(\mb{E})=\sigma_{\text{max}}^2(\mb{E})= 1$.

\section{The Alternating $\ell_p-\ell_2$ Projections Algorithm}
\label{ALPA}
\indent We adopt an {\it alternating minimization} (Alt. Min.) strategy to solve the optimization problem in \eqref{gen_MAP}. In the first step, we
fix $\h$, and optimize $F(\h,\e)$ over $\e$ (excitation optimization or \emph{e-step}). In the next step, $F(\h,\e)$ is updated with the estimate of $\e$ and then optimized over $\h$ (filter optimization or \emph{h-step}). The Alt. Min. iterations are carried out until a suitable convergence criterion is met.\\
\indent Let $\h^{(k)}$ denote the filter estimate obtained at the end of the $k^{\text{th}}$ iteration. In the \emph{e-step}, 
$F\left(\mathbf{h}^{(k)},\mathbf{e}\right)$ is  optimized with respect to $\e$  to obtain $\e^{(k+1)}$:
\begin{eqnarray}
\mathbf{e}^{(k+1)} &=& \argmin{\e} \underbrace{\|\y - \mb{H}^{(k)} \e\|_2^2 + \delta \|\e\|_p^p}_{F(\h^{(k)},\e)},\label{e-step_orig_cost}
\end{eqnarray}
where $\mb{H}^{(k)} \in \real{N \times M}$ is a linear convolution matrix  constructed from $\mb{h}^{(k)}$.  For $0 \le p < 1$, $ F (\mathbf{h}^{(k)},\mathbf{e})$ is non-convex and its gradient with respect to $\e$, denoted by $\nabla_{\e} F\left(\mathbf{h}^{(k)},\mathbf{e}\right)$ has
a discontinuity at $\e = \mb{0}$. To circumvent the discontinuity, we approximate the $\ell_p$-norm with its $\epsilon-$regularized version: $\|\e\|_{p,\epsilon}^p = \sum_{i=0}^{M-1} (e_i^2+ \epsilon)^{p/2},\, \epsilon >0$, resulting in the modified cost $F_{\epsilon}$ given by
\begin{equation}
F_{\epsilon}(\mb{h}^{(k)},\mb{e}) = \|\y-\mb{H}^{(k)}\mb{e}\|_{2}^{2} + \delta \|\e\|_{p,\epsilon}^p.
\label{epsilon_lp_function}
\end{equation}
Replacing $F(\h^{(k)},\e)$ in \eqref{e-step_orig_cost} with $F_\epsilon(\h^{(k)},\e)$, the
estimate of $\e$ in the $(k+1)^{\text{st}}$ iteration is obtained as follows:
\begin{eqnarray}
\e^{(k+1)} = \argmin{\e} F_\epsilon(\h^{(k)},\e).\nonumber
\end{eqnarray}
The cost function $F_\epsilon(\h^{(k)},\e)$ is differentiable with respect to $\e$, and has a stationary point $\tilde{\mb{e}}$, corresponding to which
\begin{equation}
{\mb{H}^{(k)}}^{ \text{\sc T}} (\mb{H}^{(k)} \tilde{\mb{e}} - \y) +\delta \mathbf{W} \tilde\e = \mb{0},
\label{approx_lp_nonlin}
\end{equation}
where $\mathbf{W}$ is a diagonal matrix with $i^{\text{th}}$ diagonal entry given by 
$p\left(\tilde{e}_i^{2}+\epsilon\right)^{p/2-1}$.
Equation \eqref{approx_lp_nonlin} is nonlinear in $\tilde{\mb{e}}$ since $\mathbf{W}$ depends on $\tilde{\mb{e}}$, which makes a closed-form solution infeasible. Hence, we estimate the stationary point via the fixed-point iteration:
\begin{equation}
{\mb{H}^{(k)}}^{ \text{\sc T}} (\mb{H}^{(k)} \tilde{\mb{e}}^{(j+1,k)} - \y) +\delta \mathbf{W}^{(j,k)} \tilde\e^{(j+1,k)} = \mb{0},
\label{fixed-point} 
\end{equation}
where $(j, k)$ indicates the $j^{\text{th}}$ iterate corresponding to the fixed-point procedure within the $k^{\text{th}}$ iteration of the Alt. Min. scheme, and $\mathbf{W}^{(j,k)}$ is a diagonal matrix whose $i^{\text{th}}$ entry is $p\left((\tilde{e}_i^{(j,k)})^2 + \epsilon \right)^{p/2-1}$. The estimate
$\tilde{\mb{e}}^{(j+1,k)}$ is computed according to the IRLS algorithm as
\begin{eqnarray}
\tilde{\mathbf{e}}^{(j+1,k)} &=& \left({\mathbf{H}^{(k)}}^\text{\sc T} \mathbf{H}^{(k)} + \delta \mathbf{W}^{(j,k)}\right)^{-1} (\mathbf{H}^{(k)})^{ \text{\sc T}} \y.
\end{eqnarray}
$\mathbf{W}^{(j,k)}$ tends to blow up for small values of $\tilde{e}_i^{(j,k)}$, which may happen as iterations progress, and might lead to ill-conditioning, causing problems in inversion. However, applying the matrix inversion lemma \cite{golub} circumvents the issue, for it gives,
\begin{eqnarray}
\tilde{\mathbf{e}}^{(j+1,k)} &=&\!{\mathbf{W}^{(j,k)}}^{-1}\big(\mathbf{I}- \mathbf{H}^{(k)} \left(\mb{I}+\mathbf{H}^{(k)}  {\delta \mathbf{W}^{(j,k)}}^{-1}{\mathbf{H}^{(k)}}^{\text{\sc T}}\right)^{-1}\mathbf{H}^{(k)}{\mathbf{W}^{(j,k)}}^{-1} {\mathbf{H}^{(k)}}^{\text{\sc T}}\big) \y,\nonumber\\
\label{e-step_alpaMI}
\end{eqnarray}
where ${\mathbf{W}^{(j,k)}}^{-1}$ is a diagonal matrix with $i^{\text{th}}$ entry $p\left((\tilde{e}_i^{(j,k)})^2 + \epsilon \right)^{1-p/2}$. 
After $J$ iterations of IRLS, we obtain the $(k+1)^{\text{st}}$ iterate for $\mb{e}$ as $\mathbf{e}^{(k+1)} = \tilde{\mathbf{e}}^{(J,k)},$ which is then used to update $\h^{(k+1)}$ in the \emph{h-step} as follows:
\begin{eqnarray}
\mb{h}^{(k+1)} &=& \argmin{\mb{h}}  F_{\epsilon}\left(\mb{h},\mb{e}^{(k+1)}\right) = \argmin{\mb{h}} \|\y-\mb{E}^{(k+1)}\mb{h}\|_2^2 = {\mb{E}^{(k+1)}}^{\dagger} \y,
\label{filter_cost}
\end{eqnarray}
where $\mb{E}^{(k+1)} \in \real{N \times L}$ is a linear convolution matrix constructed from $\mb{e}^{(k+1)}$, and $\dagger$ denotes the Moore-Penrose inverse.\\
\indent There is an inherent scale-ambiguity in the problem: if $\hat{\mb{h}}$ and $\hat{\mb{e}}$ constitute a solution pair, so do the scaled versions $\alpha\hat{\mb{h}}$ and $\hat{\mb{e}}/\alpha$, $\alpha \neq 0$. In order to overcome this ambiguity, we normalize the estimate of $\mb{h}$ in every iteration, to possess unit energy, as follows: $\h^{(k)} \leftarrow \h^{(k)}/\|\h^{(k)}\|_2$. Alternatively, one could add the regularizer $\beta\left(\|\h\|^2-1\right), \beta > 0$ to the cost $F_{\epsilon}$ in \eqref{filter_cost}, which results in the update $\mb{h}^{(k+1)} = (\mb{E}^{{(k+1)}^\tr}\mb{E}^{{(k+1)}}+\beta\mb{I})^{-1}\mb{E}^{{(k+1)}^\tr}\y.$\\
\indent The Alt. Min. scheme is summarized in Algorithm~\ref{algo1}.
\begin{algorithm}[t]      
\small
\caption{Alternating $\ell_p$-$\ell_2$ projections algorithm (\text{ALPA}) for sparse blind deconvolution.} 
\textbf{Input}: Measurement vector $\y$\\
\textbf{Initialization}: $k = 0$, $\mathbf{h}^{(0)} \in \mathbb{R}^L$, $\mathbf{H}^{(0)} = \mbox{Conv. Matrix}\left(\mathbf{h}^{(0)}\right)$,
$\mathbf{W}^{(0,0)} = \mb{I}_{M \times M}$, $\delta = 1$, set flag to FALSE.\\
\textbf{While} flag FALSE \textbf{do}
\begin{algorithmic} 
{
\STATE{Step 1:} \emph{\textbf{e-step}}:\\
\textbf{For} $j = 1 \text{ to }J$
$$\tilde{\mathbf{e}}^{(j+1,k)}\!=\!{\mathbf{W}^{(j,k)}}^{-1}\big(\mathbf{I}- \mathbf{H}^{(k)} \left(\mb{I}+\mathbf{H}^{(k)}  {\delta \mathbf{W}^{(j,k)}}^{-1}{\mathbf{H}^{(k)}}^{\text{\sc T}}\right)^{-1}\mathbf{H}^{(k)}{\mathbf{W}^{(j,k)}}^{-1} {\mathbf{H}^{(k)}}^{\text{\sc T}}\big) \y
$$
\textbf{end}
$$\mathbf{e}^{(k+1)} = \tilde{\mathbf{e}}^{(J,k)}.$$
\STATE{Step 2:} Construct $\mathbf{E}^{(k+1)} = \mbox{Conv. Matrix}\left(\mathbf{e}^{(k+1)}\right)$.
\STATE{Step 3:} \emph{\textbf{h-step}}:
$\mathbf{h}^{(k+1)} =  {\mb{E}^{(k+1)}}^{\dagger} \y$.
\STATE{Step 4:} Normalization: $\mathbf{h}^{(k+1)} \leftarrow \mathbf{{h}}^{(k+1)}/ \|\mb{{h}}^{(k+1)}\|_2.$
\STATE{Step 5:} Update $\mathbf{H}^{(k+1)} = \mbox{Conv. Matrix}\left(\mathbf{h}^{(k+1)}\right)$.
\STATE{Step 6:} Update $\mathbf{W}_{ii}^{(1,k+1)} = p \left( (e_{i}^{(k+1)})^2+\epsilon \right)^{p/2-1}$, $1 \leq i \leq N.$
\STATE{Step 7:} Stopping criterion: If $\frac{\left\|\mathbf{e}^{(k+1)}-\mathbf{e}^{(k)}\right\|_2^2}{\left\|\mb{e}^{(k)}\right\|_2^2} \leq \text{tolerance}$, then set
flag to TRUE, $\mb{e}_{\text{opt}} = \mb{e}^{(k+1)}, \mb{h}_{\text{opt}} = \mb{h}^{(k+1)}$, else $k\leftarrow k+1$.
}
\end{algorithmic}
\textbf{end while}\\
\textbf{Outputs}: $\mb{e}_{\text{opt}}$ and $\mb{h}_{\text{opt}}$.
\label{algo1}
\end{algorithm}
\normalsize

\section{Error Reduction Properties}
\label{conv}
\begin{figure}[t]
\centering
\includegraphics[width=2.3in]{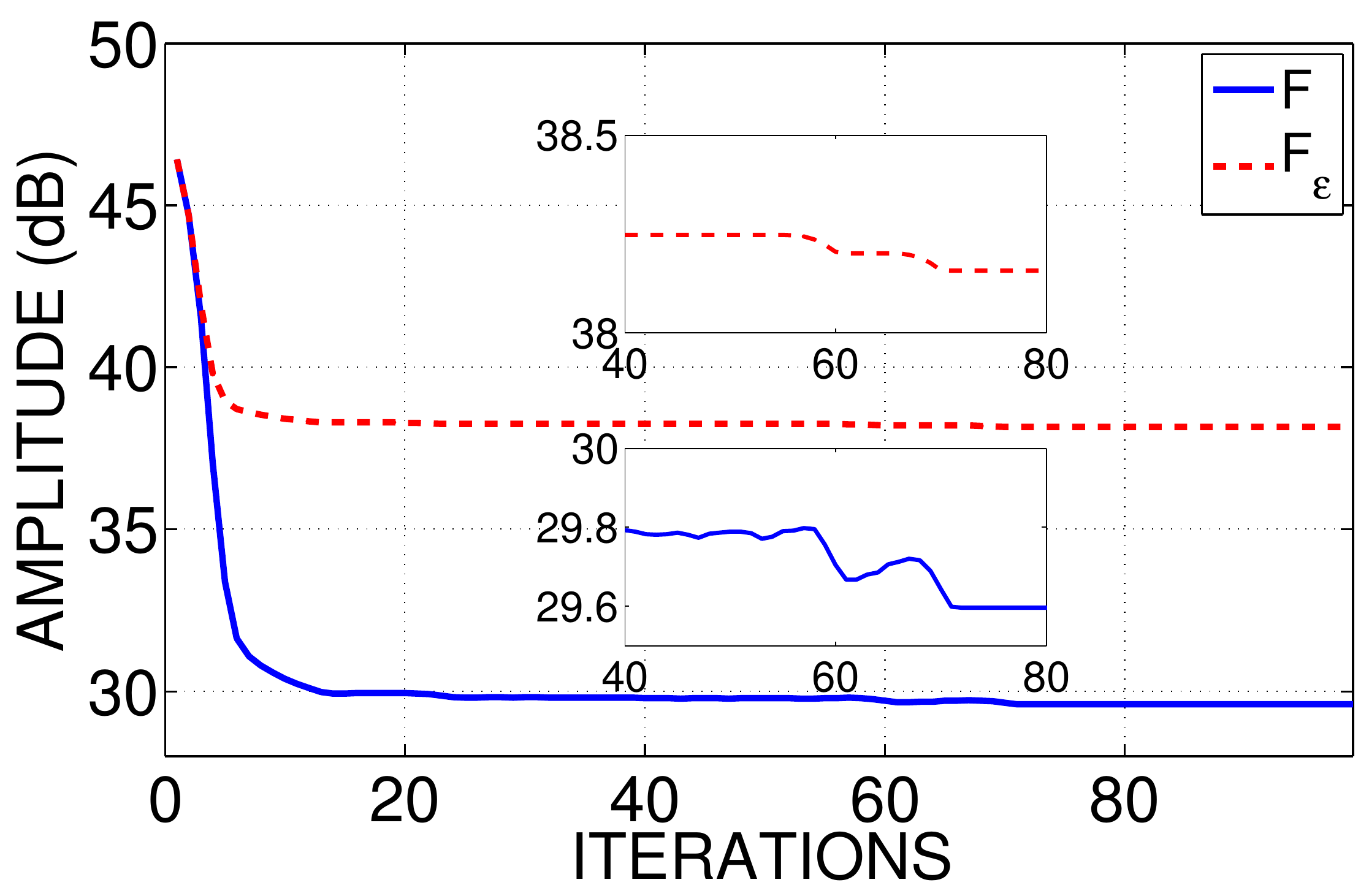}
\caption{(Color online) The cost functions $F$ and $F_{\epsilon}$ (with $\epsilon = 10^{-6}$) pertaining to the example considered in Section~\ref{conv}.}
\label{cost_function_compare}
\end{figure}

\indent We next establish that, after every update of $\h^{(k)}$ and $\e^{(k)}$, the cost $F\left(\h^{(k)},\e^{(k)}\right)$ is upper-bounded by a non-increasing cost $F_\epsilon\left(\h^{(k)},\e^{(k)}\right)$, $\epsilon > 0$. We first consider the behavior of the cost functions $F$ and $F_{\epsilon}$ for a fixed $\h^{(k)}$, but with the excitation estimated (Lemma~\ref{lemma1}), then with a fixed $\e^{(k)}$ and the filter estimated (Lemma~\ref{lemma2}). Finally, we combine the two results to get Lemma~\ref{lemma3}.
\begin{lem}
\label{lemma1}
Let $\mb{e}^{(k+1)}$ be the minimizer of $F_{\epsilon}(\mb{h}^{(k)},\mb{e})$ defined in \eqref{epsilon_lp_function} after the $(k+1)^{\text{st}}$ iteration, for a fixed $\mb{h}^{(k)}$. Then, $F_{\epsilon}$ satisfies the descent property
\begin{equation*}
F_{\epsilon}\left(\mb{h}^{(k)},\mb{e}^{(k+1)}\right) \le F_{\epsilon}\left(\mb{h}^{(k)},\mb{e}^{(k)}\right).
\end{equation*}
\end{lem}
The proof is given in Appendix~\ref{appendix1}.
\begin{lem}
\label{lemma2}
Let $\mb{h}^{(k+1)}$ be the minimizer of $F_{\epsilon}\left(\mb{h},\mb{e}^{(k+1)}\right)$ defined 
in \eqref{filter_cost} after the $(k+1)^{\text{th}}$ iteration, for a given $\mb{e}^{(k+1)}$. Then, $F_{\epsilon}$ satisfies the descent property
\begin{equation*}
F_{\epsilon}\left(\mb{h}^{(k+1)},\mb{e}^{(k+1)}\right) \le F_{\epsilon}\left(\mb{h}^{(k)},\mb{e}^{(k+1)}\right).
\end{equation*}
\end{lem}
\begin{proof} From the analysis given in Section~\ref{sec:sufficient_cond}, we know that $\mb{E}^{(k+1)}$ consists of linearly independent
columns. The Hessian of the cost function $F_\epsilon(\h, \e^{(k+1)}) = \|\y-\mb{E}^{(k+1)}\mb{h}\|_2^2$ is ${\mb{E}^{(k+1)}}^{\text{\sc T}}
\mb{E}^{(k+1)}$, which is a positive-definite matrix. Consequently, $F_\epsilon(\h, \e^{(k+1)})$ is strictly convex, and hence $F_\epsilon(\h, \e^{(k+1)})$ has
a unique minimizer, which we denote as $\h^{(k+1)}$. Thus, $F_{\epsilon}(\mb{h}^{(k+1)},\mb{e}^{(k+1)}) \le
F_{\epsilon}\left(\mb{h}^{(k)},\mb{e}^{(k+1)}\right).$
\end{proof}\\
\indent We haven't considered normalization of the filter estimate in the above proof. However, one could establish a similar property where the normalization is enforced via the regularizer $\beta\left(\|\h\|^2-1\right)$ alluded to at the end of Section~\ref{ALPA}.\\
\indent Combining Lemmas~\ref{lemma1} and \ref{lemma2} gives the following result pertaining to the descent of the cost $F_{\epsilon}$ after updating both filter and excitation.
\begin{lem}
\label{lemma3}
Suppose $\mb{e}^{(k+1)}$ and $\mb{h}^{(k+1)}$ are the minimizers in \eqref{epsilon_lp_function} and \eqref{filter_cost}, respectively. After the $(k+1)^{\text{st}}$ iteration of ALPA,
\begin{equation*}
F_{\epsilon}(\mb{h}^{(k+1)},\mb{e}^{(k+1)}) \le F_{\epsilon}\left(\mb{h}^{(k)},\mb{e}^{(k)}\right).
\end{equation*}
\end{lem}
The following proposition establishes that, in every iteration, the difference between $F_{\epsilon}$ and $F$ is bounded by a function of $\epsilon$, which can be made arbitrarily small.
\begin{propos}
\label{proposition1}
In every iteration of ALPA, the surrogate cost $F_{\epsilon}$ and the actual cost $F$ satisfy the inequality:
\begin{equation*}
0 < F_{\epsilon}\left(\mb{h}^{(k)},\mb{e}^{(k)}\right)-F\left(\mb{h}^{(k)},\mb{e}^{(k)}\right) \le M\epsilon^{p/2}.
\end{equation*} 
\end{propos}
\begin{proof}
Consider the difference between the surrogate cost $F_{\epsilon}$ and the actual cost $F$:
\begin{eqnarray}
F_{\epsilon}\left(\mb{h}^{(k)},\mb{e}^{(k)}\right) - F\left(\mb{h}^{(k)},\mb{e}^{(k)}\right) = \sum_{j=0}^{M-1} \underbrace{ \left(\left(e_j^{(k)}\right)^2 + \epsilon\right)^{p/2} -
\left(\left(e_j^{(k)}\right)^2\right)^{p/2}}_{g_j^{(k)}(\epsilon)}.
\label{eps_orig_cost}
\end{eqnarray}
The function ${g_j^{(k)}(\epsilon)}$ is symmetric in $e_j^{(k)}$ and has a maximum value of $\epsilon^{p/2}$ at $e_j^{(k)} = 0$, and a minimum value of $0$ as $\left|e_j^{(k)}\right| \rightarrow \infty$. Therefore, $0 \le g_j^{(k)}(\epsilon) \le \epsilon^{p/2}$, which leads to the inequality
$0<F_{\epsilon}\left(\mb{h}^{(k)},\mb{e}^{(k)}\right) - F\left(\mb{h}^{(k)},\mb{e}^{(k)}\right) \le  M\epsilon^{p/2}.$
\end{proof}
\\
\indent For illustration, consider a synthetic signal $y(n)$, which is the output of a filter excited by the sparse sequence of length 200 samples with Kronecker impulses at some randomly selected indices $[10, 62, 85, 100, 150, 182]$. The impulse response of the filter is chosen to be a sum of exponentially decaying sinusoids: $\displaystyle h(n) = \sum \limits_{k=1}^{3} e^{-\alpha_k n} \cos(\omega_k n) \, u(n), \,\, n =1, 2, \dots, 100; [\alpha_1, \alpha_2, \alpha_3] = [0.01, 0.014, 0.025],$ and $[\omega_1, \omega_2, \omega_3] = [0.075, 0.138, 0.375]$. The observed signal $y(n)$ was deconvolved using ALPA with parameters chosen as $\lambda = 1$, $p = 0.1$, and $\epsilon = 10^{-6}$. The plots of the resulting cost functions $F$ and $F_\epsilon$ are shown in Figure~\ref{cost_function_compare}. Observe that the surrogate cost $F_\epsilon$ is non-increasing and upper-bounds $F$, which is not monotonic in general --- this is an experimental confirmation of the result established in Lemma~\ref{lemma3}. The local variations in $F$ are bounded as indicated by the following proposition.
\begin{propos}
\label{proposition2}
The ALPA, which minimizes $F_{\epsilon}(\mb{h},\mb{e})$, leads to a sequence $\quad \quad$ $F\!\left(\mb{h}^{(k)},\mb{e}^{(k)}\right)$ such that
 \begin{equation*}
 F\left(\mb{h}^{(k+1)},\mb{e}^{(k+1)}\right)  - F\left(\mb{h}^{(k)},\mb{e}^{(k)}\right) \le M\epsilon^{p/2}.
 \end{equation*} 
 \end{propos}
\begin{proof}
Using Lemma~\ref{lemma3} and (\ref{eps_orig_cost}), we get that 
\[F\left(\mb{h}^{(k+1)},\mb{e}^{(k+1)}\right) < F\left(\mb{h}^{(k)},\mb{e}^{(k)}\right)
+ \underbrace{\sum_{j=0}^{M-1}\left(g_j^{(k)}(\epsilon) - g_j^{(k+1)}(\epsilon)\right)}_{\psi_{k+1}(\epsilon)}.\] 
The function $\psi_{k+1}(\epsilon)$ is continuous in $\epsilon$ and $\text{lim}_{\epsilon \rightarrow 0^{+}}\psi_{k+1}(\epsilon) = 0$. Also, from the proof of Proposition~\ref{proposition1}, both $g_j^{(k)}(\epsilon)$ and {$g_{j}^{(k+1)}(\epsilon)$} are bounded as $0 \le g_j^{(k)}(\epsilon) \text{, } g_j^{(k+1)}(\epsilon) \le \epsilon^{p/2}.$ Hence, $\psi_{k+1} \leq \left|\psi_{k+1}\right| \le M\epsilon^{p/2}$, which establishes the result.
\end{proof}

\section{Regularized Least-Squares Initialization and Concentration of Error}
\label{prob_guarantees}
\indent Since the cost $F_\epsilon(\h^{(k)},\e)$ in the \emph{e-step} has local minima, initialization becomes important. In the blind case, the algorithm is initialized with a filter that we denote by $\htil$. The weight matrix is initialized to
$\mb{W}^{(0,0)} = \mb{I}_{M \times M}$, which results in the regularized LS estimate for the excitation. The error between the estimate and the true excitation depends on measurement noise. In the following analysis, we quantify the \emph{concentration} of the estimate about the true value. The mean-absolute error (MAE) between $\mb{x} \in \mathbb{R}^{M}$ and its estimate $\hat{\mb{x}}$ is defined as 
\begin{align}
\text{MAE}=\frac{1}{M} \|\mb{x}-\mb{\hat{x}}\|_1.
\label{maemse}
\end{align}

\indent Consider $\tilde{\mb{{h}}}$, an estimate of $\mb{h}^*$, which has the error $\Delta \mb{h} =
\mb{h}^*-\tilde{\mb{h}}$. The convolution matrix constructed from $\htil$ is given by $\Htil =
\Hs+\dH$, where $\dH$ is the (error) convolution matrix corresponding to $\Delta \h$ and hence the Frobenius norm of $\dH$ is related as $\|\dH\|_F = \sqrt{M}\|\Delta \h\|_2$. Let $\|\dH\|_2$ and $\|\Hs\|_2$ denote the matrix 2-norms -- these equal the largest singular values of the corresponding matrices. Using the matrix-norm equivalence $\|\dH\|_2 \le \|\dH\|_F$, we have $\|\dH\|_2 \le \sqrt{M}\|\Delta \h\|_2$. We define \emph{quasi-condition-number}\footnotemark\footnotetext{Note that the denominator in $\kappa_q(\mb{H})$ is the square of the smallest singular value.} as $ \kappa_q(\mb{H}) = {\sigma_{\text{max}}(\mb{H})}/\left(\sigma^2_{\text{min}}(\mb{H})+\delta\right)$. If we initialize
$\mb{W}^{(0,0)} = \mb{I}_{M \times M}$, then we get the reg. LS solution $\ebrls = (\tilde{\mb{H}}^{\text{\sc T}} \tilde{\mb{H}} + \delta \mb{I})^{-1} \tilde{\mb{H}}^{\text{\sc T}} \y$, which satisfies the equation: $(\Htil^\text{\sc T} \Htil + \delta \mb{I}) \ebls = \Htil^\text{\sc T} \y$. In terms of the ground-truth quantities $\Hs$ and $\es$, we can write
\begin{eqnarray}
\left((\Hs+ \dH)^\text{\sc T}(\Hs+ \dH) + \delta \mb{I}\right)(\es+\dbrls) = (\Hs+\dH)^\text{\sc T} \left(\mb{\Hs \es + \w}\right),
\label{dbrls_expand}
\end{eqnarray}
where $\dbrls = \es - \ebrls$ is the estimation error. An upper bound on MAE can be derived from \eqref{dbrls_expand} by rearranging the terms and employing properties of norms such as the triangle inequality and compatibility of norms. A detailed calculation is presented in Appendix~\ref{Appendix_BRLS_expand}. The final expression for the bound turns out to be
\begin{align}
\frac{1}{M}\|\dbrls\|_1 \le& \frac{1}{\sqrt{M}\left(1\!-\!2C_{\Delta \h}\right)}\Bigg(\left(\kappa_q(\Hs) \!+\! C_{\Delta \h} \right)  \|\w\|_2 \!+\! \left(\delta \!+\!C_{\Delta \h}  \right)\|\es\|_2 \Bigg).
\label{dbrls_ineq}
\end{align}
The MAE between $\ebrls$ and $\e^*$ is concentrated as follows.

\begin{propos}
\label{BRLS_prop}
Let $\htil$ be an estimate of $\hs$ such that $\|\htil-\hs\|_2 = \|\dsh\|_2 < 1/(2\sqrt{M}\kappa_q(\Hs))$ and let the reg. LS estimate of $\es$ be denoted as $\ebrls = (\Htil^\text{\sc T} \Htil + \delta \mb{I})^{-1} \Htil^\tr \y$. The MAE defined as $\frac{1}{M} \|\es-\ebrls\|_1 = \frac{1}{M} \|\dbrls\|_1$ is upper bounded as in \eqref{dbrls_ineq} and is concentrated as follows
\begin{align}
\mathcal{P}\left(\frac{1}{M}\|\dbrls\|_1 > \xi\right) \le \frac{\sigma^2 (\kappa_q(\Hs)+C_{\Delta \h})^2}{\left(\sqrt{M} (1-2C_{\Delta\h})\xi - \left(\delta + C_{\Delta \h}\right)\|\es\|_2\right)^2},
\label{Markov_upperbound_bd}
\end{align}
where $C_{\Delta\h} = \sqrt{M} \kappa_q(\Hs) \|\dsh\|_2$.
\end{propos}
\indent The proof is given in Appendix~\ref{appendix_BRLS_prop} and is a consequence of the Markov inequality.\\
\indent If the bound in \eqref{Markov_upperbound_bd} exceeds unity, it becomes trivial and noninformative. A nontrivial bound (i.e., a bound lesser than unity) is obtained when
\begin{align}
\|\dsh\|_2 \le \left(\frac{\xi/(\sigma \kappa_q(\Hs))-1}{(\sigma + 2) \sqrt{M} + \|\es\|_2 }\right), 
\end{align}
with $\xi \ge \sigma\kappa_q(\Hs).$\\     
\indent The bound in \eqref{Markov_upperbound_bd} requires that the first- and second-order moments of noise be finite. In addition, if the noise is bounded, one can provide sharper bounds using the Hoeffding inequality stated below (recalled from Ch.2, pp. 34 of \cite{lugosi}).
\begin{theorem} (Hoeffding's inequality): Suppose that $X_1, X_2,\dots, X_M$ are $M$ independent random variables with $\exptn{X_i} = \mu_i$ and $\mathcal{P}\left(X_i \in [a_i,b_i]\right) = 1$ with $a_i,b_i \in \mathbb{R}$, then
\begin{equation}
\mathcal{P}\left(\frac{1}{M}\sum_{i = 1}^{M} (X_i-\mu_i) > \xi\right) \le \exp{\left(-\frac{2M^2\,\xi^2}{\sum_{i=1}^{M}(a_i-b_i)^2}\right)}, \quad \forall \xi > 0.
\nonumber
\end{equation}
For the i.i.d. case, $X_i \in \mathcal{B}_{-a,a}$, $\exptn{X_i} = \mu$, the bound gets simplified to
\begin{equation}
\mathcal{P}\left(\frac{1}{M}\sum_{i = 1}^{M} X_i-\mu > \xi\right) \le \exp{\left(-\frac{M\xi^2}{2a^2}\right)}, \quad \forall \xi > 0.
\label{Hoeffding_bound}
\end{equation}
\end{theorem}

Applying the Hoeffding inequality gives rise to the following result.
\begin{propos}
\label{BRLS_prop_hoeff}
Let $\w$ be an i.i.d. random vector with $w_i \in \mathcal{B}_{-a,a}$, and $\exptn{w_i^2} = \sigma^2$. Then, for $\xi>0$, the average error $\frac{1}{M}\|\dbrls\|_1$ is concentrated as
\begin{align}
\mathcal{P}\Bigg(\frac{1}{M}\|\dbrls\|_1 \!&>\! \xi \Bigg)  \nonumber\\
&\le \exp\left(\frac{-M}{2a^2}
\!\left(\!\!\left(\frac{\sqrt{M}(1\!-\!2 C_{\Delta \h}) \xi \!-\! \left(\delta \!+\! C_{\Delta \h}\right)\|\es\|_2}{\kappa_q(\Hs)+C_{\Delta \h}}\right)^2\!\!\!-\!\sigma^2\right)^2\right)\!.
\label{prop_BRLS_bound_hoeff}
\end{align}
\end{propos}
\begin{proof}
Consider \eqref{Markov_for_w} and let $$t = \frac{1}{M}\Bigg(\frac{\sqrt{M}(1-2C_{\Delta\h}) \xi - \left(\delta + C_{\Delta \h}\right)\|\es\|_2}{\kappa_q(\Hs)+ C_{\Delta \h}}\Bigg)^2 - \mu.$$
Employing the Hoeffding bound gives
\begin{equation*}
\mathcal{P}\left(\frac{1}{M}\|\w\|^2_2  - \mu >  t \right) \le \exp{\left(\frac{Mt^2}{2a^2}\right)},
\end{equation*}
from which \eqref{prop_BRLS_bound_hoeff} follows.
\end{proof}\\
\indent To gain some insights into the bound, consider the simpler case of non-blind deconvolution without regularization (i.e., $C_{\Delta\h} = 0, \delta = 0$), corresponding to which the error in excitation is denoted as $\dls$
\begin{align}
&\mathcal{P}\left(\frac{1}{M}\|\dls\|_1 \!>\! \xi \right)  \le \exp\left(-\frac{M^3}{2\,a^2}\left(\frac{\xi^2}{\kappa_q(\Hs)^2}-\frac{\sigma^2}{M}\right)^2\right).
\end{align}
For $\xi = n\sigma, n>0$, we get
\begin{align}
&\mathcal{P}\left(\frac{1}{M}\|\dls\|_1 \!>\! n\sigma \right)  \le \exp\left(-\frac{M^3\sigma^4}{2\,a^2}\left(\frac{n^2}{\kappa_q(\Hs)^2}-\frac{1}{M}\right)^2\right).
\end{align}
For a random variable bounded over $[-a, a]$, the maximum variance is $\sigma^2=a^2$ \cite{lugosi}. Taking this into account, the worst-case dependence on noise variance is expressed as
\begin{align}
\mathcal{P}\left(\frac{1}{M}\|\dls\|_1 \!>\! na \right)  &\le \exp\left(-\frac{M^3\,a^2}{2}\left(\frac{n^2}{\kappa_q(\Hs)^2}-\frac{1}{M}\right)^2\right).
\end{align}

\section{Application to Speech Deconvolution}
\label{voiced_speech}
\indent We next consider an application of the ALPA deconvolution technique to speech signals. Considering the LSI model for speech production \cite{rabiner,Makhoul}, a speech signal $y(n)$ can be expressed as a convolution of the vocal-tract impulse response $h(n)$ and excitation $e(n)$, as follows:
\begin{equation}
y(n) = (h \ast e)(n).
\label{convo}
\end{equation}
Depending on whether the speech segment is voiced or unvoiced, the excitation $e(n)$ is assumed to be a quasi-periodic impulse train or white noise, respectively  \cite{Makhoul}. In the case of voiced sounds, the impulses are placed at the instants of significant excitation (referred to as {\it epochs} or glottal closure instants (GCIs)) \cite{Yegna,Drugman_epoch_review,Ravi_epoch}, which depends on the pitch of the speaker. The vocal-tract configuration in producing a certain sound determines the frequency response of the filter. The vocal-tract impulse response is a convolution of exponentially decaying sinusoids, one corresponding to each resonance, and the excitation is sparse with one impulse per pitch cycle. The estimation of $h(n)$ and $e(n)$, given $y(n)$, is the problem of blind deconvolution and has widespread applications in speech analysis, coding, and recognition \cite{rabiner}. The {\it de facto} standard for speech deconvolution is based on linear prediction (LP), which relies on an autoregressive model for the vocal-tract filter, whose coefficients are estimated by minimizing the $\ell_2$-norm of the prediction error (also known as the {\it residue}) \cite{Makhoul}. The residue has embedded in it information about epochs and pitch of the speaker \cite{Anantha, LPC1,Yegna}.

\begin{figure}[t]
\centering
$
\begin{array}{ccc}
\subfigure[]{
\includegraphics[width=.3\linewidth]{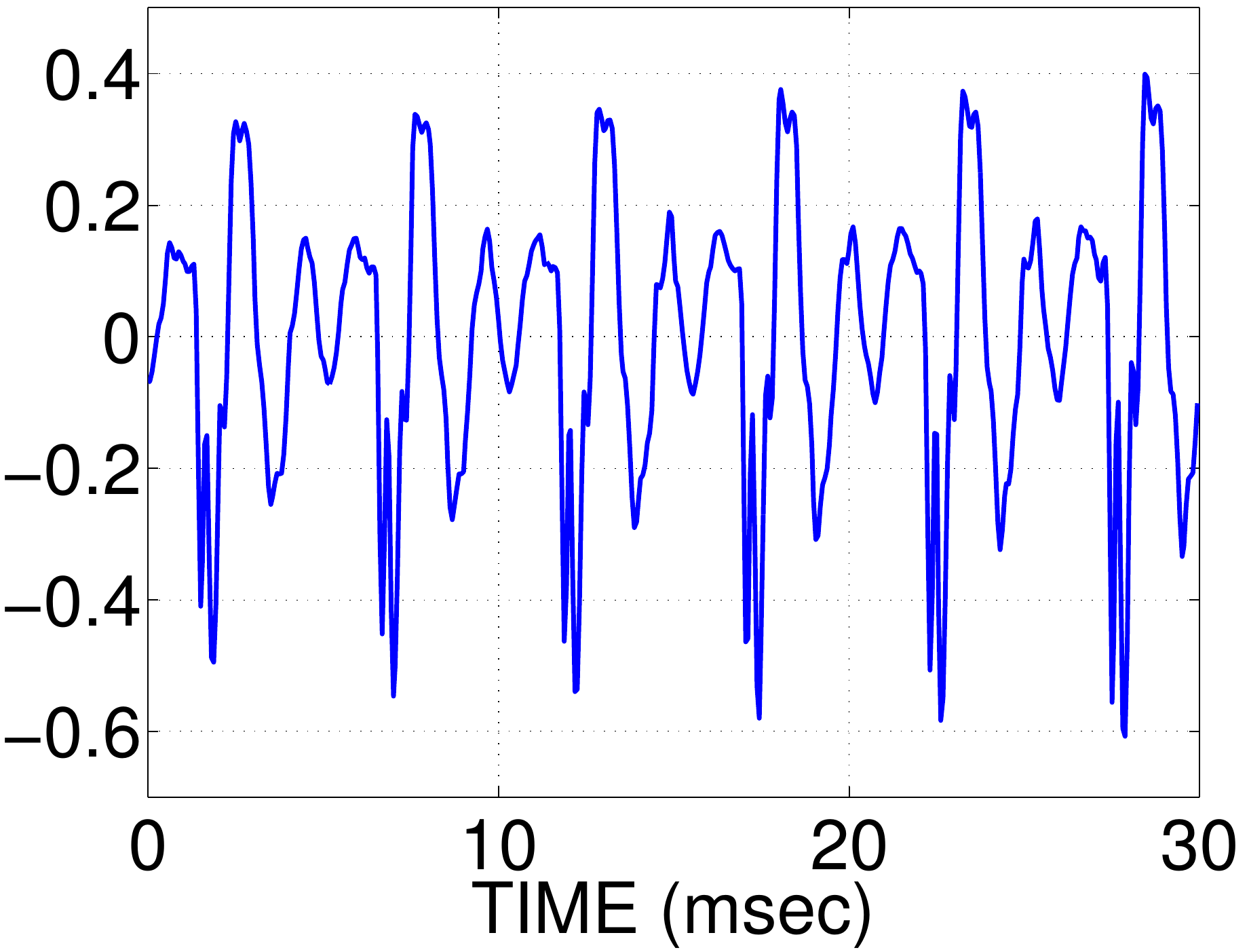}
\label{fig:Speech_orig_sig_clean}
}
\subfigure[]{
        \includegraphics[width=.3\linewidth]{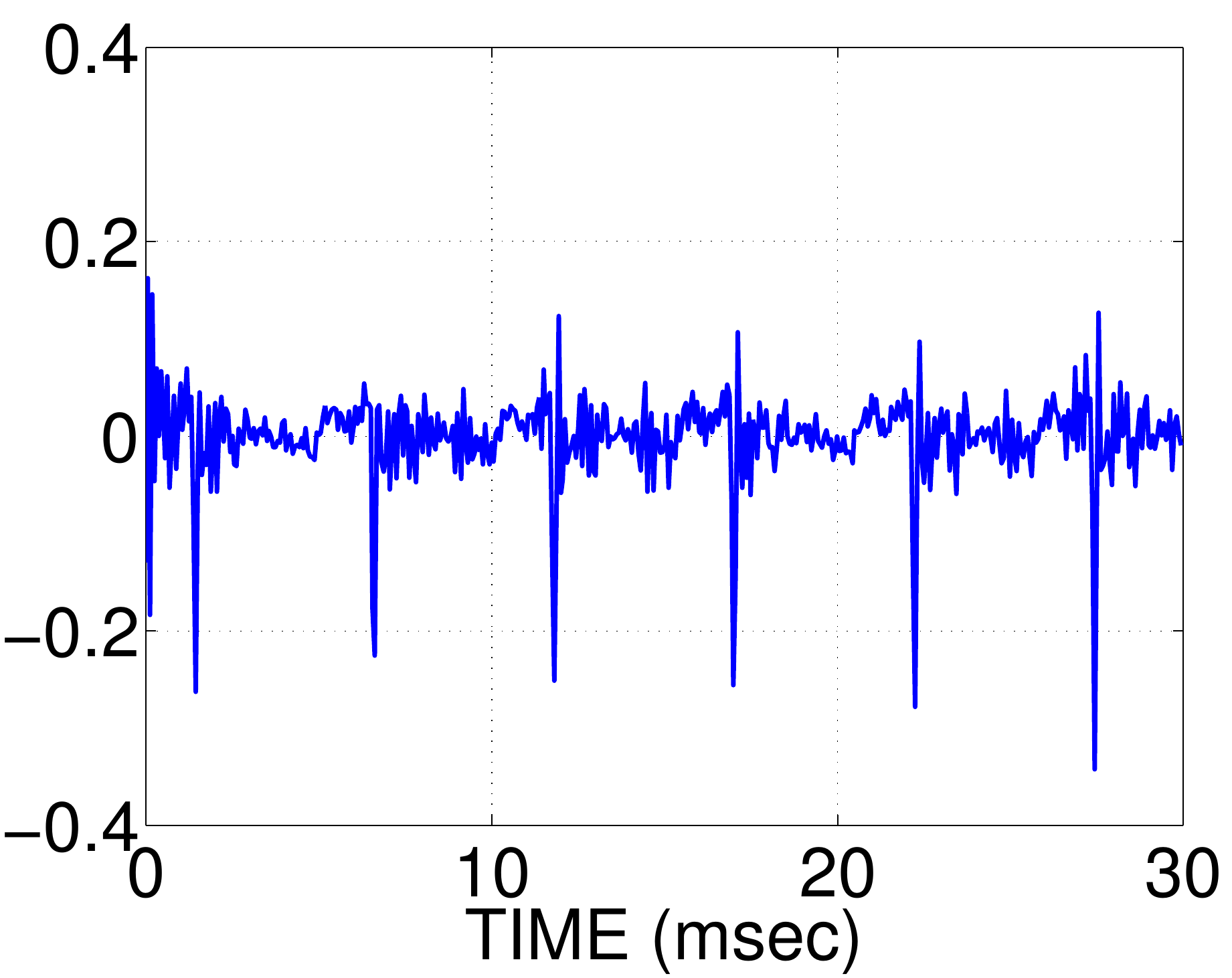}
        \label{fig:Speech_LPres_clean}
        } 
\subfigure[]{
                
\includegraphics[width=.3\linewidth]{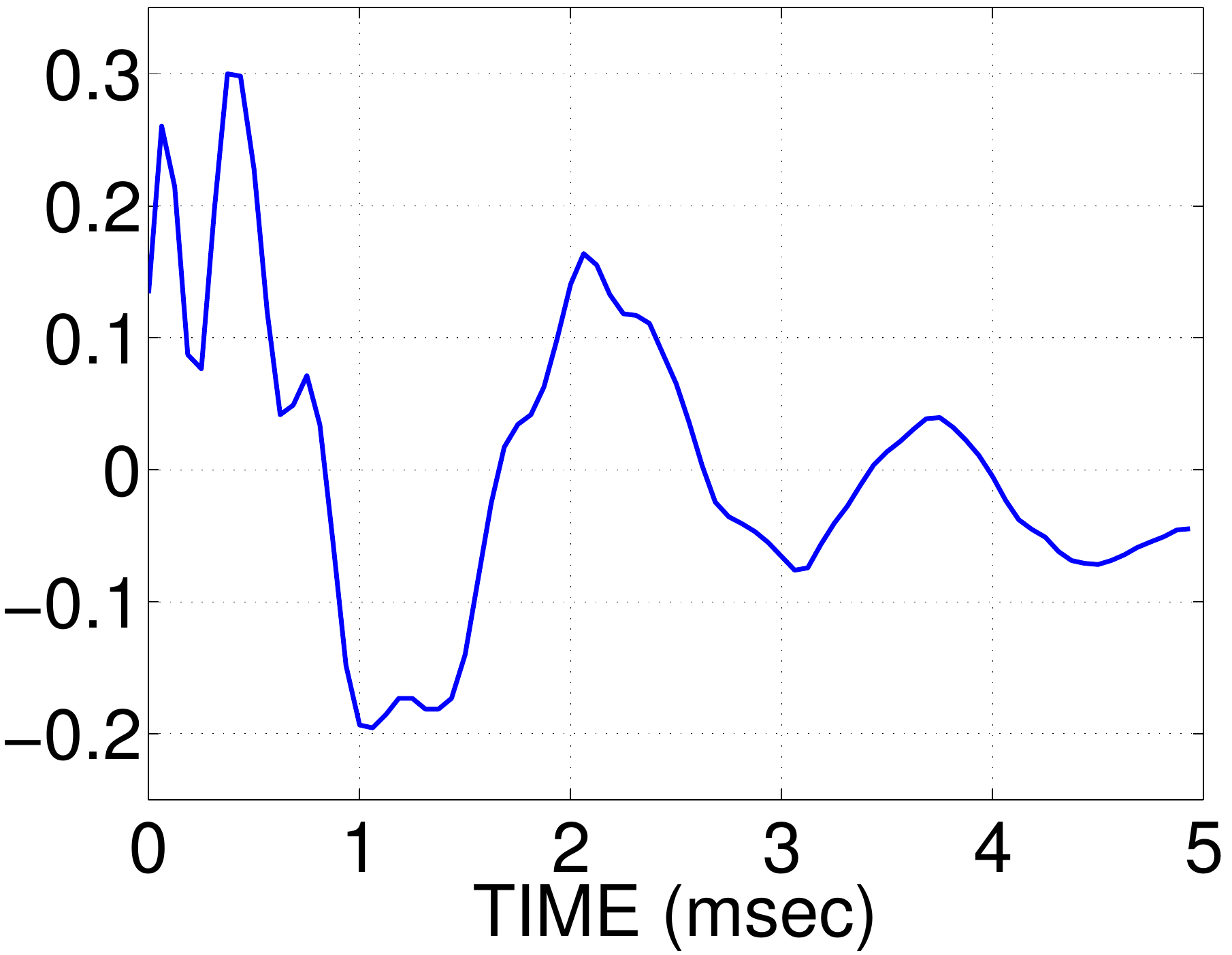}
                \label{fig:Speech_est_filt_clean}
                }\\
                \subfigure[]{
                \includegraphics[width=.3\linewidth]{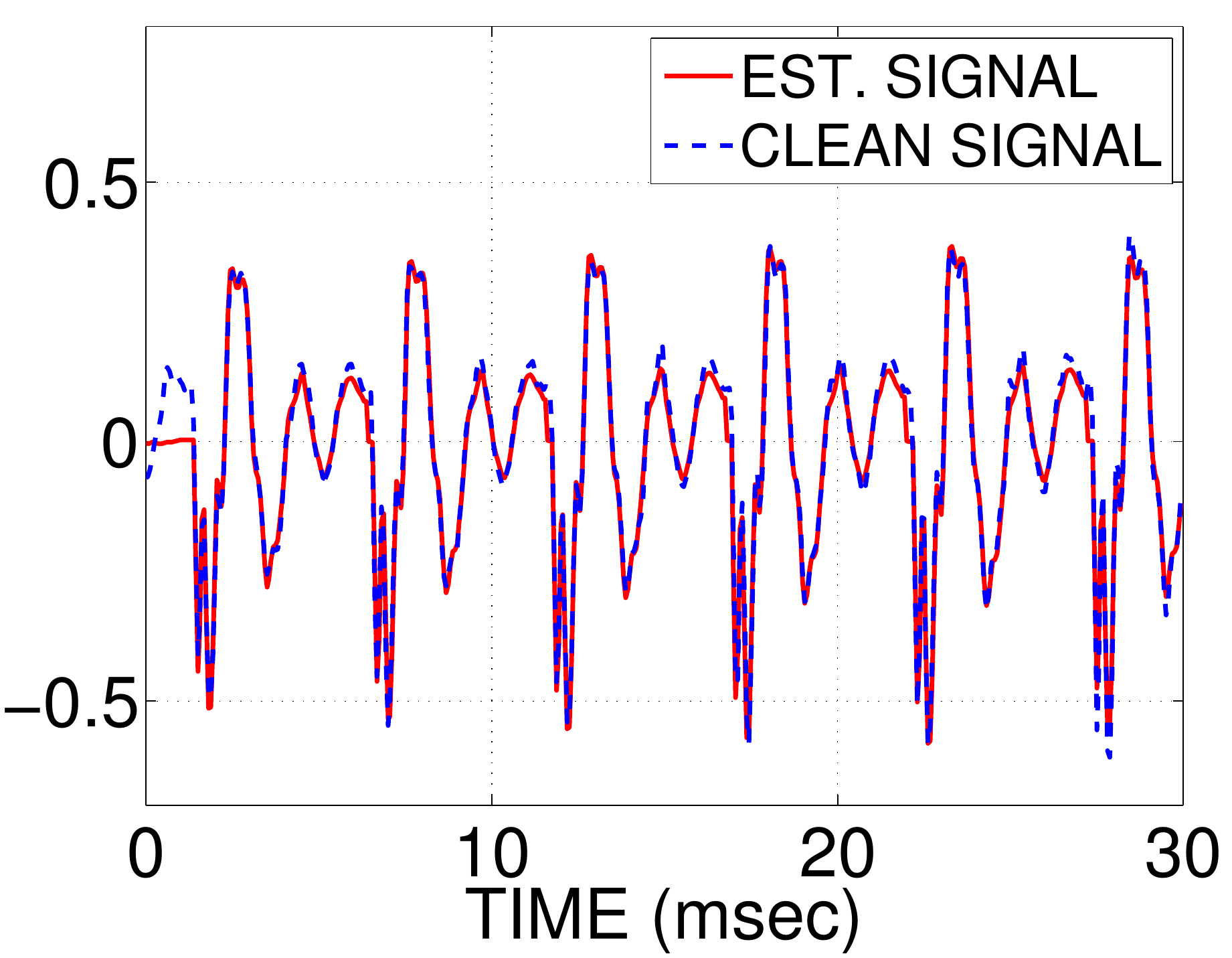}
                                \label{fig:Speech_est_sig_clean}
                                }
\subfigure[]{
\includegraphics[width=.3\linewidth]{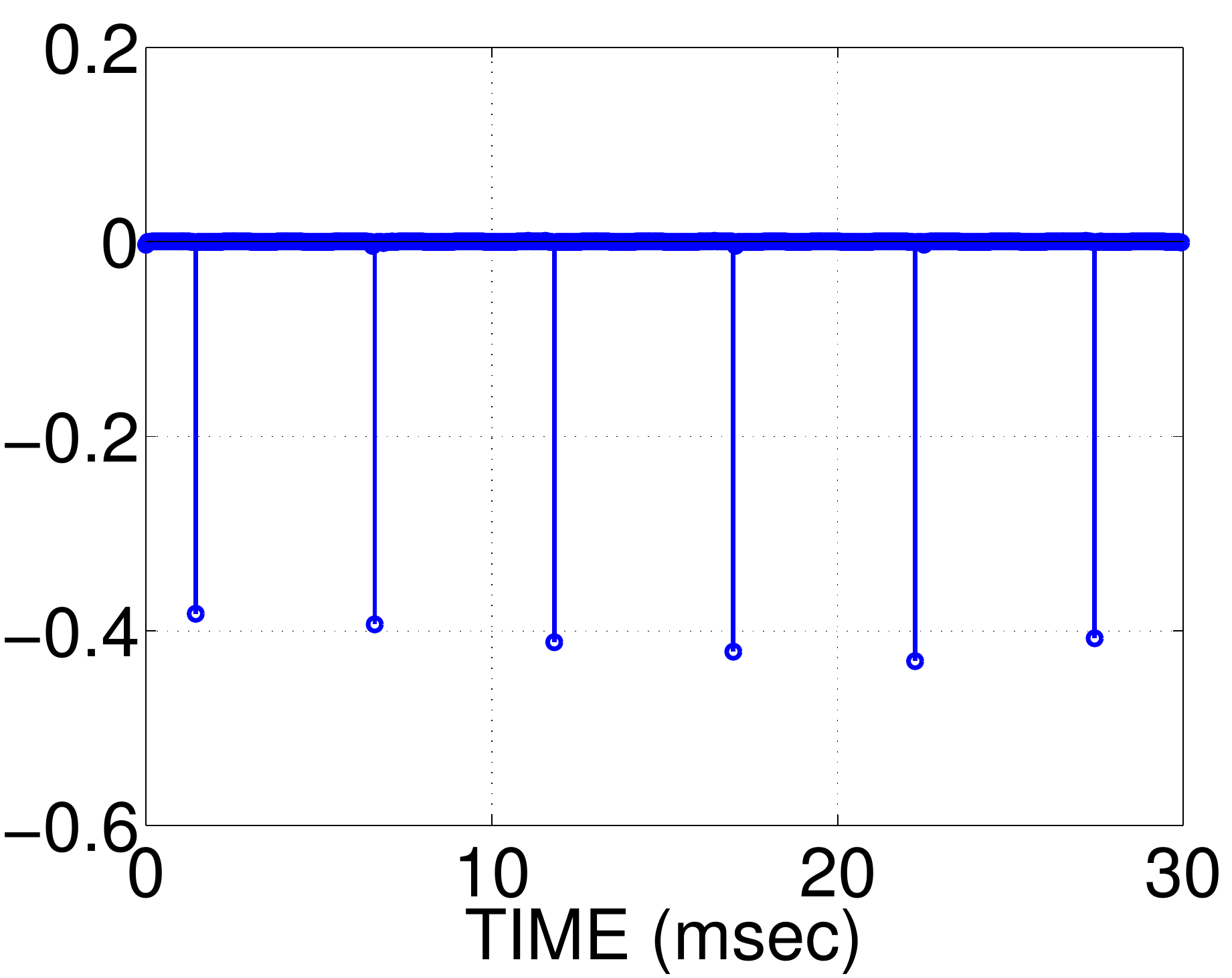}
\label{fig:Speech_est_exc_clean}
}
\subfigure[]{
        \includegraphics[width=.3\linewidth]{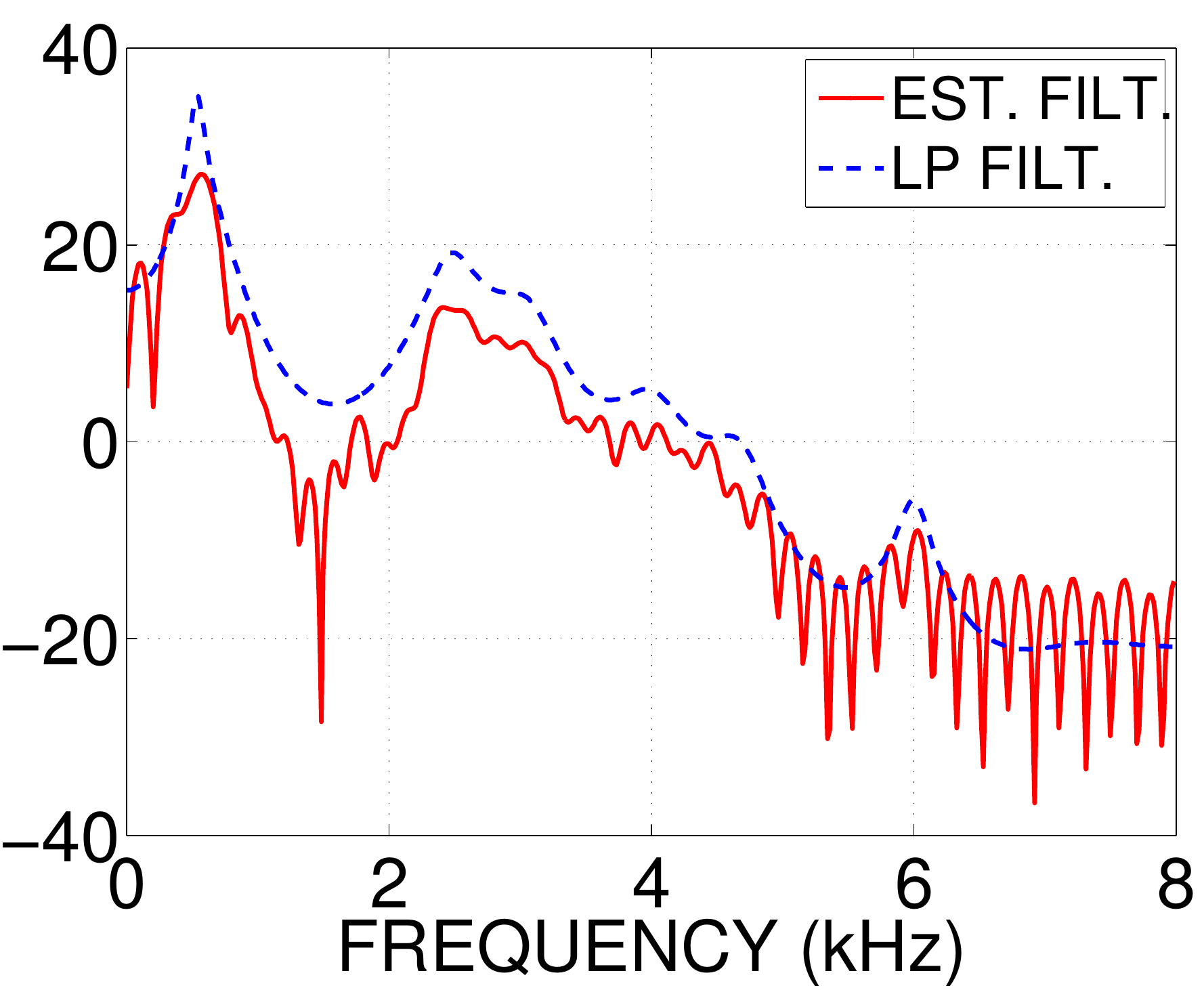}
        \label{fig:Speech_filtresp_clean}
        }
\end{array}
$
\caption{(Color online) (a) A voiced segment /\ae/ (female speaker) of length 480 samples (sampling rate 16 kHz); (b) LP residue (model order 20); (c) ALPA
estimate of the filter; (d) comparison between the original vowel segment shown in (a) and that synthesized based on the estimated filter and excitation; (e) ALPA estimate of the excitation; and (f) frequency response of the estimated filter.}
\label{results_female_clean}
\end{figure}

\begin{figure}[t]
\centering
$
\begin{array}{cc}
\subfigure[]{
\includegraphics[width=.3\linewidth]{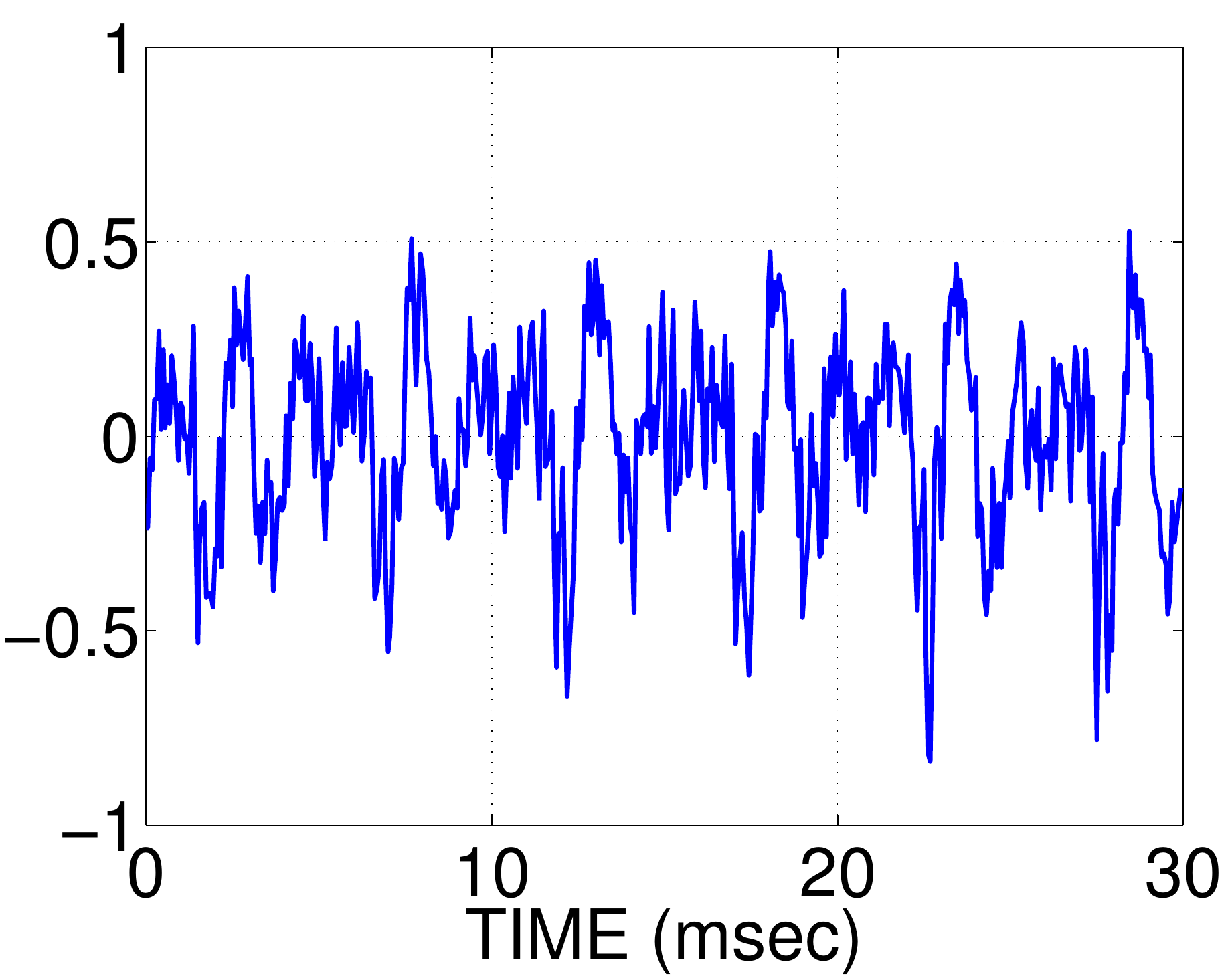}
\label{fig:Speech_orig_sig_5db}
}
\subfigure[]{
        \includegraphics[width=.3\linewidth]{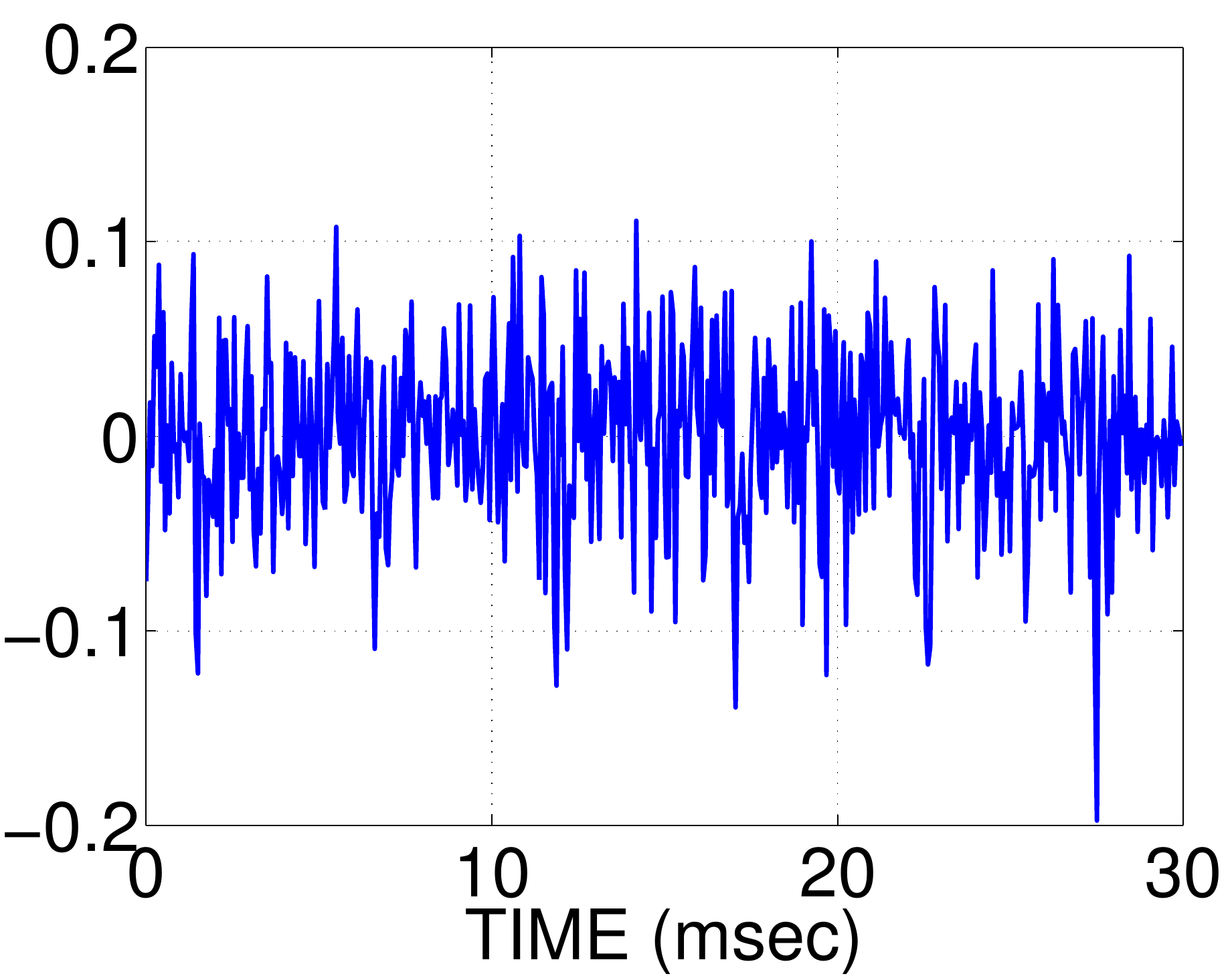}
        \label{fig:Speech_LPres_5db} 
}
\subfigure[]{
        \includegraphics[width=.3\linewidth]{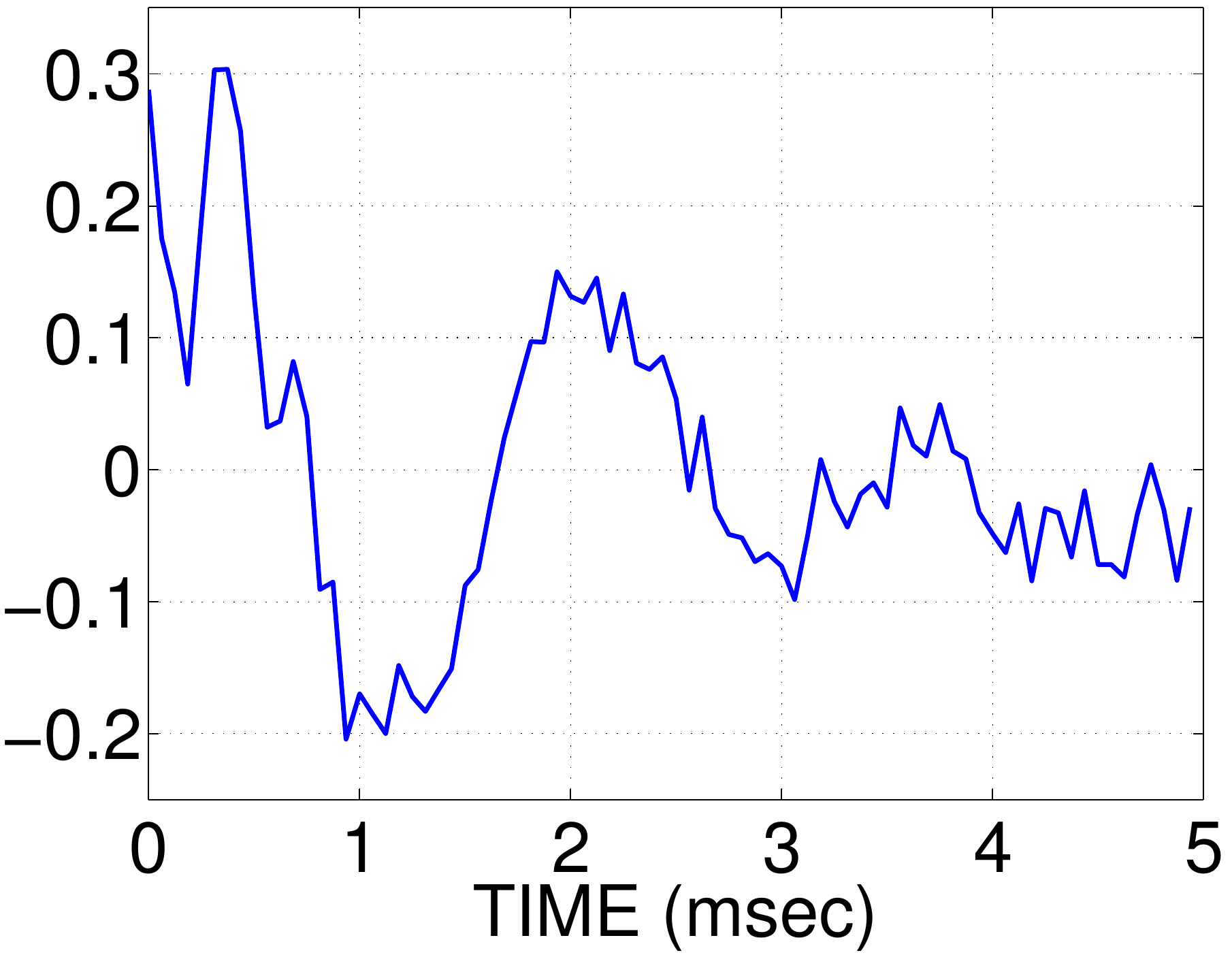}
        \label{fig:Speech_est_filt_5db}
        }\\
          \subfigure[]{
        \includegraphics[width=.3\linewidth]{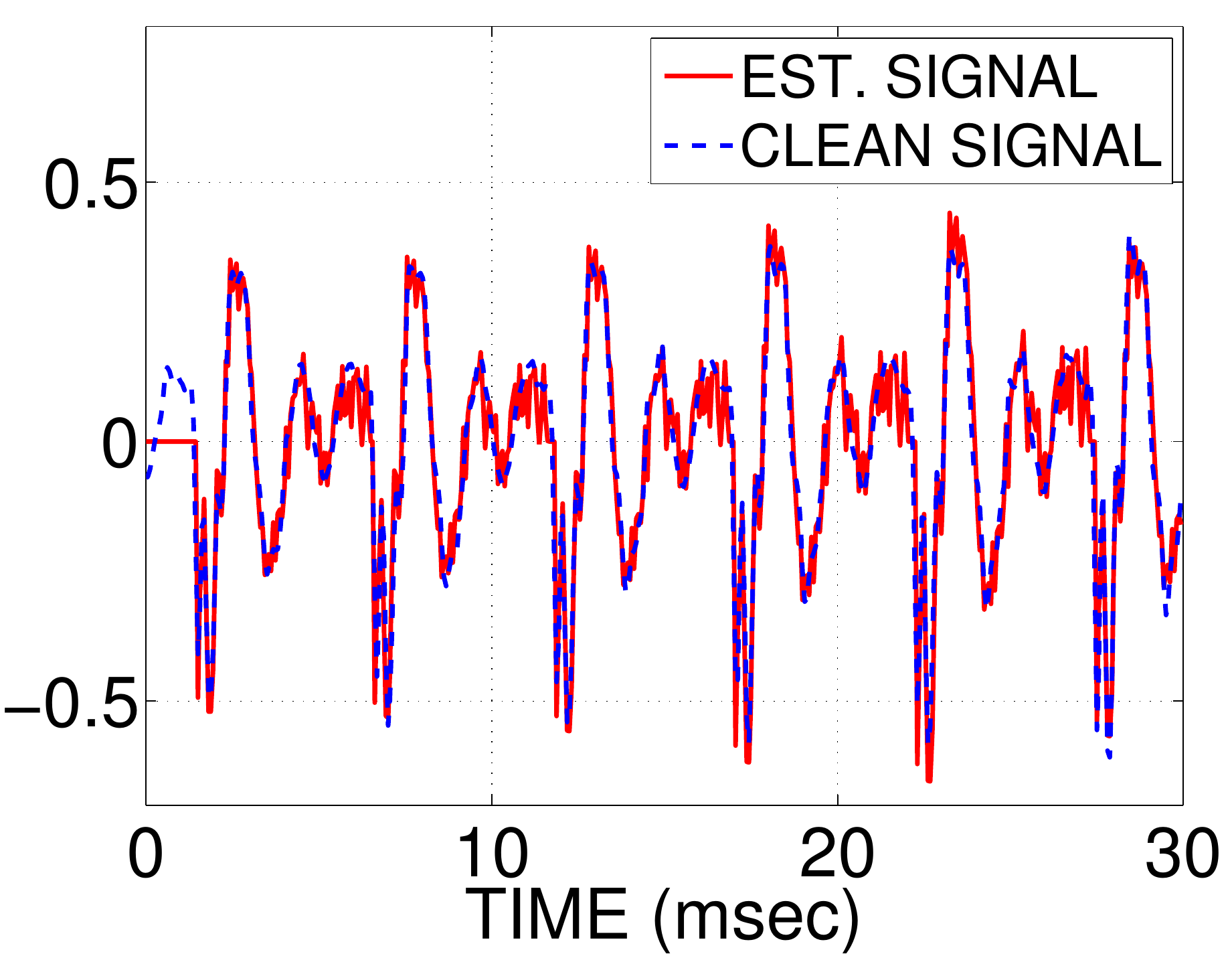}
                        \label{fig:Speech_est_sig_5db}
                        }
\subfigure[]{
\includegraphics[width=.3\linewidth]{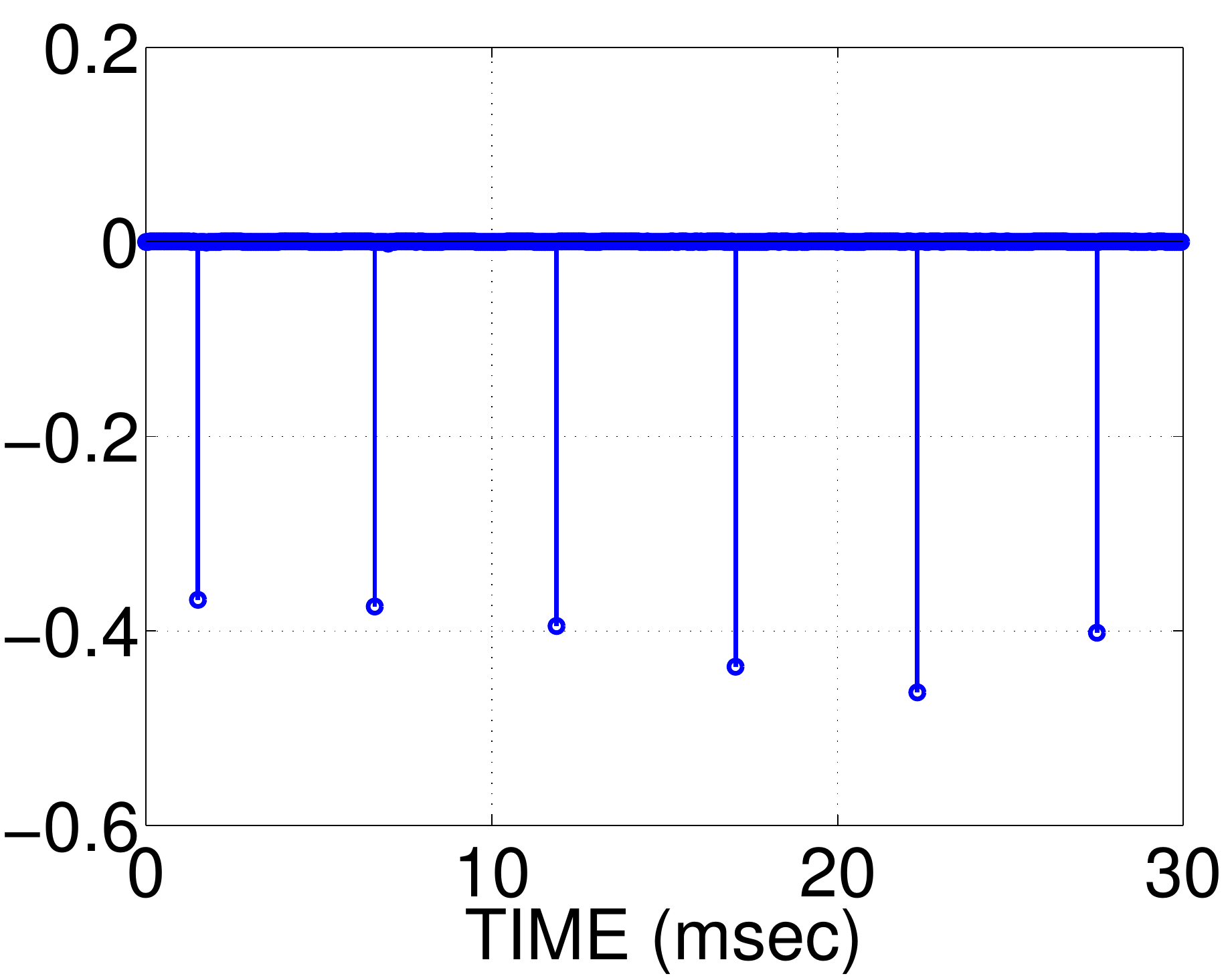}
\label{fig:Speech_est_exc_5db}
}
\subfigure[]{
        \includegraphics[width=.3\linewidth]{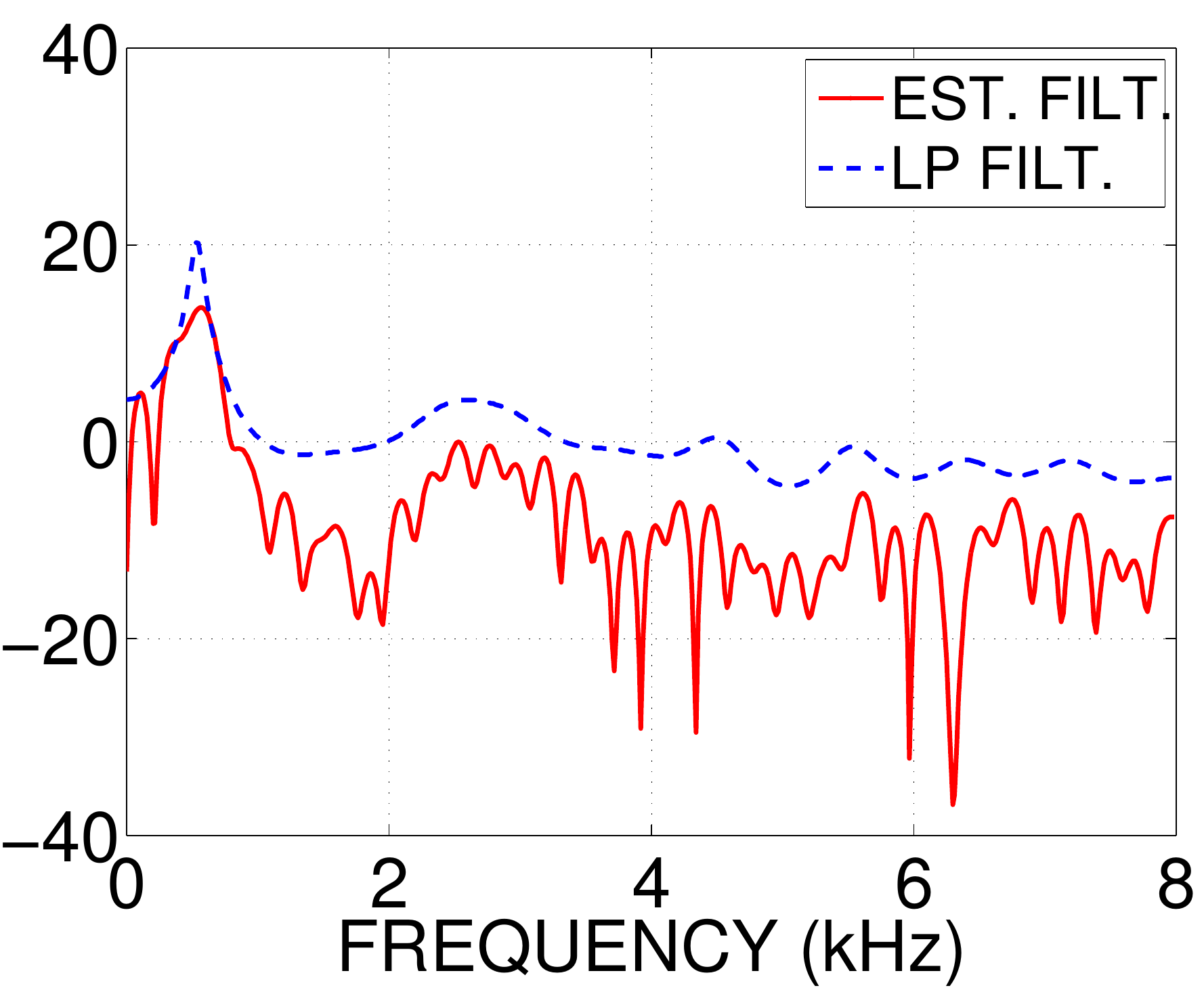}
        \label{fig:Speech_filtresp_5db}
        }
\end{array}
$
\caption{(Color online) (a) A voiced segment /\ae/ (female speaker) of length 480 samples (sampling rate 16 kHz, SNR = 5 dB); (b) LP residue (model order 20);
(c) ALPA estimate of the filter; (d) comparison between the original clean vowel segment shown in Fig~\ref{fig:Speech_orig_sig_clean} and that synthesized based on the estimated filter and excitation. The improvement in SNR is 4.5 dB. (e) ALPA estimate of the excitation; and (f) frequency response of the estimated filter.}
\label{results_female_5db}
\end{figure}

\indent Following the convolutional matrix notation established in Section~\ref{sec:sufficient_cond}, a vectorial representation of \eqref{convo} is given by
\begin{equation}
\mb{y} = \mb{H}\mb{e} = \mb{E}\mb{h},
\end{equation}
where $\mb{y} \in \mathbb{R}^N$, $\mb{h} \in \mathbb{R}^L$, and $\mb{e} \in \mathbb{R}^M$ are the speech, vocal-tract filter, and glottal excitation vectors, respectively. We employ ALPA to deconvolve the filter and the excitation (both of which are unknown) from the speech signal.\\
\indent For experiments, we use speech utterances from the Western Michigan University database \cite{Vowel_WMD} (sampling frequency of 16 kHz). We excised a 30 ms long
vowel segment corresponding to /\ae/ from the utterance ``had,'' of a female speaker. ALPA was initialized with the LP filter and the excitation
corresponding to model order 20. The parameters $\lambda$ and $p$ were set to $1$ and $0.1$, respectively, based on experimentation. The stopping criterion was typically met in about 20 iterations. The deconvolution results {\it
vis-\`a-vis} LP estimates are shown in Figure~\ref{results_female_clean}. The excitation estimated by ALPA is sparse and quasi-periodic as opposed to the standard LP, which does not have a sparsity promoting regularizer. Even in the presence of AWGN (5 dB signal-to-noise ratio (SNR\footnotemark\footnotetext{For the model $\y = \x + \w$, where $\x$ denotes the signal (in $\mathbb{R}^N$) and $\w \sim \mathcal{N}(0,\sigma_w^2\,\mathbf{I})$, the SNR is defined as $\text{SNR} = 10\,\log_{10}\left(\frac{\|\x\|^2}{N\sigma_w^2}\right)$ dB.}), ALPA is robust (cf. Figure~\ref{results_female_5db}) and results in a sparse excitation (compare Figure~\ref{fig:Speech_est_exc_5db} with Figure~\ref{fig:Speech_est_exc_clean}, in particular). The estimated filter was found to contain some noise, but lower than that present in the signal. The
signal synthesized using the estimated excitation and the filter gave an SNR improvement of 4.5 dB, which shows that ALPA performs implicit denoising, which is an important feature of sparsity promoting formulations.\\
\indent The excitation estimated by ALPA (cf. Figures~\ref{fig:Speech_est_exc_clean} and \ref{fig:Speech_est_exc_5db}) is sparser than that estimated by LP (cf. Figures~\ref{fig:Speech_LPres_clean} and \ref{fig:Speech_LPres_5db}). Although a sparse excitation model is used to motivate the LP formulation, the estimated excitation does not actually turn out to be sparse, primarily because the standard LP formulation does not explicitly incorporate any sparsity promoting constraints. This drawback was recently overcome by Giacobello et al. \cite{gia} who developed sparse LP. We shall next compare against this technique as well as the other sparse deconvolution techniques reported in the literature.


\subsection{Comparisons With Sparse Deconvolution Methods}
\label{sec:compare}
\indent We next compare the performance of ALPA with a recently
proposed smoothed one-over-two norm (SOOT)
penalty-based blind deconvolution algorithm \cite{Euclid_taxicab_repetti}, and the {\it sparse linear prediction} (SLP) technique \cite{gia}. Further,
we compare the sparse excitation estimated by ALPA with that obtained using the MM-based sparse deconvolution algorithm (SDMM) \cite{SDMM}, which is a
non-blind deconvolution algorithm. We briefly review the three techniques, before reporting performance comparisons.

\subsubsection{The SOOT algorithm} The \loot function, which is a ratio of the $\ell_1$ and $\ell_2$ norms, is scale-invariant and has been employed as a sparsity-promoting prior in the blind deconvolution of natural images \cite{l1l2_krishnan}. However, since \loot is non-convex and non-smooth, the blind deconvolution problem becomes difficult to solve. Repetti \etal \cite{Euclid_taxicab_repetti} incorporated  a smooth approximation of the \loot function (smoothed one over two (SOOT)), which is the logarithm of the quotient of the smoothed $\ell_1$ and $\ell_2$ norms. An Alt. Min. approach combined with proximal methods is employed to optimize the cost function. In each step, Repetti \etal perform quadratic majorization
of the smooth, non-convex cost function and minimize it using weighted proximal operators. The algorithm has been developed in the context of blind deconvolution of seismic signals and a MATLAB toolbox has been provided by the authors.
\subsubsection{Sparse linear prediction (SLP)}
\indent Giacobello \etal \cite{gia} introduced sparsity constraints within the LP framework. For a speech
segment of length $N$, the model is represented in matrix form as $\mathbf{y} = \mathbf{Ya+r},$ where
\begin{equation*}
\mathbf{y} = \begin{bmatrix}
        y(N_1)\\
        \vdots\\
        y(N_1+N)
\end{bmatrix}, \,\,
\mathbf{Y} = \begin{bmatrix}
y(N_1 - 1) & \cdots & y(N_1 - P)\\
\vdots & \ddots & \vdots\\
y(N_1+N - 1) & \cdots & y(N_1+N - P)
\end{bmatrix},
\end{equation*} 
and $P$ is the order of the predictor. They have proposed multiple formulations that yield either sparse residue or predictor coefficients or both. In particular, an $\ell_p$-norm criterion is considered and the predictor coefficients corresponding to a sparse residue are obtained as: $\mathbf{a}^{*} = \mbox{arg }\underset{\mathbf{a}}{\mbox{min }}
\lim\limits_{p\rightarrow 0}\|\mathbf{y-Ya}\|^p_p.$ The cost is minimized using iteratively reweighted $\ell_1$ minimization technique (IRL1) \cite{candes}, where, in the $k^{\text{th}}$ iteration, the predictor coefficient vector is $\mathbf{a}^{(k)} = \mbox{arg }\underset{\mathbf{a}}{\mbox{min
}} \|\mb{W}^{(k-1)}\left(\mathbf{y-Ya}\right)\|_1,$ where $\mb{W}^{(k-1)} = \text{diag}(|\mb{y-Y}\mb{a}^{(k-1)}|+0.01)^{-1}$ (cf. Algorithm~1 in \cite{gia}). Typically, the IRL1 convergence criterion was met in five iterations.

\subsubsection{MM-based sparse deconvolution (SDMM)}
\indent In SDMM \cite{SDMM}, one assumes that the speech signal $\y$ is the output of an LP filter (finite-length approximation $\h$), excited by a sparse sequence $\e$. The
excitation is obtained as a solution to the LASSO: 
\begin{equation}
\e^* = \argmin{\e} \|\y - \mb{H} \e\|_2^2 + \delta \|\e\|_1,
\label{LASSO}
\end{equation}
computed using the IRLS approach, wherein the update utilizes the banded structure of the convolution matrix $\mb{H}$ for efficient matrix inversion. Based on the optimality criterion satisfied by the minimizer of \eqref{LASSO}, a lower bound on the regularization parameter was derived in \cite{SDMM, Selesnick} as $\delta \ge 3 \sigma \|\mb{h}\|_2$, where $\sigma$ is the noise variance.  

\subsubsection{Deconvolution results}
\indent We generated a synthetic vowel (/\ae/, fundamental frequency F0 = 200 Hz) speech segment (30 ms duration) using standard speech processing software accompanying
\cite{rabiner}. We considered a 100-tap FIR filter, and prediction order 20 for SLP and SDMM. In the SOOT toolbox, Repetti \etal
\cite{Euclid_taxicab_repetti}
consider a fixed set of noise standard deviations (0.01, 0.02, and 0.03) and optimized the regularization parameters accordingly. To ensure a fair
comparison, we used the same noise conditions and parameter settings. The experiments were performed on an iMac with
Intel{\textregistered}~Core$^{\mbox{TM}}$ i5, 3.2 GHz, four-core processor.\\
\indent The MSE and MAE (defined in \eqref{maemse}) computed from the estimates of the excitation, filter, and the signal, averaged over 500 realizations of noise for the three noise variances
considered are provided in Tables~\ref{table_comp_all_MSE} and~\ref{table_comp_all_MAE}. To facilitate comparison, shifts in the estimated excitation and filter are compensated for by
using the cyclic permutation operator. ALPA turned out to be consistently better than the other
techniques in approximating the excitation and the filter. At higher SNR, SDMM is able to estimate the excitation with high accuracy, however, since
no refinement is involved in the LP filter, it is unable to provide a good approximation to the ground-truth filter. A lower MAE in the estimation of the excitation in the case of both ALPA and SDMM indicates that the estimates are sparse and more accurate than SOOT. However, ALPA gives sharper peaks than SDMM. The SLP did not estimate the excitation accurately.\\
\indent The results for a natural vowel segment $/\ae/$ of 30 ms duration uttered by a female speaker under clean and noisy conditions (SNR = 10 dB) are shown
in Figures~\ref{fig:female_comp_deconv_clean} and \ref{fig:female_comp_deconv_noise}. To remove variability due to scale across the techniques and facilitate fair comparison, the excitations shown in Row 3 of Figures~\ref{fig:female_comp_deconv_clean} and \ref{fig:female_comp_deconv_noise} have been rescaled to possess unit energy. \\
\indent We observe that ALPA, SOOT, and SDMM yield sparse excitations in both clean and noisy conditions, with ALPA resulting in the sparsest excitation. The spectral estimates of the
vocal-tract filter are similar for ALPA and SOOT. In the case of SLP, we observe that the excitation is \emph{peakier} than the LP residue, but not as
sparse as what the other techniques give, and especially under noisy conditions, the uncorrelated noise appears in the residue. 
\begin{table}[t]
\footnotesize
\caption{Comparison of MSE in the estimation of excitation, filter, and signal obtained using ALPA, SOOT, SLP, and SDMM.}
\centering
\begin{tabular}{c|c|c|c|c}
\hline  \hline
\multicolumn{2}{  c|  }{Noise standard deviation $\rightarrow$} & 0.01 & 0.02 & 0.03\\
\hline \hline
\multirow{4} {*} {MSE in excitation (dB)} & ALPA & $-$17.4 & $-$10.7 & $-$8.3\\
\cline{2-5}
& SOOT& $-$2.0 & $-$2.1 & $-$2.2\\ \cline{2-5}
& SLP & $-$0.04 & 0.46 & 0.74 \\ \cline{2-5}
& SDMM & $-$22.6 & $-$11.0 & $-$3.7 \\ 
\hline \hline
\multirow{4} {*} {MSE in filter (dB)} & ALPA & $-$15.0 & $-$14.3 & $-$10.0 \\ \cline{2-5}
& SOOT& $-$10.6 &$-$10.0 & $-$8.6 \\ \cline{2-5}
& SLP & $-$11.3 & $-$7.5 & $-$6.3 \\ \cline{2-5}
& SDMM & $-$10.9 & $-$6.0 & $-$4.1 \\ \cline{2-5}
 \hline \hline
\multirow{4} {8em} {MSE in reconstruction (dB)} & ALPA & $-$21.6 & $-$17.5 & $-$14.1 \\ \cline{2-5}
& SOOT& $-$25.0 & $-$19.6 & $-$16.3\\ \cline{2-5}
& SLP & $-$15.6 & $-$12.0  & $-$9.2 \\ \cline{2-5}
& SDMM& $-$12.5 & $-$6.8 & $-$4.6 \\ \hline \hline 
\multirow{4} {8em} {Average time (sec.)} & ALPA & 0.1 & 0.1 & 0.2 \\ \cline{2-5}
& SOOT& 1.3 & 1.3 & 1.3 \\ \cline{2-5}
& SLP & 2.7 & 2.8 & 2.8\\ \cline{2-5}
& SDMM & 0.16 & 0.17 & 0.17 \\ \hline \hline
\end{tabular}
\label{table_comp_all_MSE}
\end{table}
\normalsize
\begin{table}[t]
\footnotesize
\centering
\caption{Comparison of MAE in the estimation of excitation, filter, and signal obtained using ALPA, SOOT, SLP, and SDMM.}
\begin{tabular}{c|c|c|c|c}
\hline  \hline
\multicolumn{2}{  c|  }{Noise standard deviation $\rightarrow$} & 0.01 & 0.02 & 0.03\\
\hline \hline
\multirow{4} {*} {MAE in excitation (dB)} & ALPA & $-$6.8 & $-$4.0 & $-$1.4\\
\cline{2-5}
& SOOT& 3.8 & 3.6 & 3.4\\ \cline{2-5}
& SLP & 10.6 & 11.0 & 11.2 \\ \cline{2-5}
& SDMM & $-$8.7 & $-$2.3 & 1.7 \\ 
\hline \hline
\multirow{4} {*} {MAE in filter (dB)} & ALPA & 0.9 & 1.1 & 3.6 \\ \cline{2-5}
& SOOT& 2.6 & 3.1 & 3.8 \\ \cline{2-5}
& SLP & 2.0 & 4.0 & 4.4 \\ \cline{2-5}
& SDMM & 3.0 & 5.5 & 6.3 \\ \cline{2-5}
 \hline \hline
\multirow{4} {8em} {MAE in reconstruction (dB)} & ALPA & $-$0.3 & 2.5 & 4.2\\ \cline{2-5}
& SOOT& $-$1.2 & 1.6 & 3.3\\ \cline{2-5}
& SLP & 3.7 & 5.5  & 7.0 \\ \cline{2-5}
& SDMM& 5.7 & 8.5 & 9.5 \\ \hline \hline 
\end{tabular}
\label{table_comp_all_MAE}
\end{table}
\normalsize
\newcommand{\figsiz}{0.88in}
\begin{figure}[h!]
\centering
$\begin{array}{ccccc}
                                \text{Original Signal} & \text{ALPA} & \text{SOOT} & \text{SLP} & \text{SDMM}\\
                                \includegraphics[width=\figsiz]{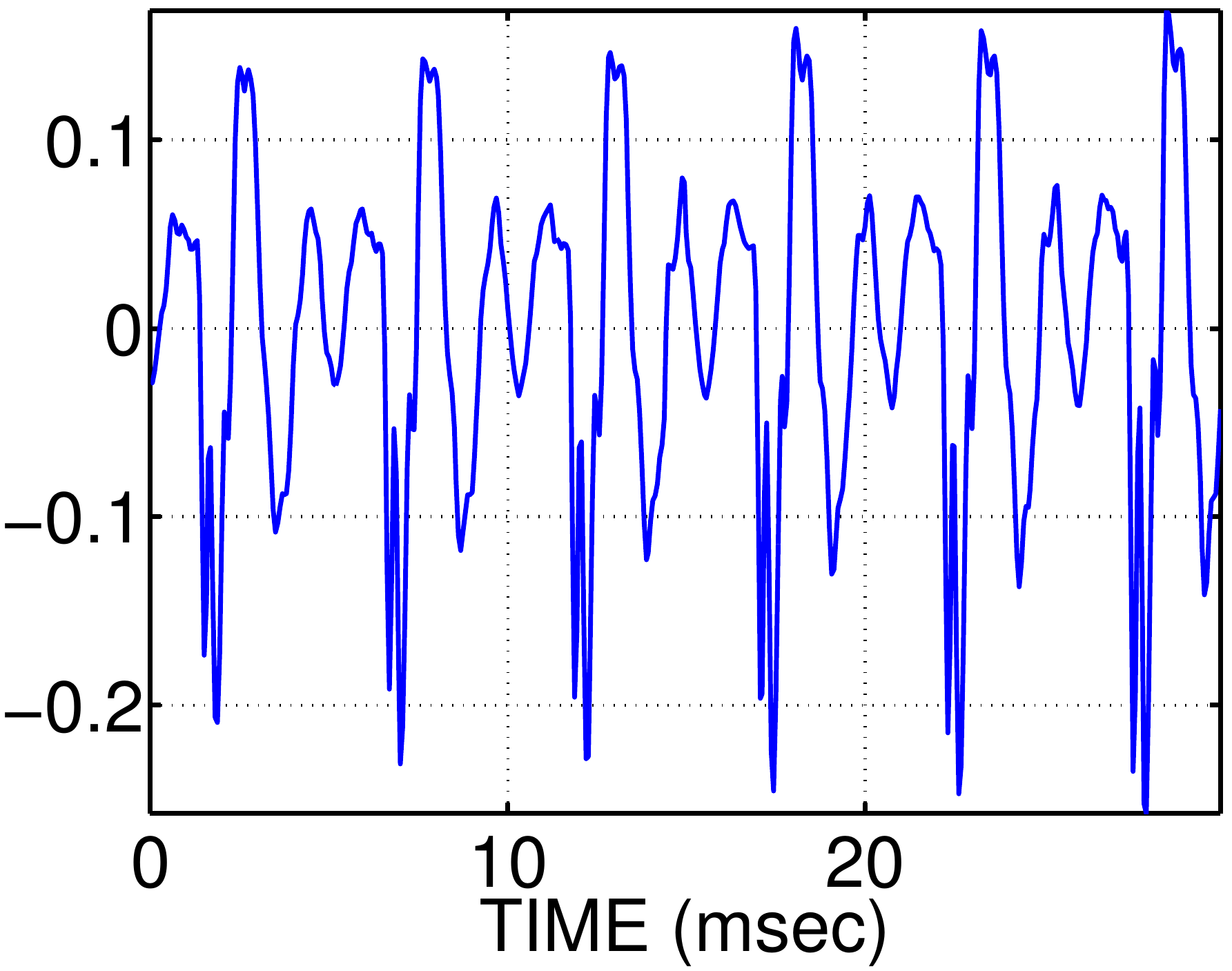}
                                                                \label{fig:female_speech_Infdb}
                                &\includegraphics[width=\figsiz]{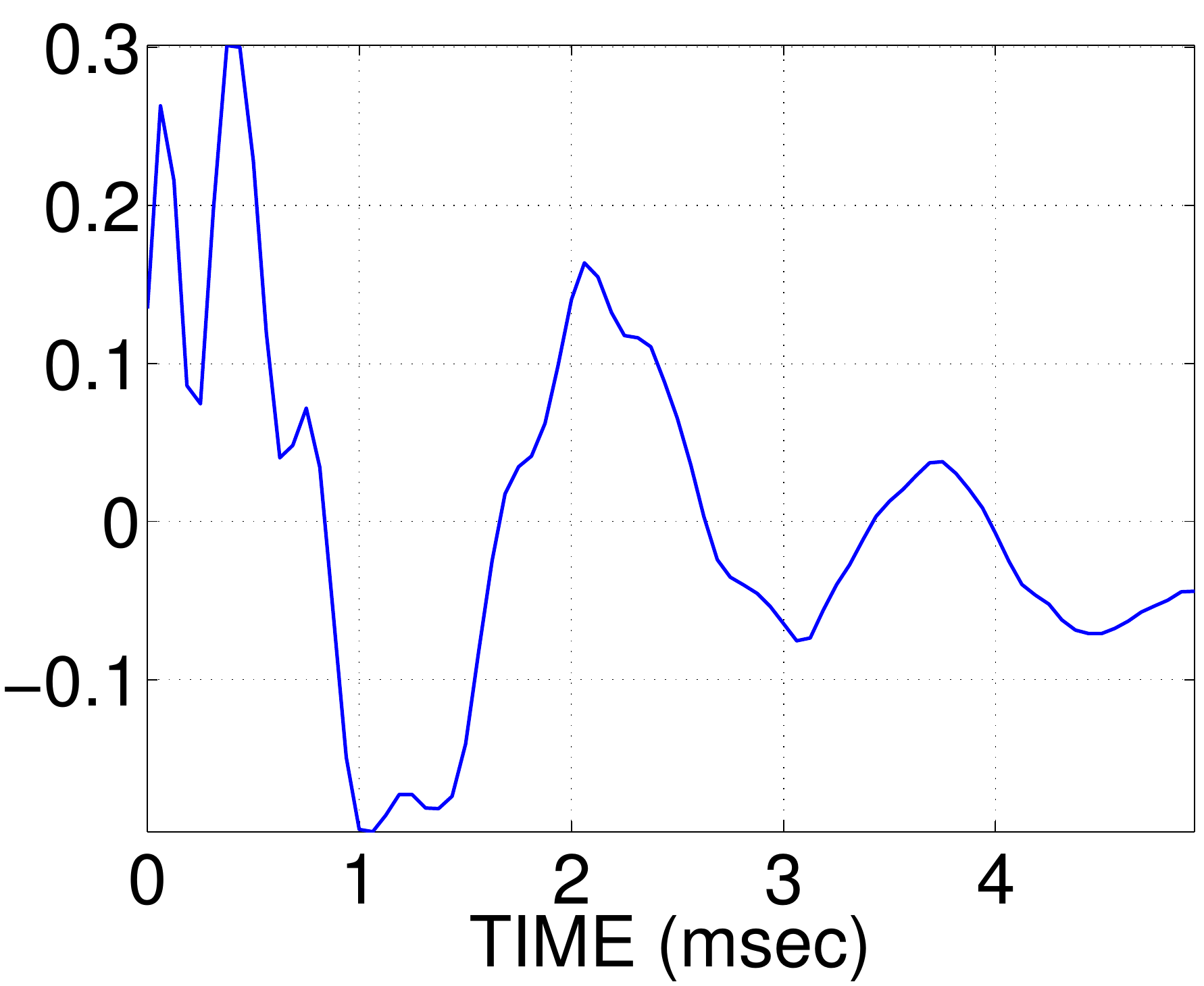}
                                \label{fig:female_alpa_filter_Infdb}
                                &\includegraphics[width=\figsiz]{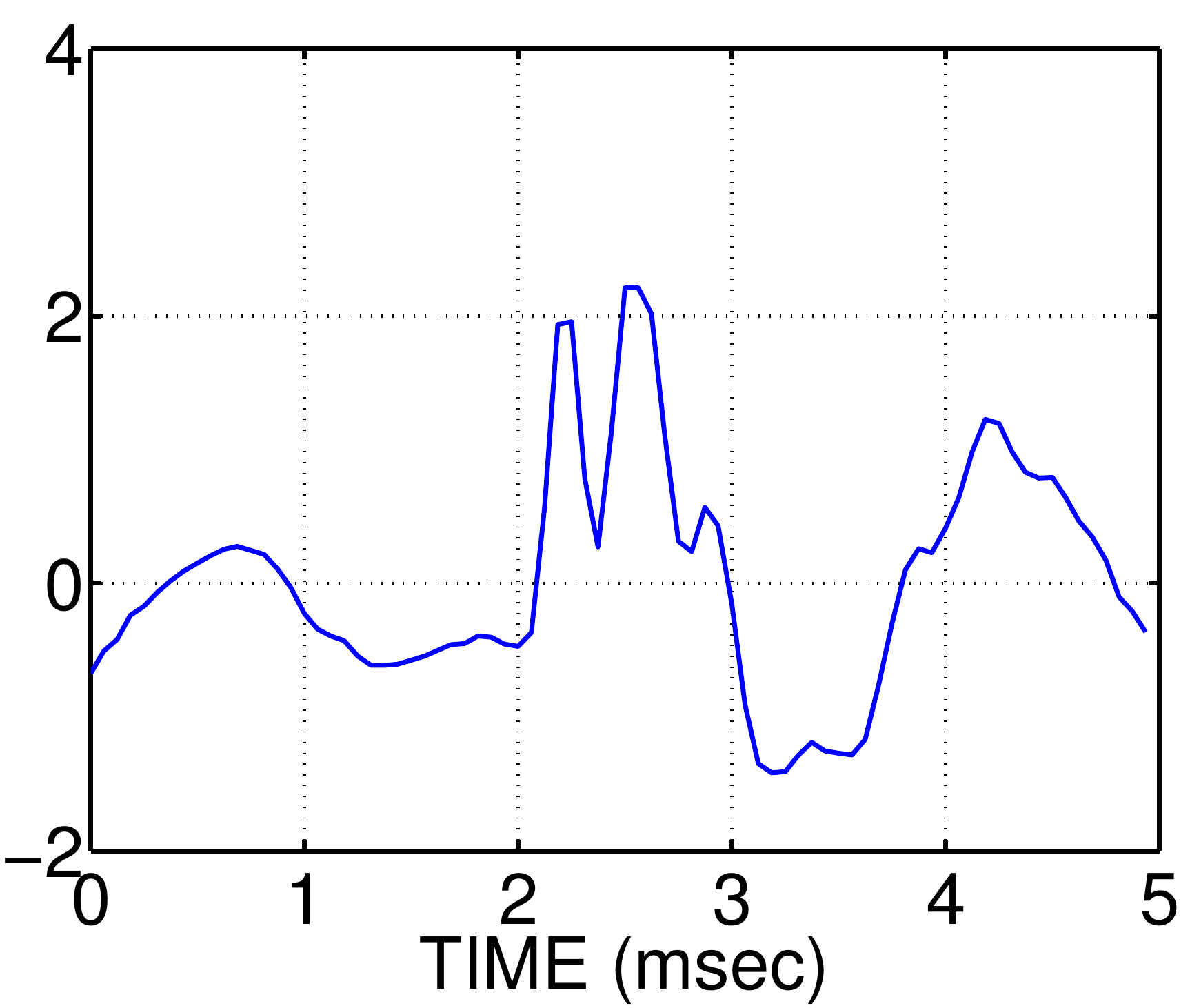}
                                                                \label{fig:female_SOOT_filter_Infdb}
                                & \includegraphics[width=\figsiz]{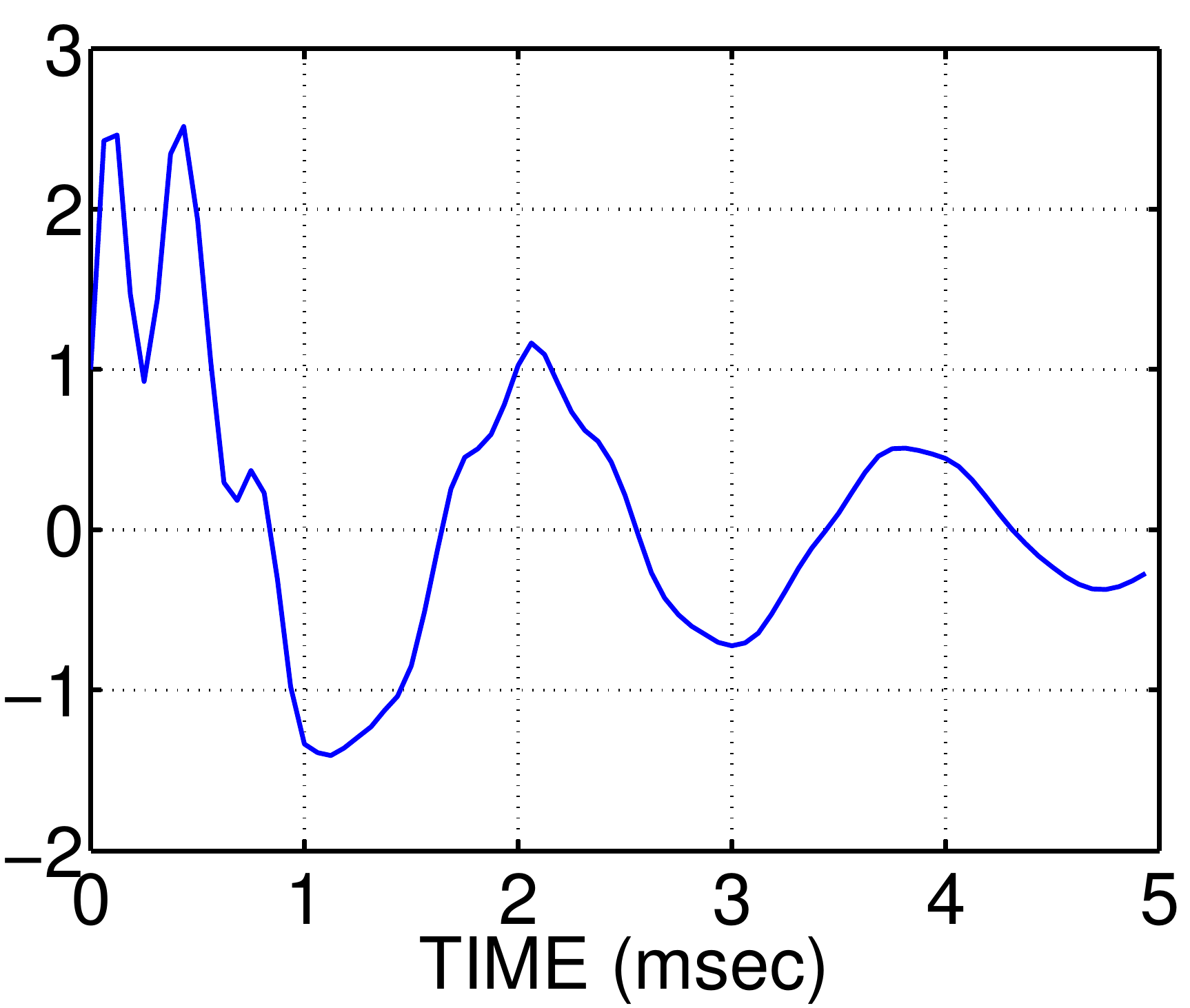}
                                \label{fig:female_Gia_filter_Infdb}
                                &\includegraphics[width=\figsiz]{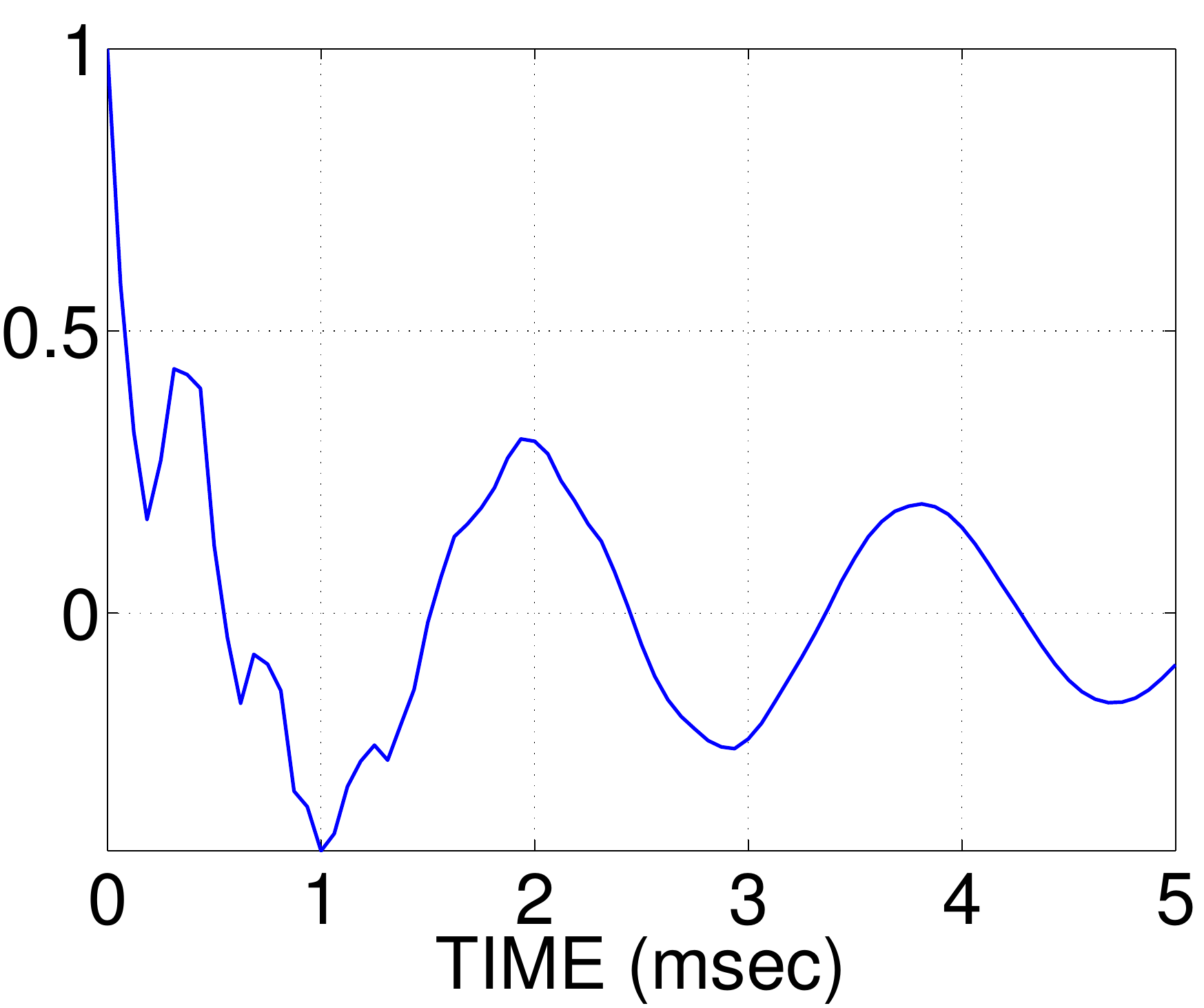}
                                \label{fig:female_SDMM_filter_Infdb}\\
 								 \includegraphics[width=\figsiz]{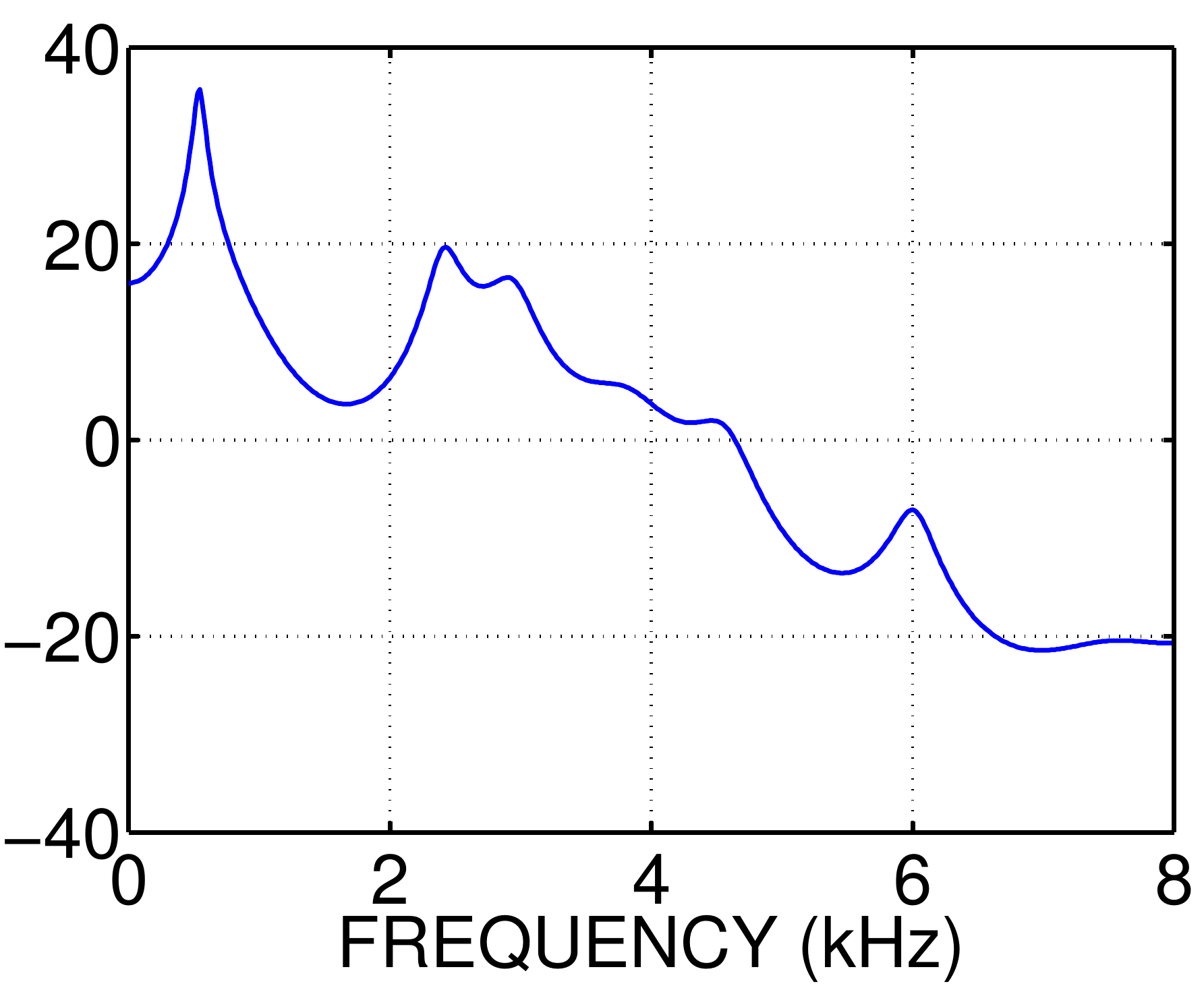} \label{fig:LP_female_freqresp_Infdb}                                                
                                &\includegraphics[width=\figsiz]{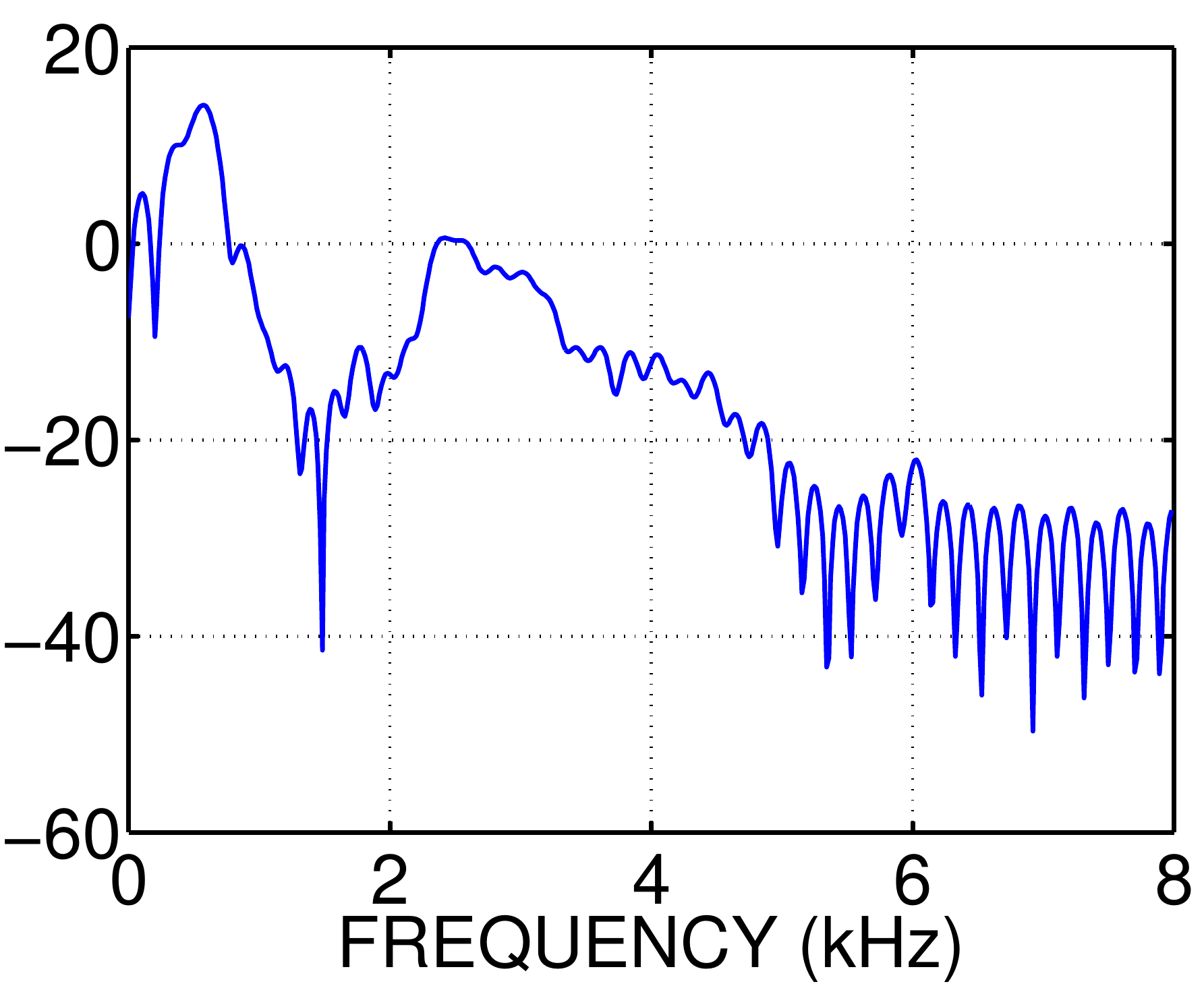}\label{fig:female_alpa_freqresp_Infdb}                                                                      
                                &\includegraphics[width=\figsiz]{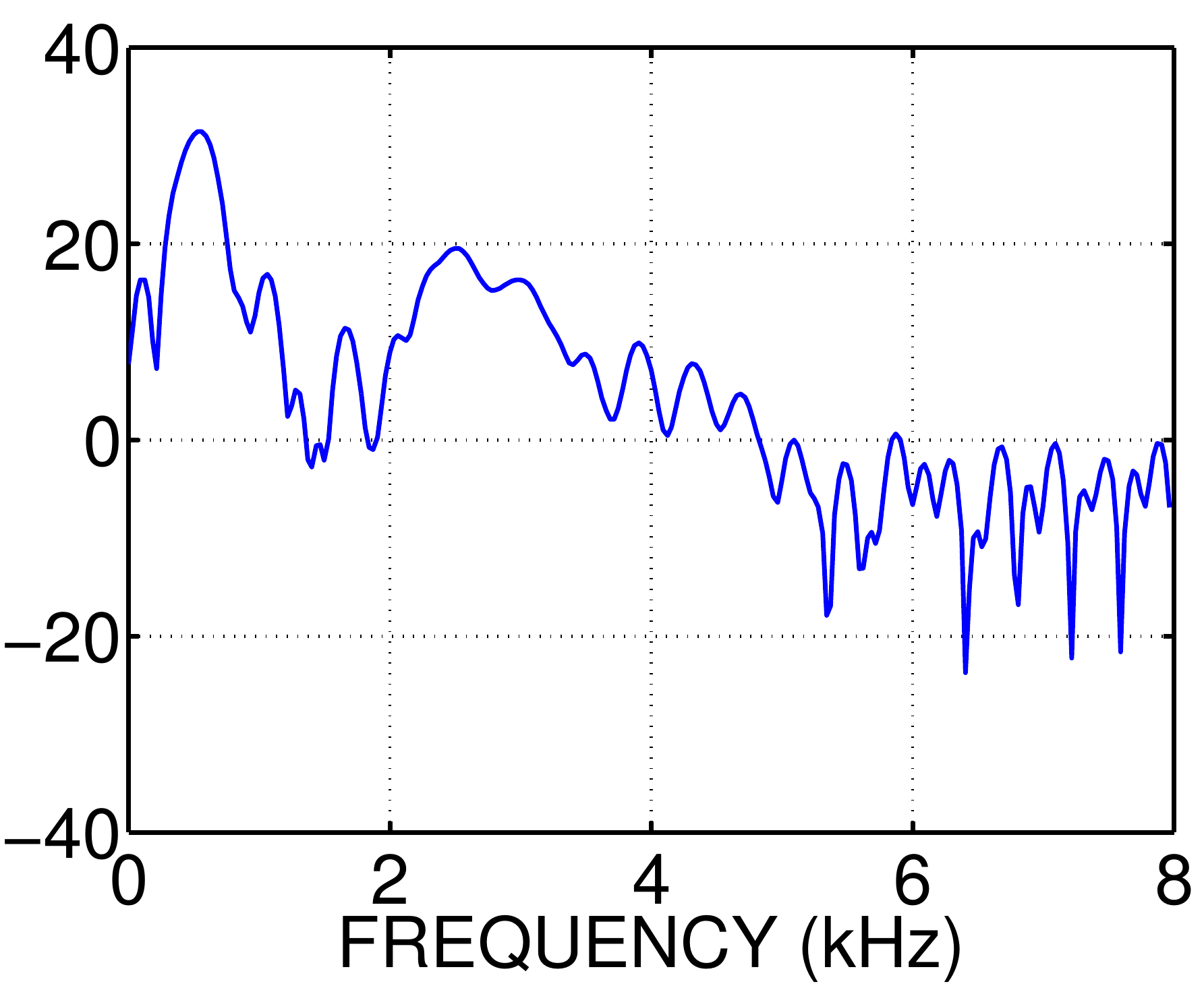}\label{fig:female_SOOT_freqresp_Infdb}
                                &\includegraphics[width=\figsiz]{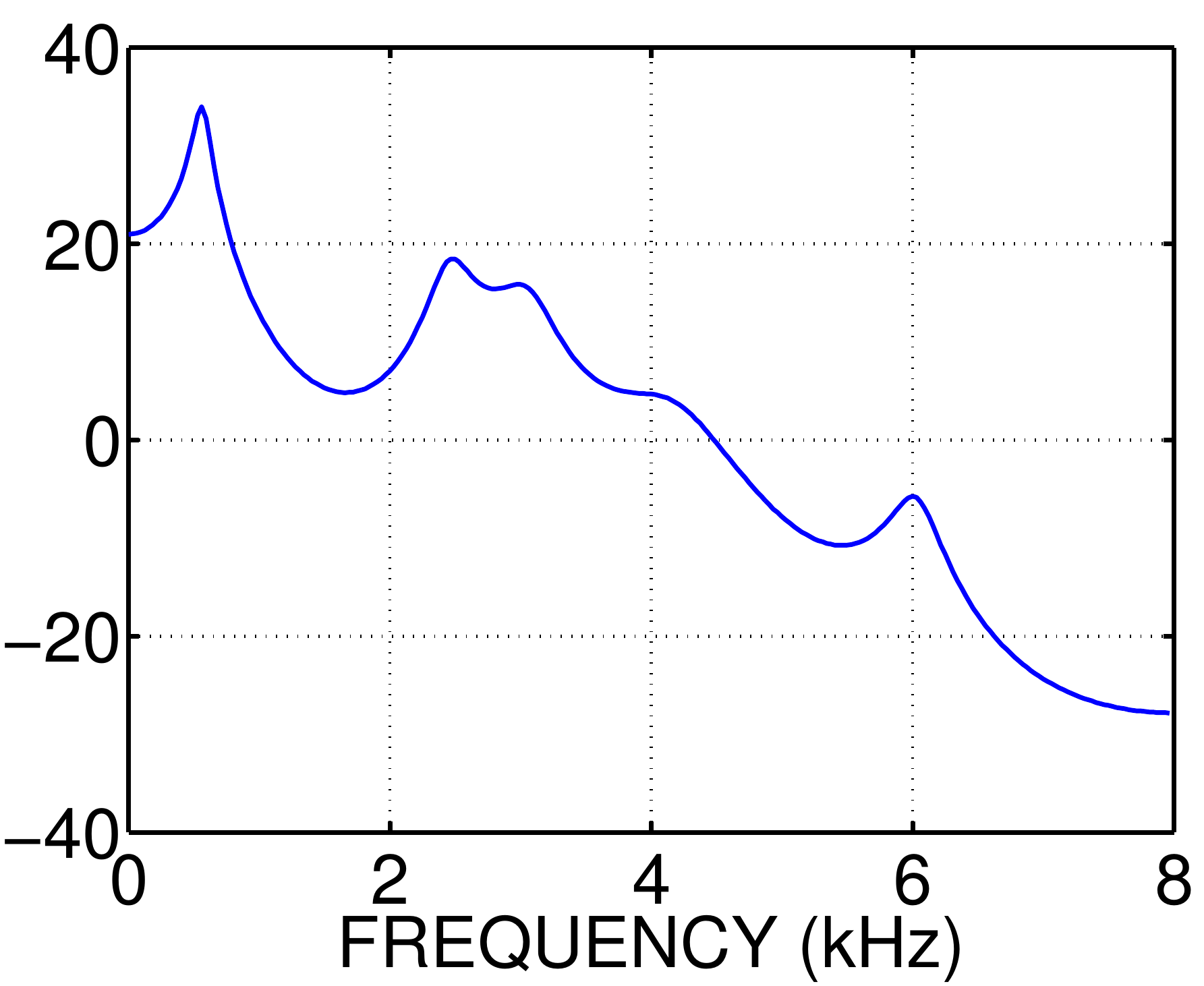} \label{fig:female_Gia_freqresp_Infdb}
                                &\includegraphics[width=\figsiz]{./chapter3fig/LP_Infdb_female_freqresp_MO16.pdf}\label{fig:SDMM_alpa_freqresp_Infdb}\\
                                \includegraphics[width=\figsiz]{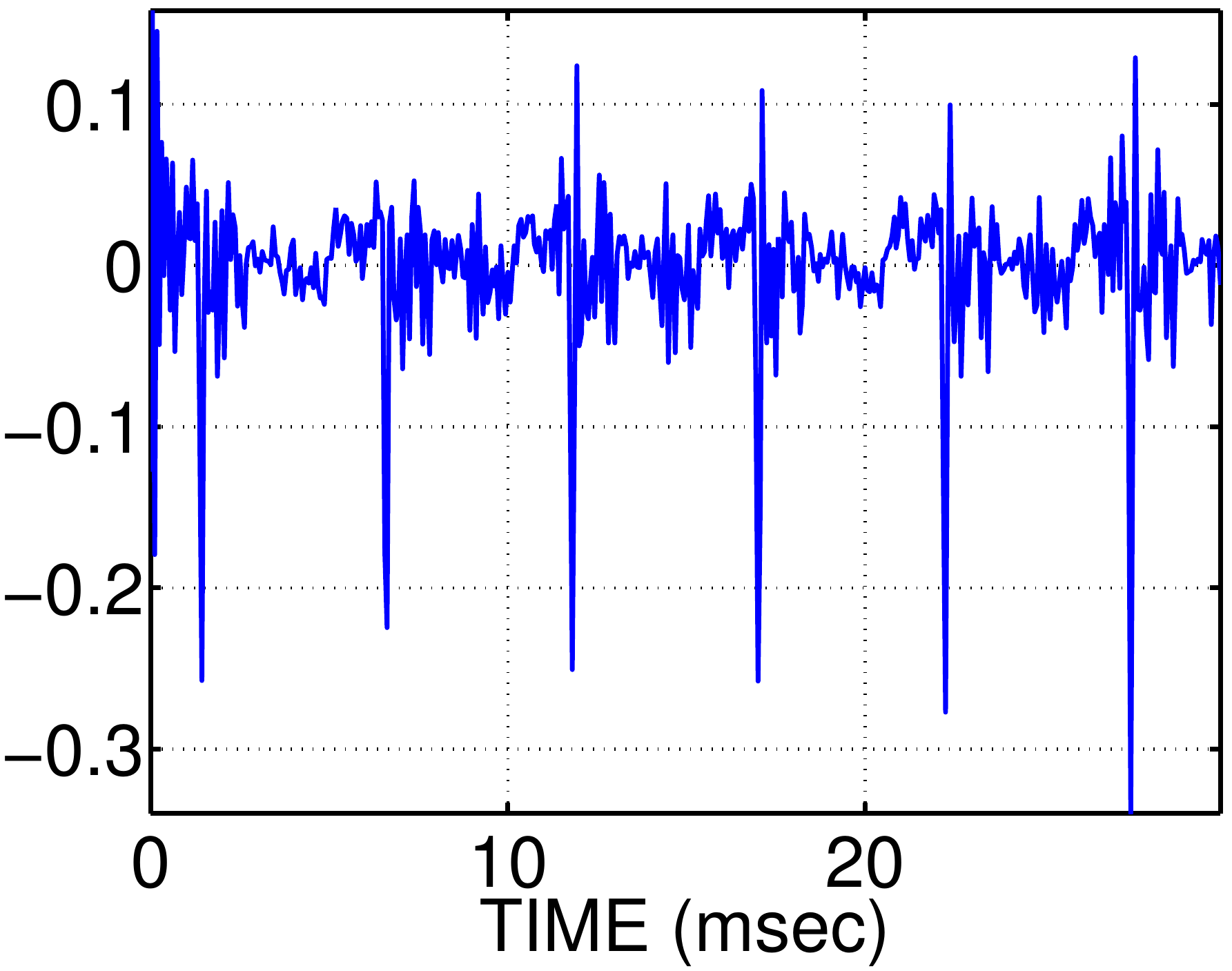}\label{fig:female_residue_Infdb}
                                &\includegraphics[width=\figsiz]{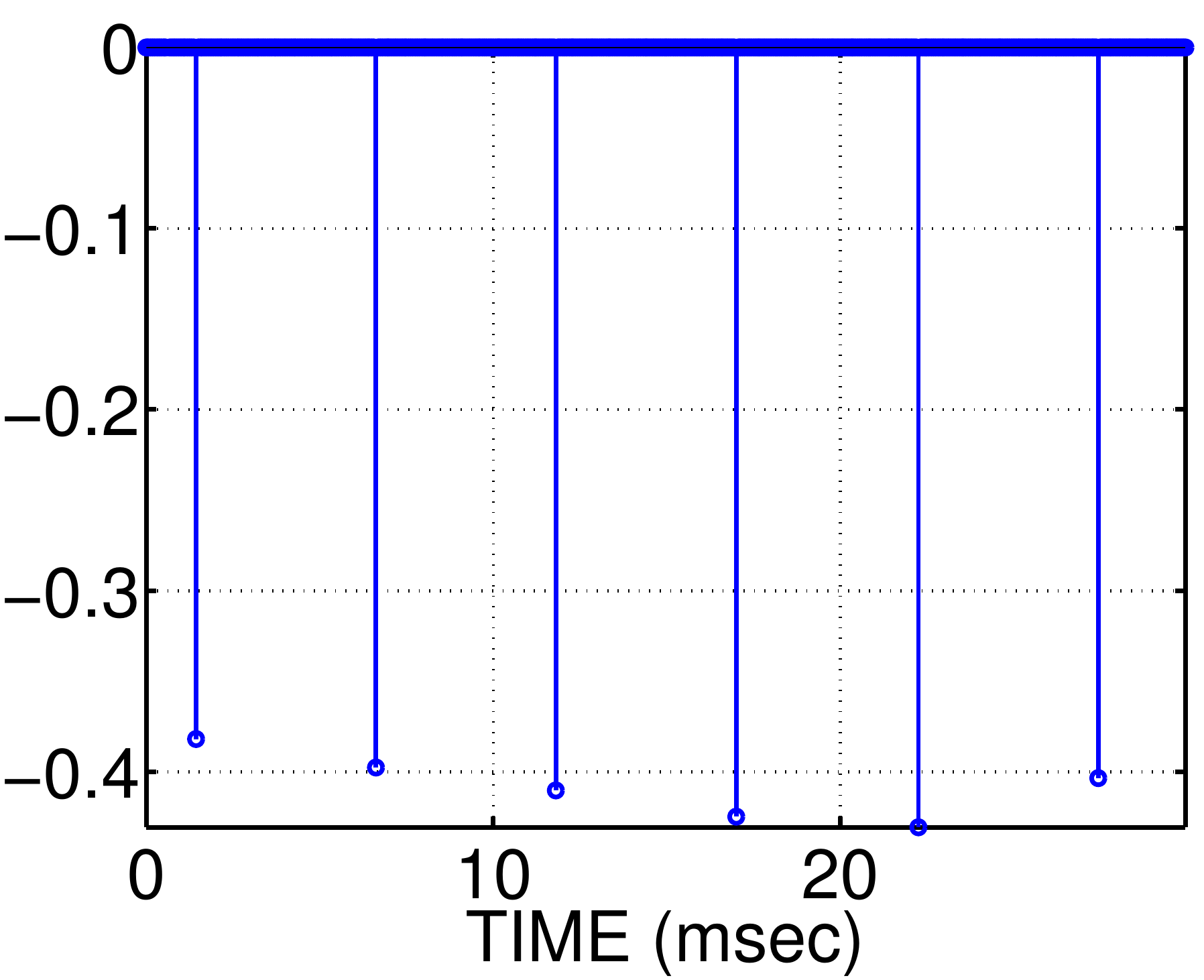} \label{fig:female_alpa_residue_Infdb}
 								&\includegraphics[width=\figsiz]{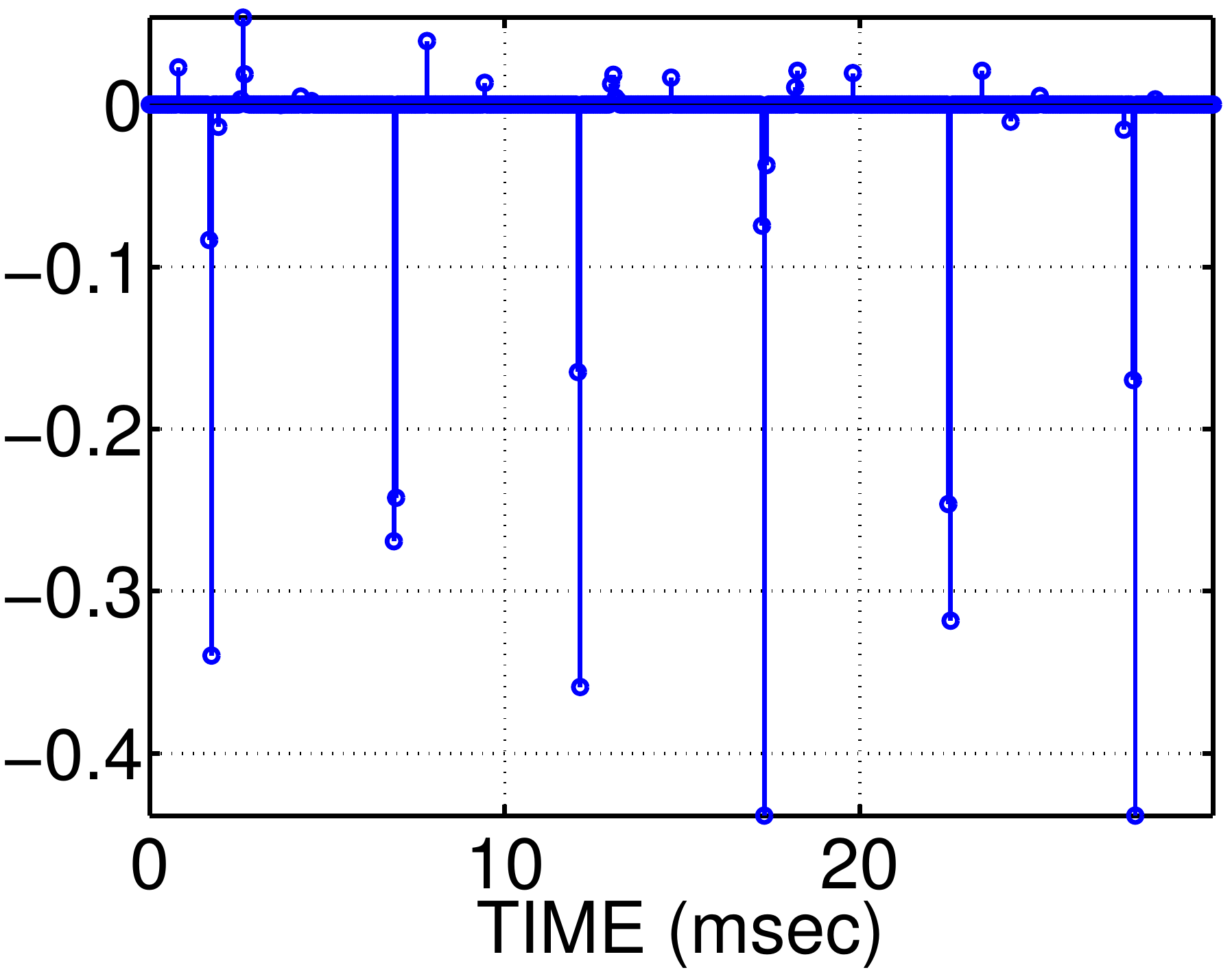}\label{fig:female_SOOT_residue_Infdb}
  								&\includegraphics[width=\figsiz]{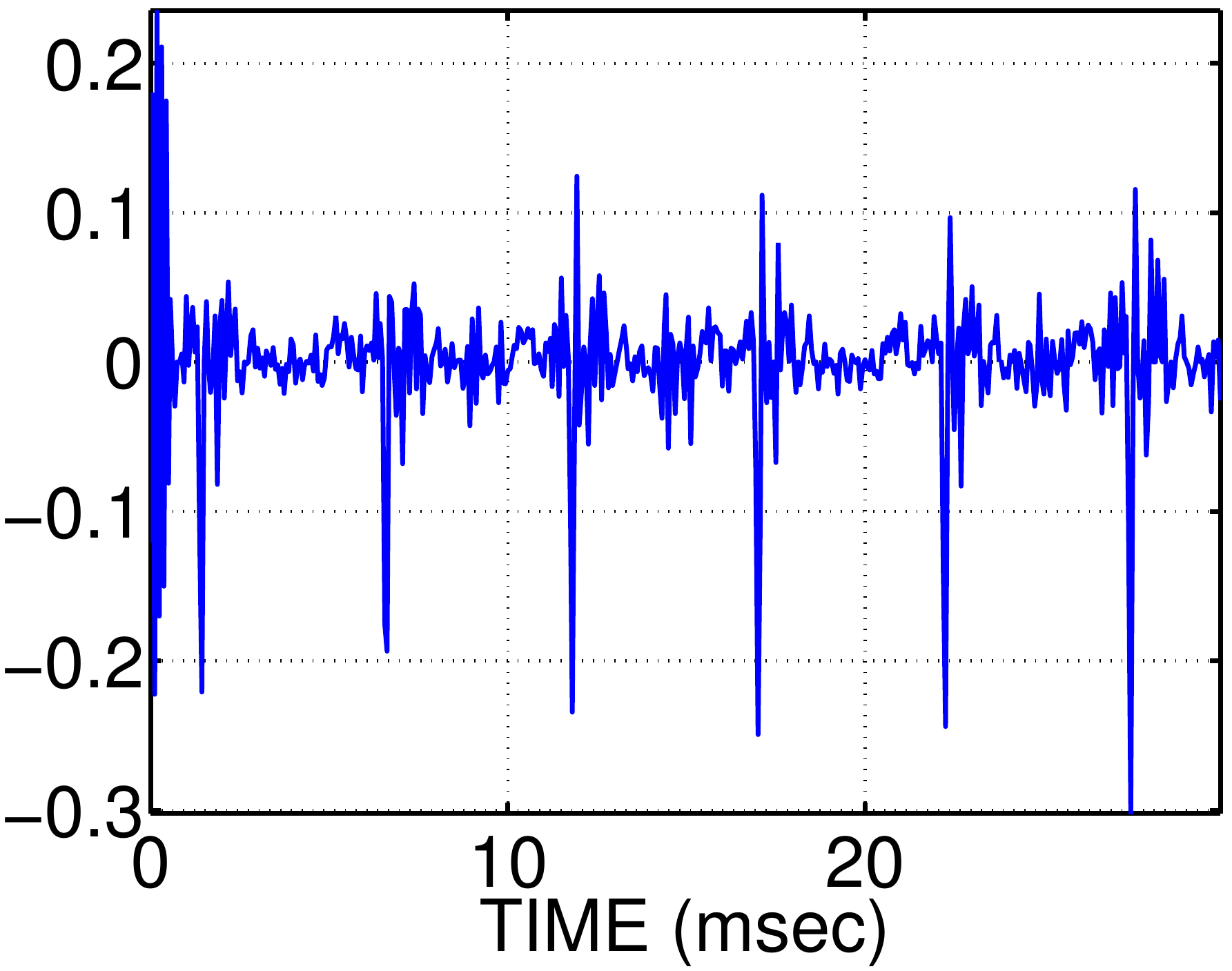}\label{fig:female_Gia_residue_Infdb}
                                &\includegraphics[width=\figsiz]{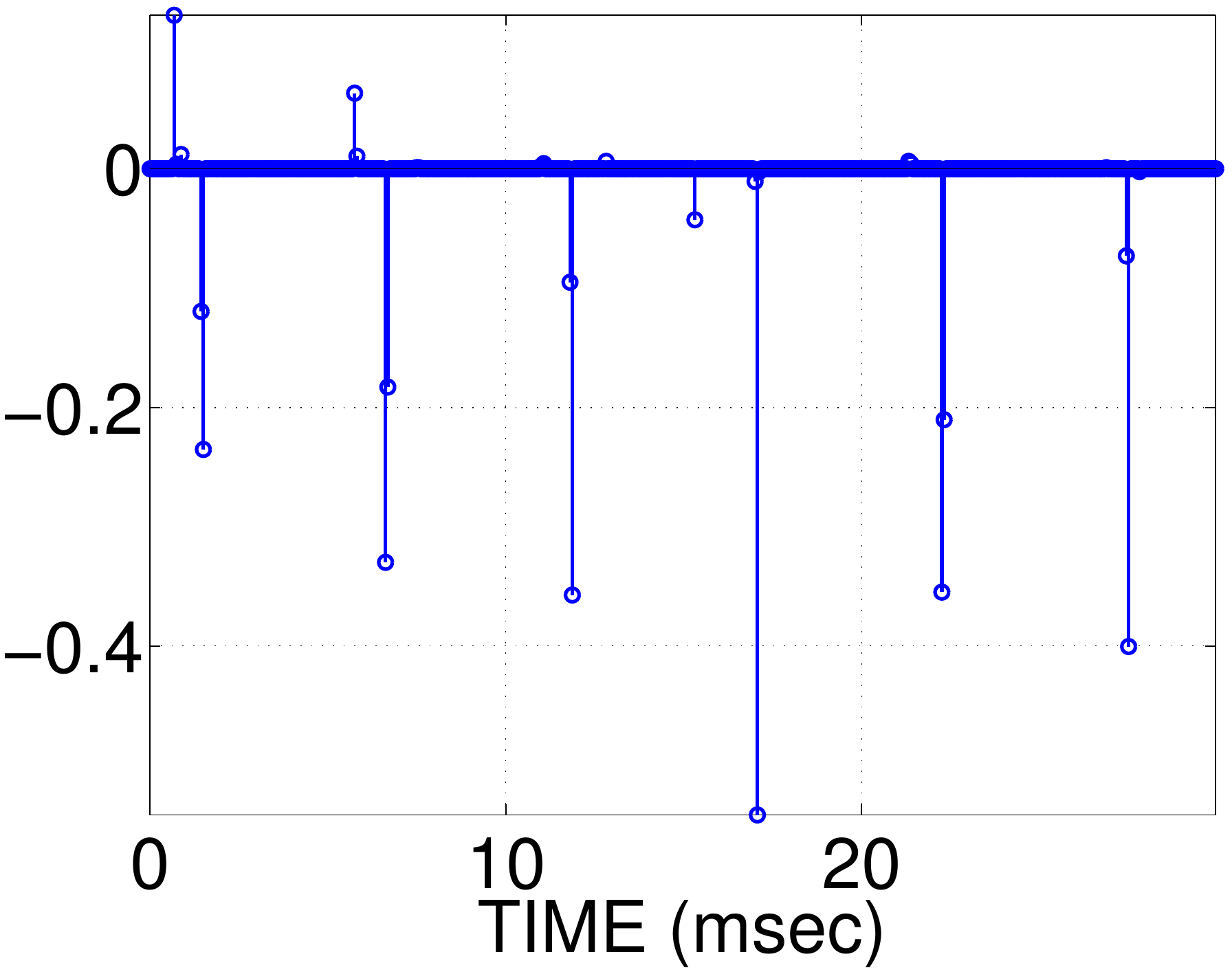}\label{fig:female_SDMM_residue_Infdb}
                                \end{array}$
                                \caption{(Color online) A comparison of sparse deconvolution methods for clean speech: Rows 1, 2, and 3 corresponding to Column 1 show the speech signal, frequency response of the LP filter, and the LP residue, respectively. For Columns 2$-$5, Rows 1, 2, and 3 show estimates of the filter, its frequency response, and the excitation, respectively.}
                                                                \label{fig:female_comp_deconv_clean}
                                \end{figure}
 \begin{figure}[h!]
 \centering
$\begin{array}{ccccc}
                                \text{Original Signal} & \text{ALPA} & \text{SOOT} & \text{SLP} & \text{SDMM}\\
                                \includegraphics[width=\figsiz]{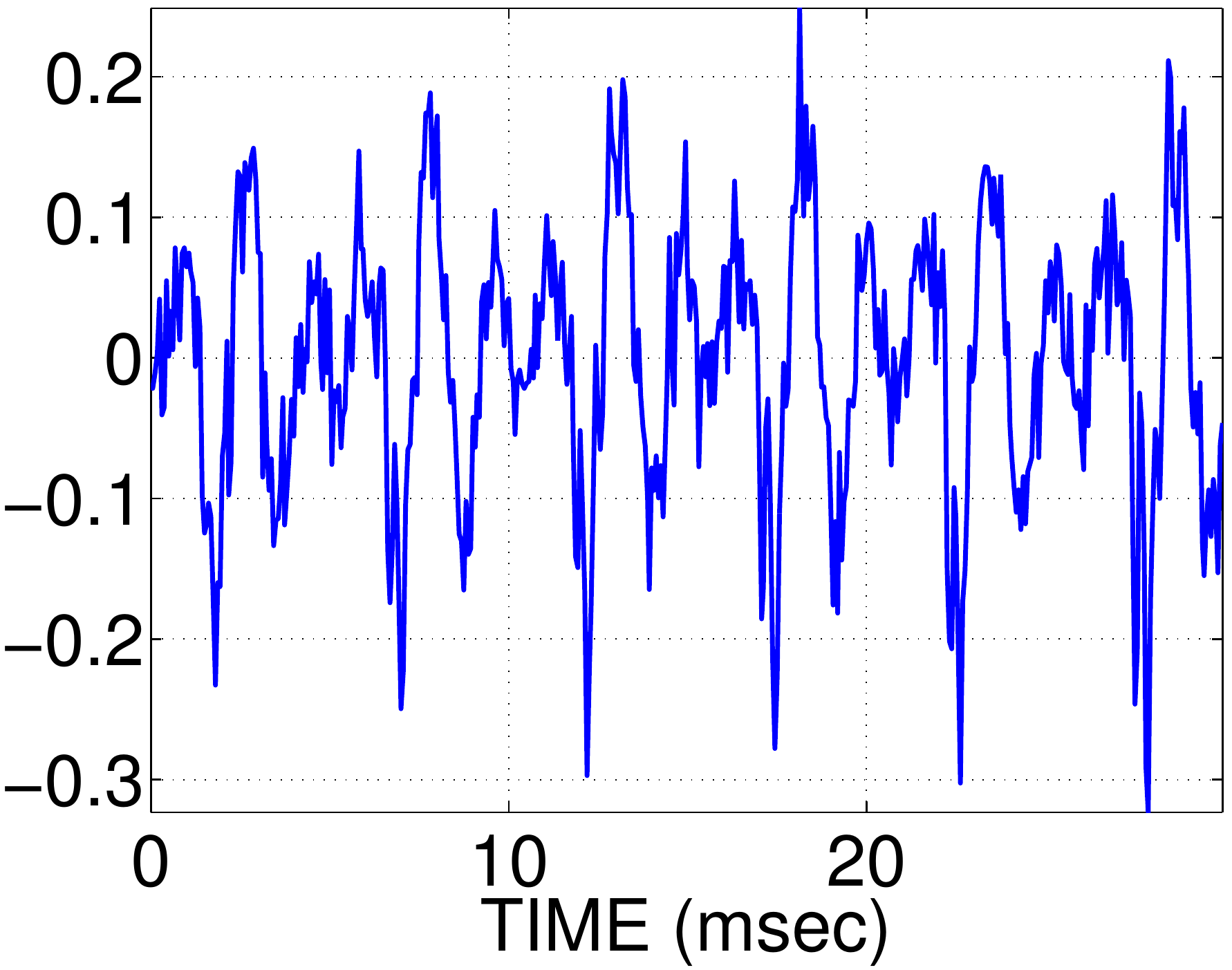}
                                                                \label{fig:female_speech_10db}
                                &\includegraphics[width=\figsiz]{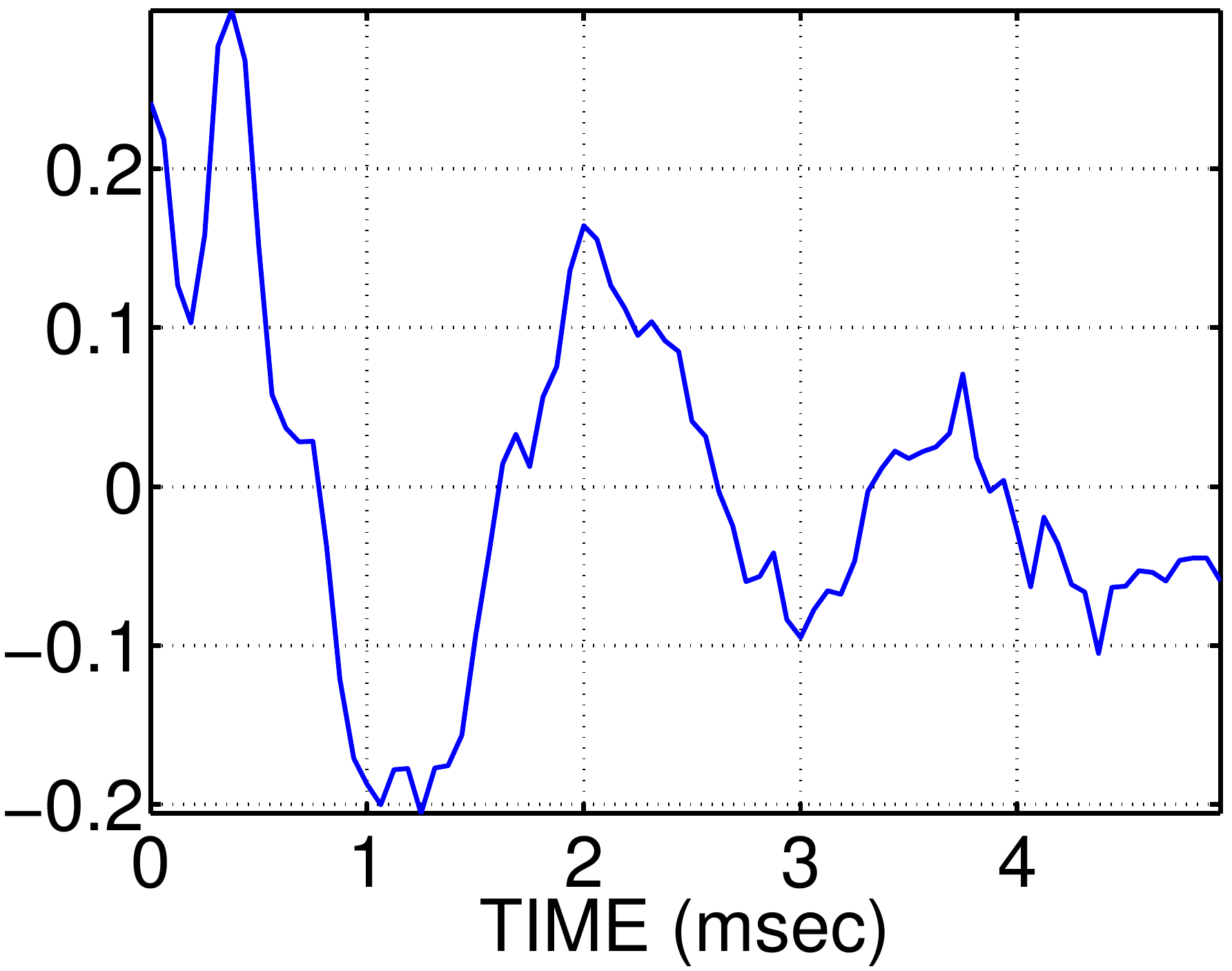}
                                \label{fig:female_alpa_filter_10db}
                                &\includegraphics[width=\figsiz]{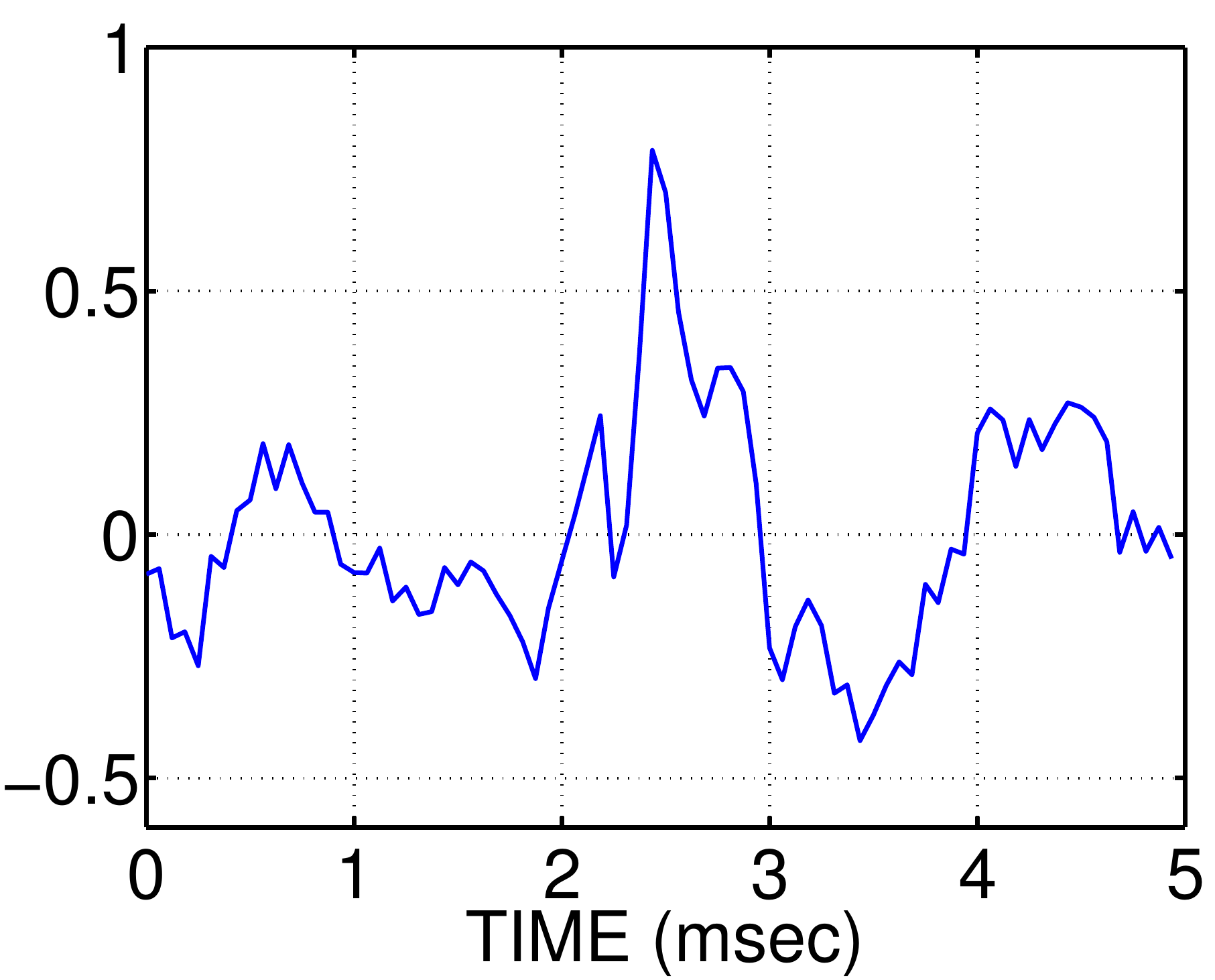}
                                                                \label{fig:female_SOOT_filter_10db}
                                & \includegraphics[width=\figsiz]{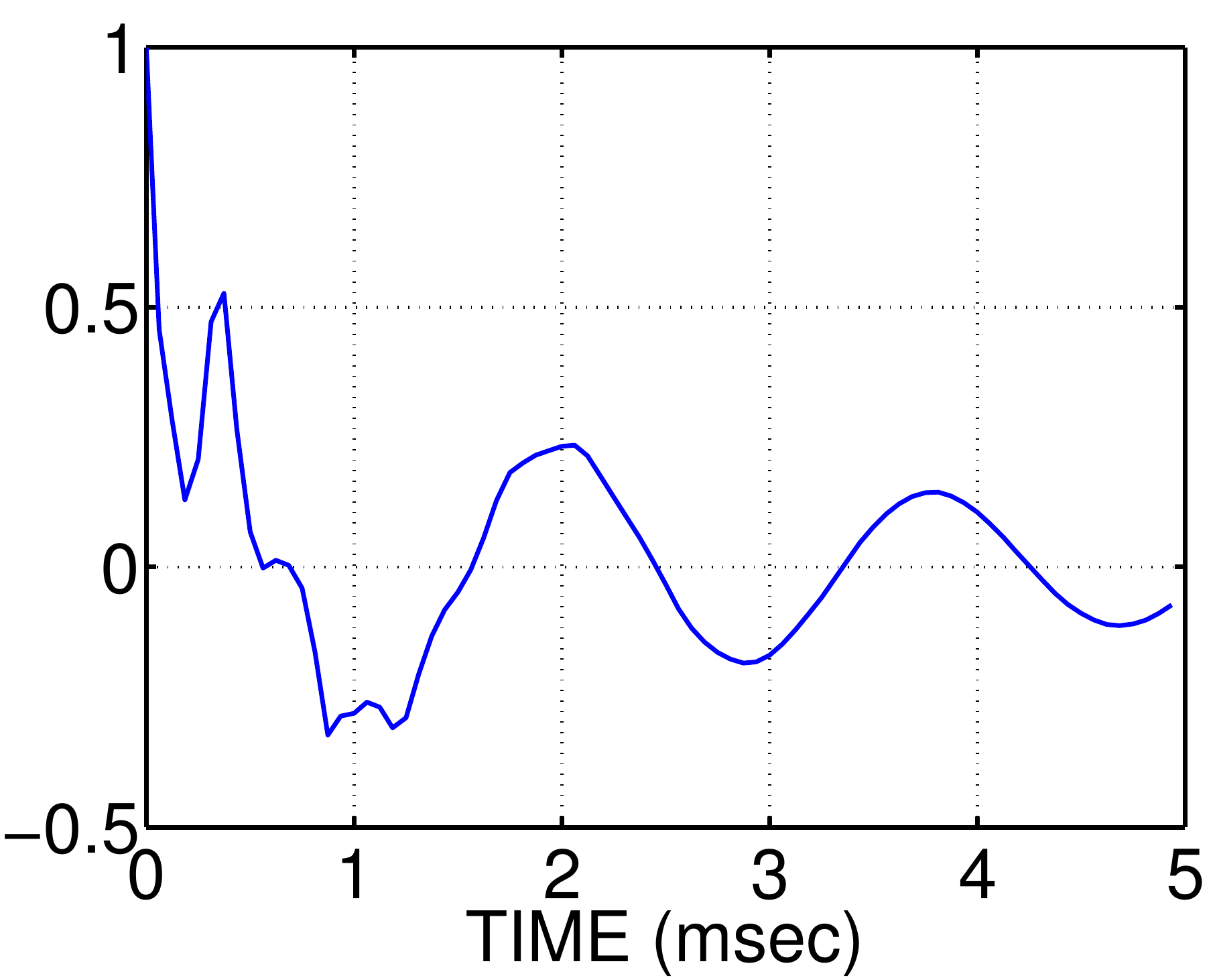}
                                \label{fig:female_Gia_filter_10db}
                                &\includegraphics[width=\figsiz]{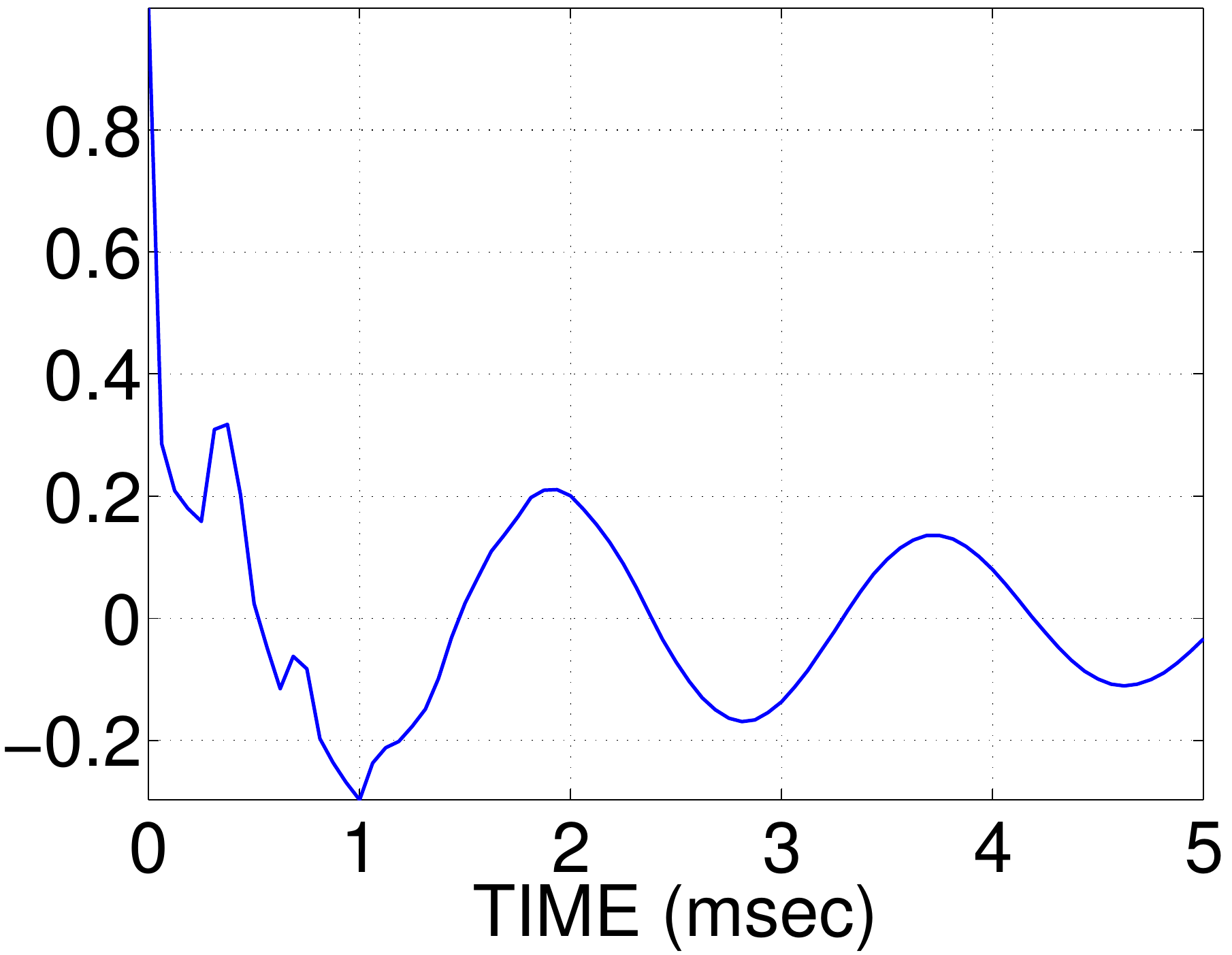}
                                \label{fig:female_SDMM_filter_10db}\\
                                                                 \includegraphics[width=\figsiz]{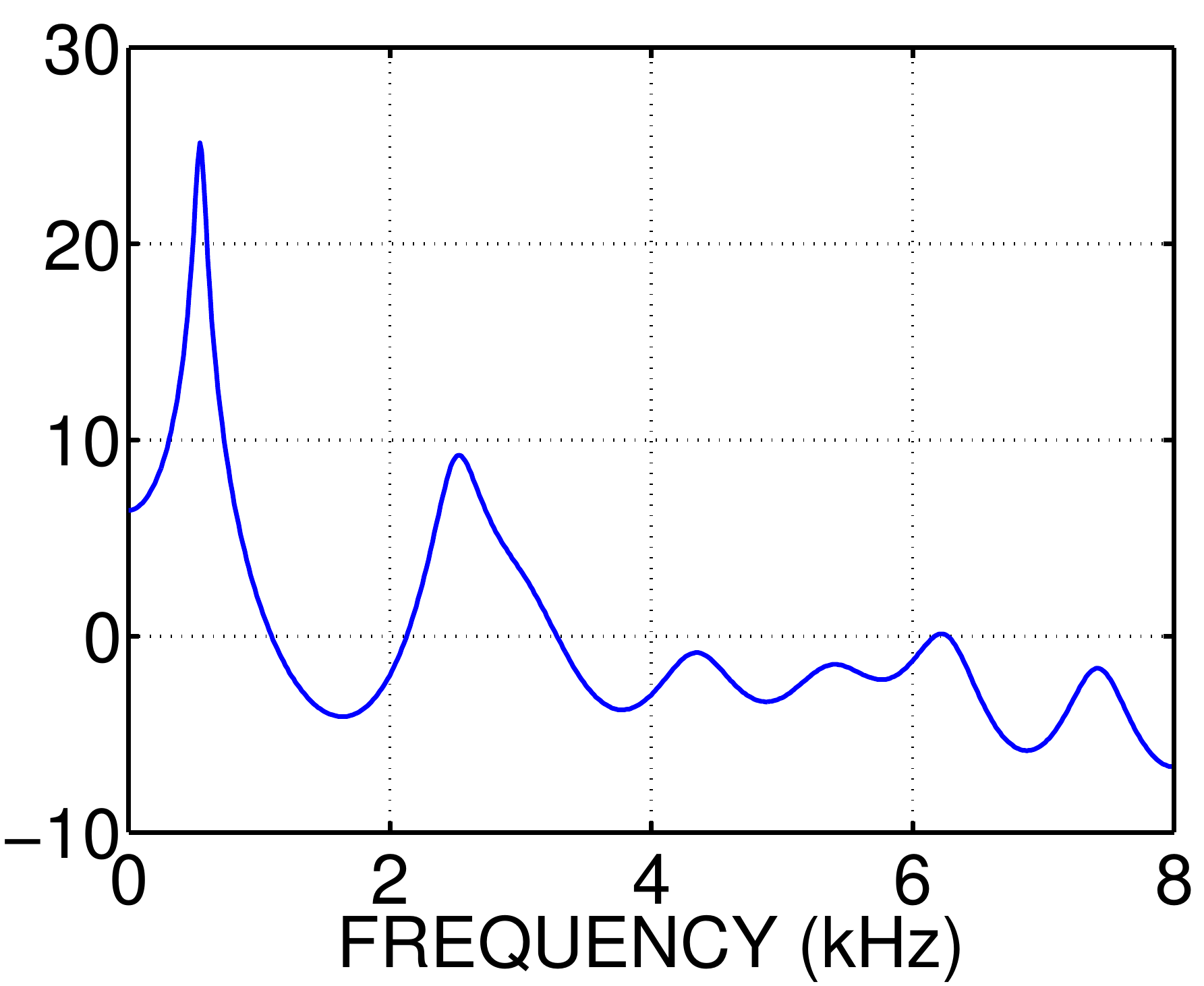}
\label{fig:LP_female_freqresp_10db}                                                
                                &\includegraphics[width=\figsiz]{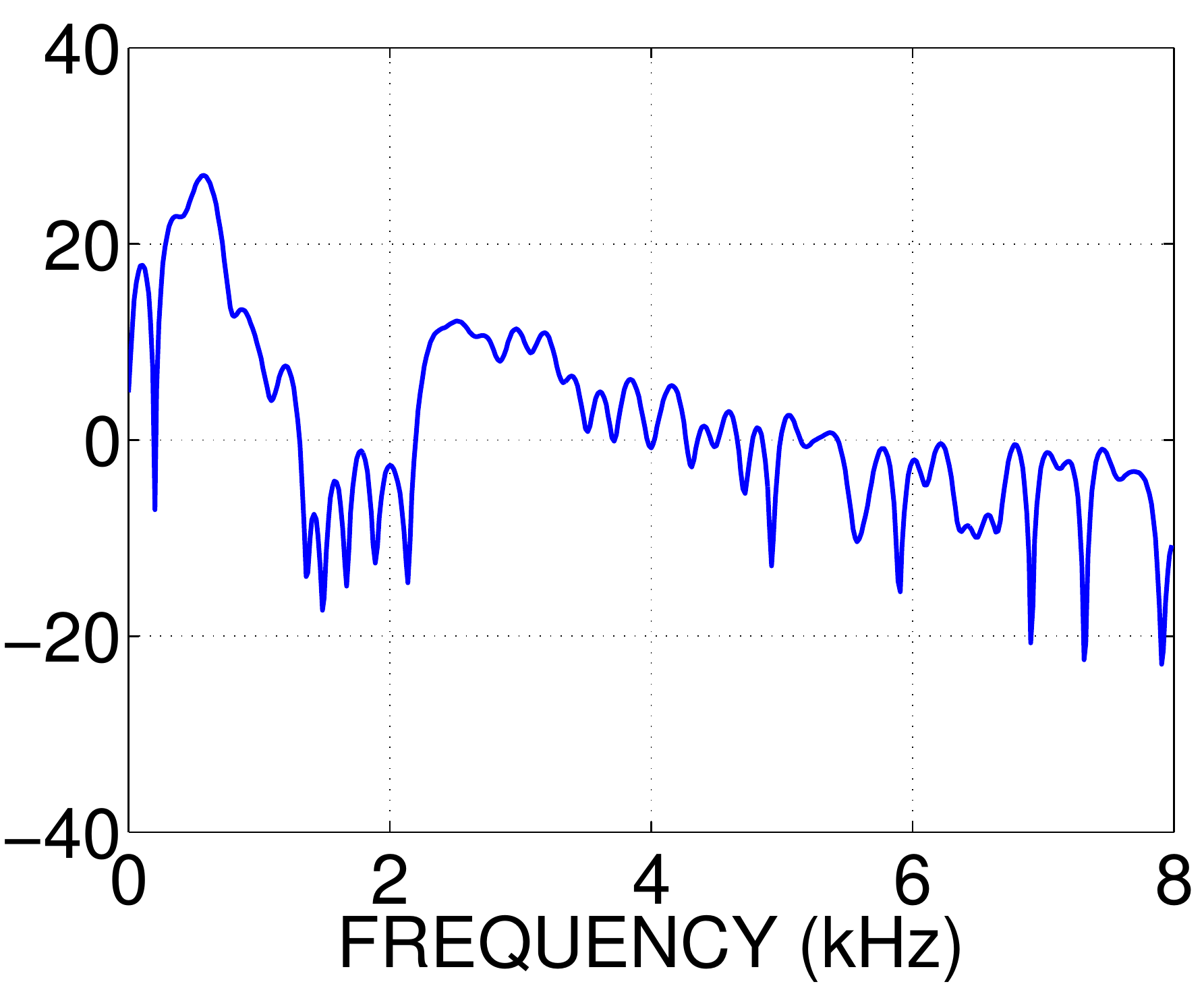}\label{fig:female_alpa_freqresp_10db}  
                                                                   
                                &\includegraphics[width=\figsiz]{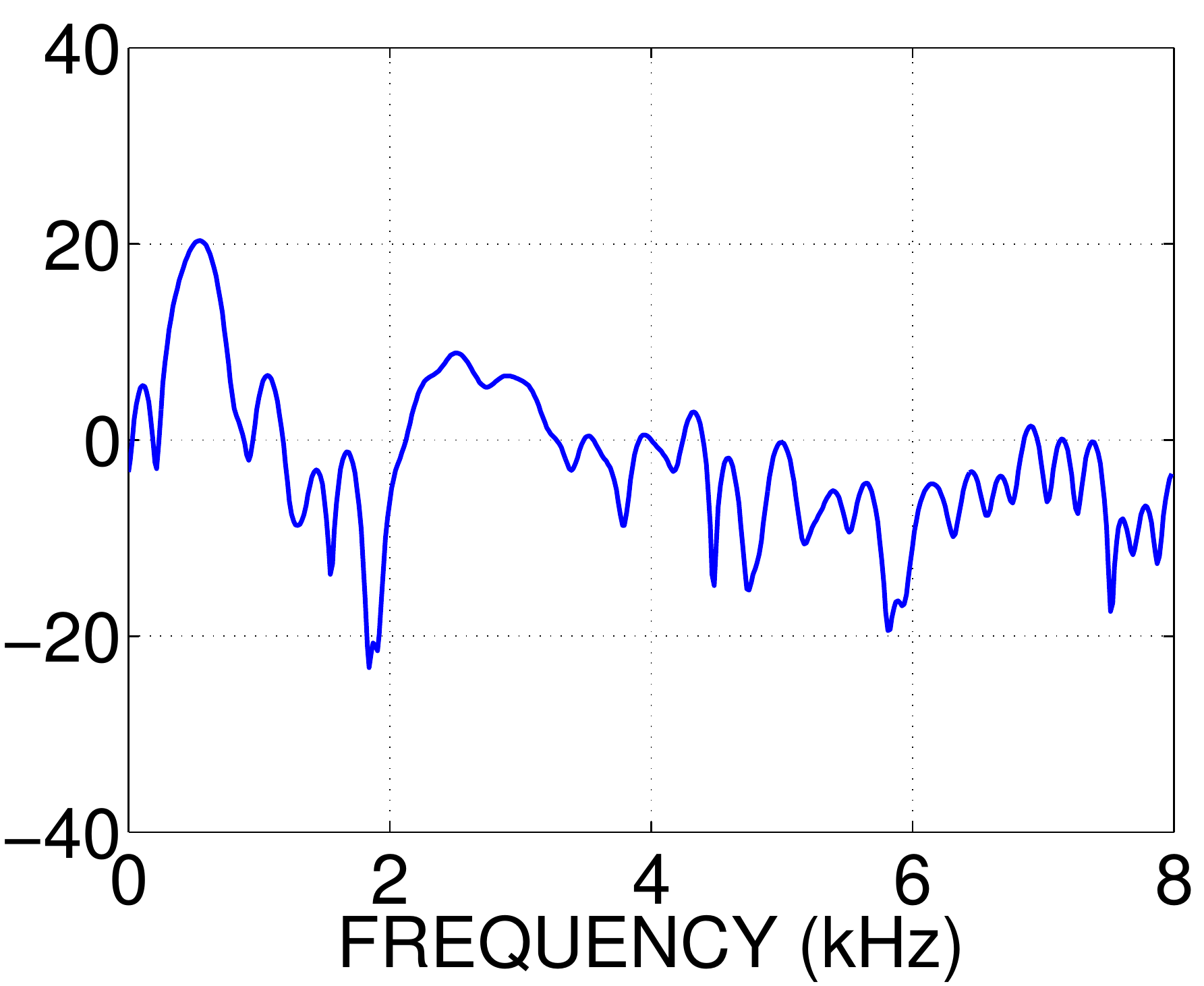}\label{fig:female_SOOT_freqresp_10db}
                                &\includegraphics[width=\figsiz]{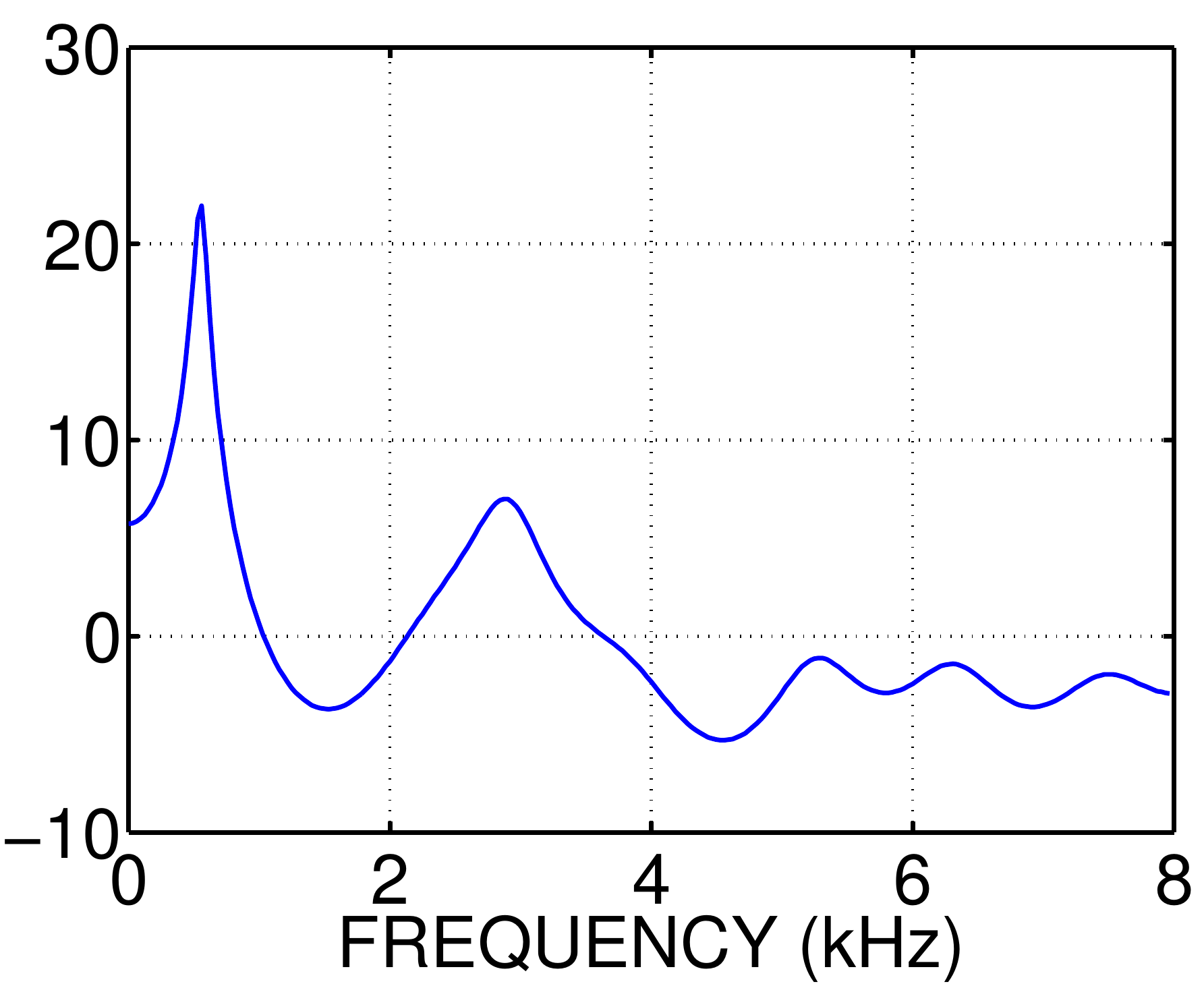}
\label{fig:female_Gia_freqresp_10db}
                               
&\includegraphics[width=\figsiz]{./chapter3fig/LP_10db_female_freqresp_MO16.pdf}\label{fig:SDMM_alpa_freqresp_10db}\\
                                \includegraphics[width=\figsiz]{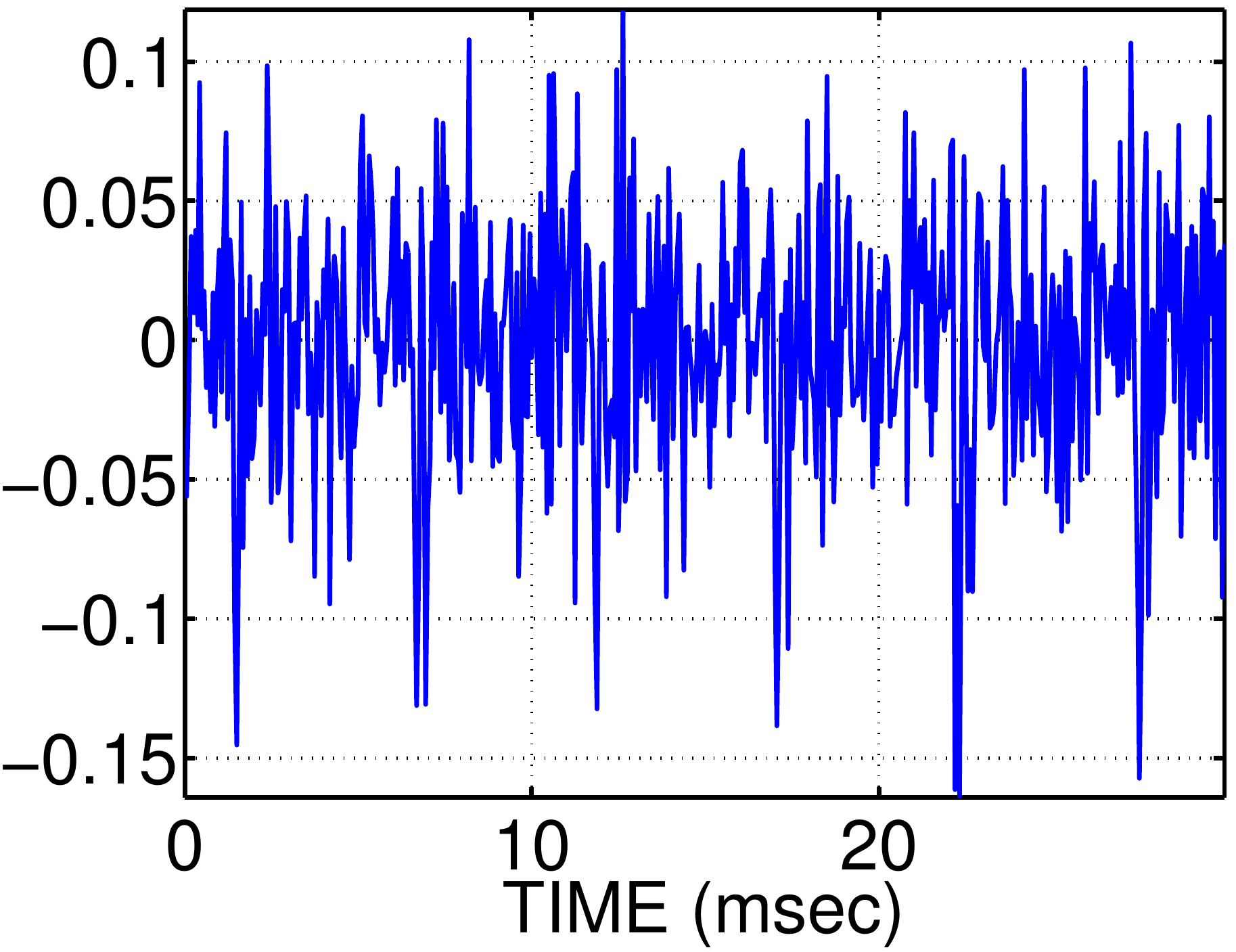}\label{fig:female_residue_10db}
                                &\includegraphics[width=\figsiz]{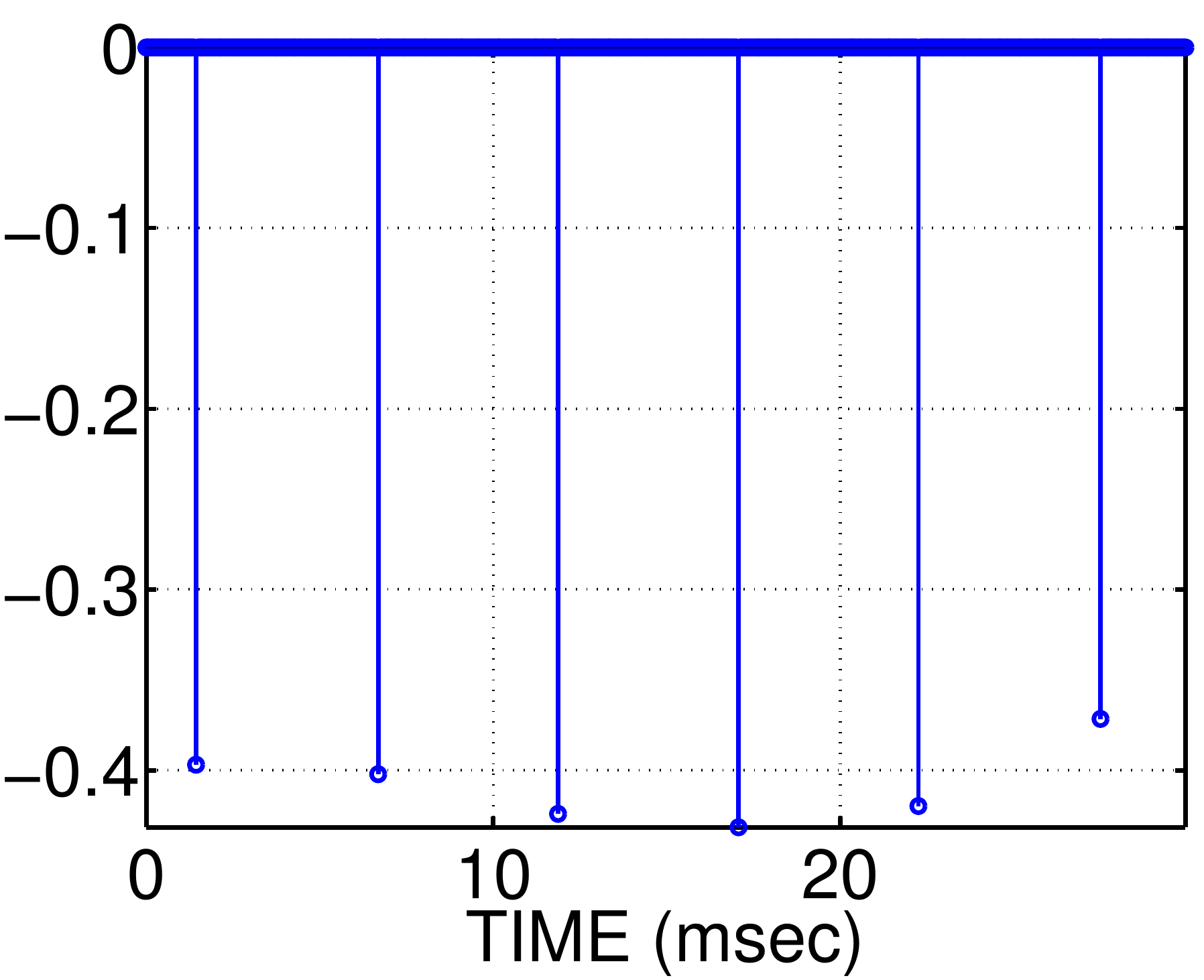}
\label{fig:female_alpa_residue_10db}
                                                               
&\includegraphics[width=\figsiz]{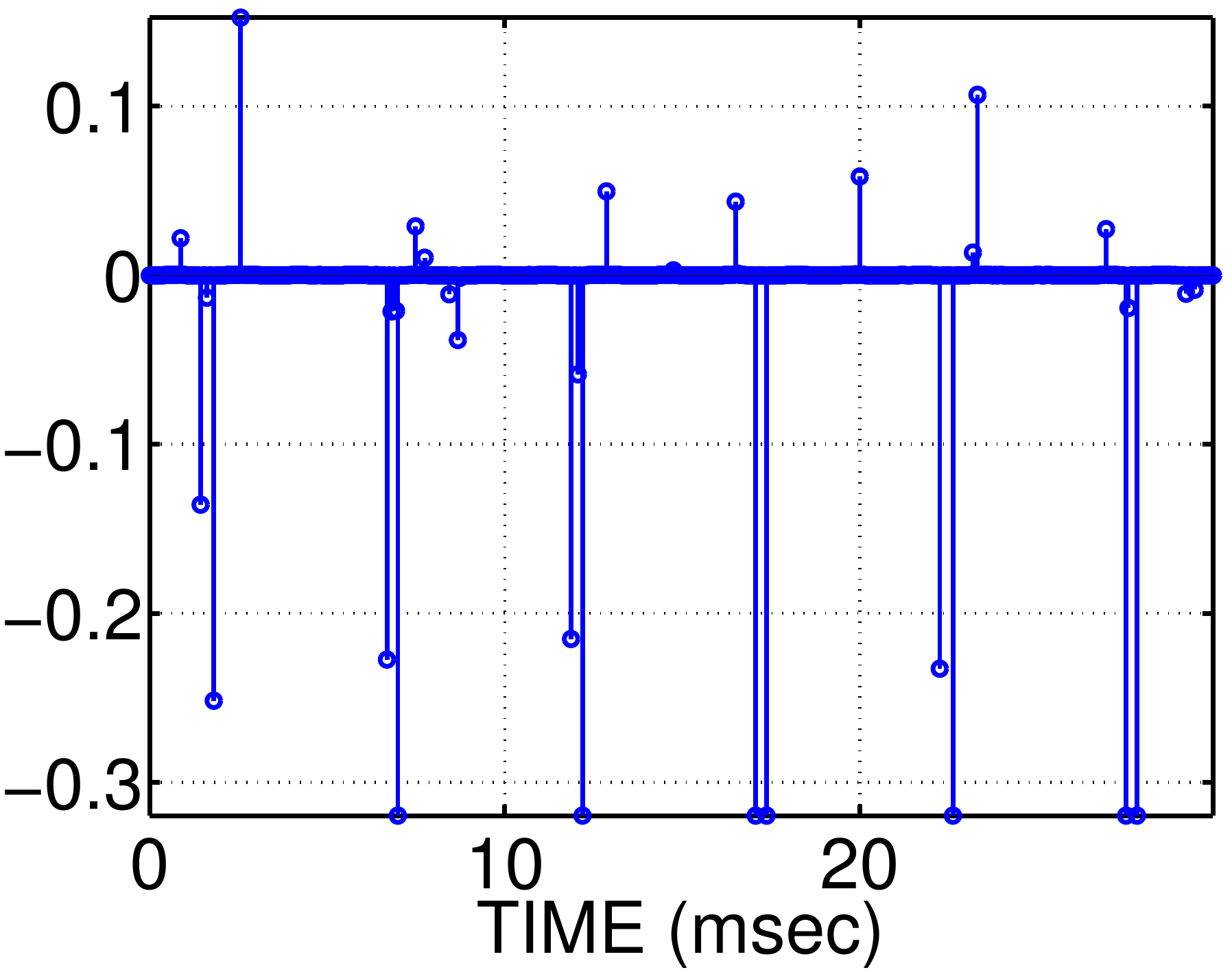}\label{fig:female_SOOT_residue_10db}
                                                               
&\includegraphics[width=\figsiz]{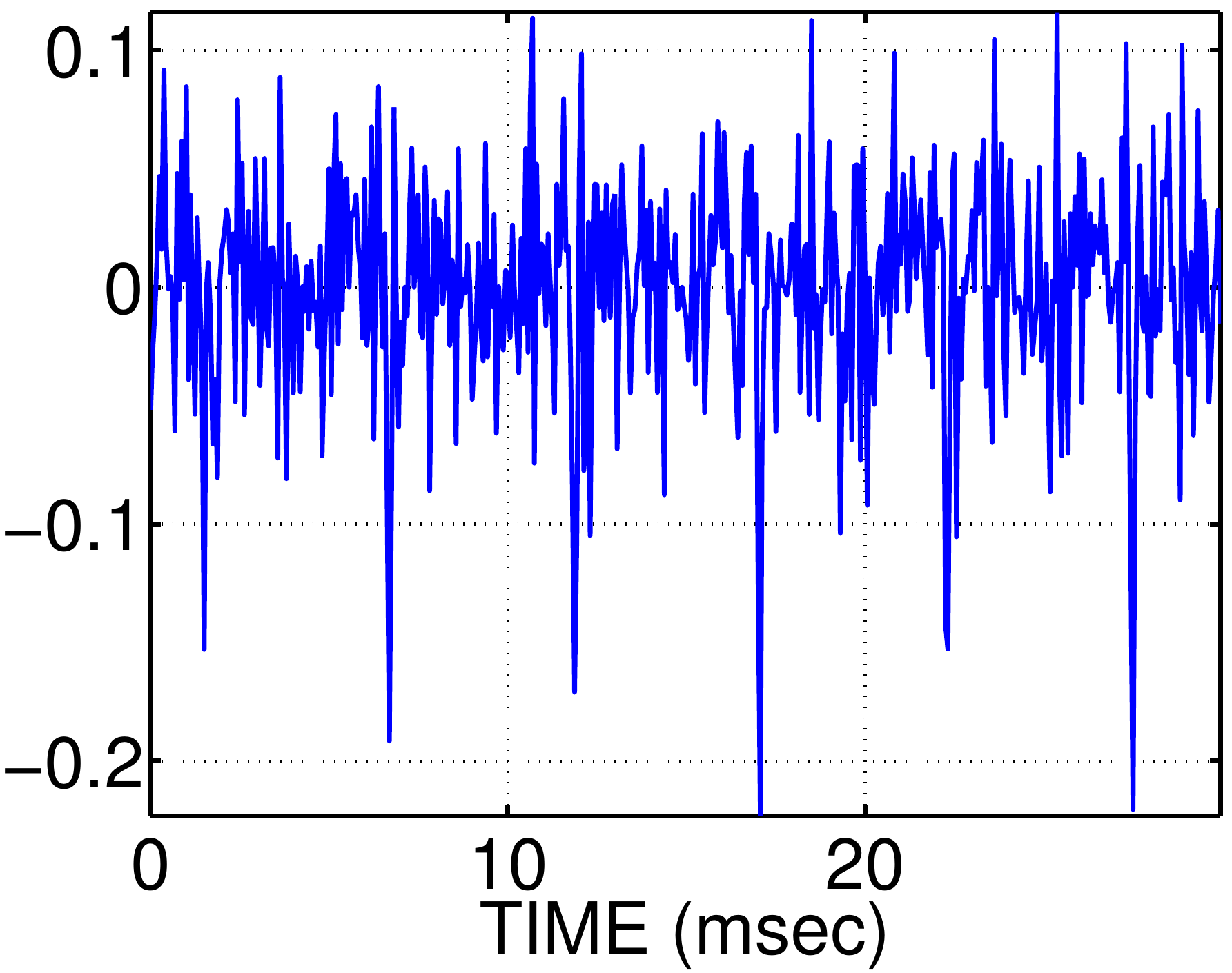}\label{fig:female_Gia_residue_10db}
                                &\includegraphics[width=\figsiz]{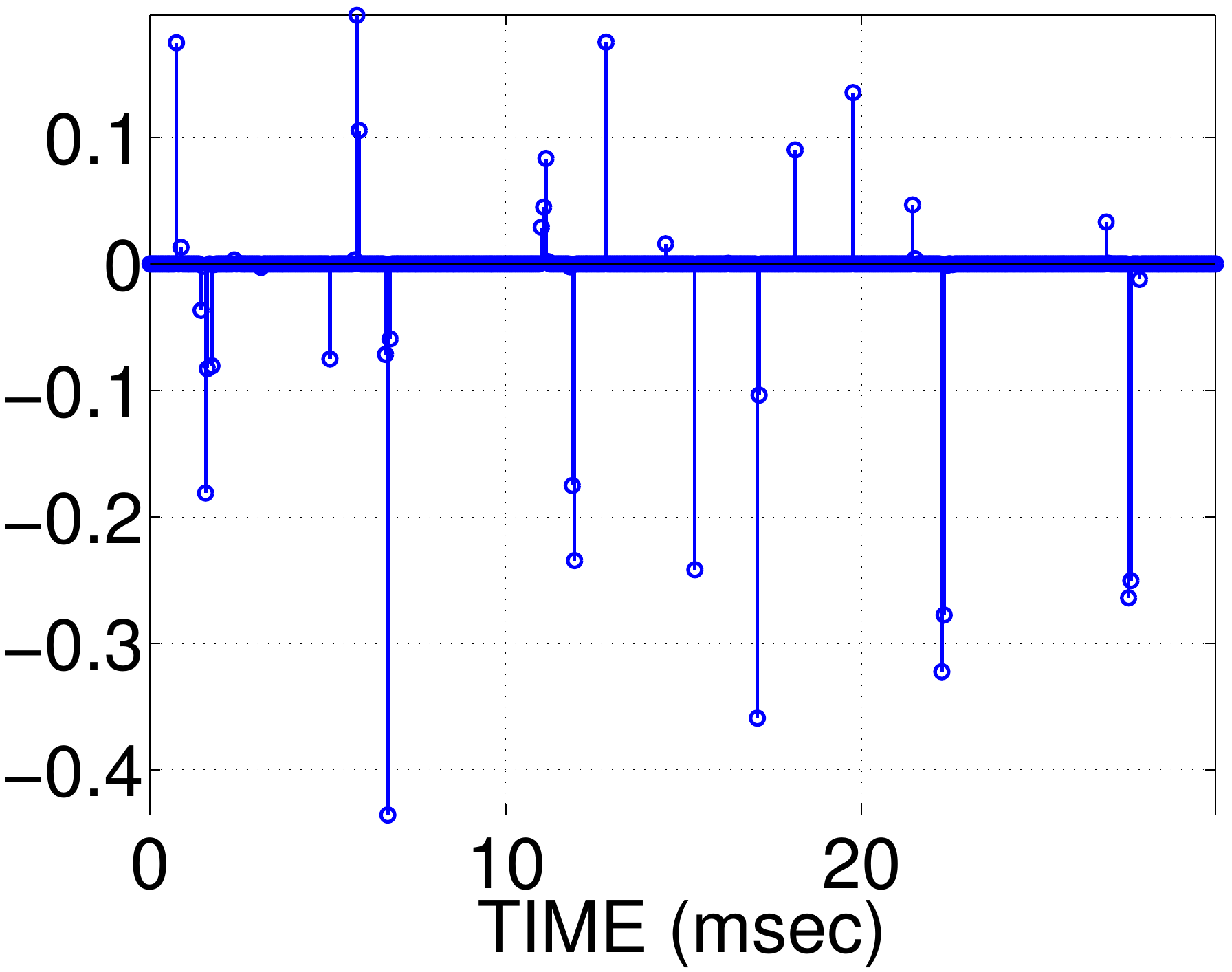}\label{fig:female_SDMM_residue_10db}
                               \end{array}$
                                \caption{(Color online) A comparison of sparse deconvolution methods for noisy speech (SNR = 10 dB): Rows 1, 2, and 3 corresponding
to Column 1 show the speech signal, frequency response of the LP filter, and the LP residue, respectively. For Columns 2$-$5, Rows 1, 2, and 3 show
estimates of the filter, its frequency response, and the excitation, respectively. Comparing the excitations, ALPA yields the sparsest excitation.}
                                                                \label{fig:female_comp_deconv_noise}
                                \end{figure}
                                \section{Conclusions}
                                \label{concl}
                                
We considered the problem of blind deconvolution of signals obtained as the output of a smooth filter excited with a sparse sequence. The sparseness of the excitation has been incorporated in the formulation by modeling it as a random vector with i.i.d. entries coming from a heavy-tailed gpG distribution. In the presence of AWGN, the cost function turned out to be non-convex and non-smooth, to optimize which we relied on an Alt. Min. scheme. The proposed algorithm ALPA optimizes a smooth, non-convex proxy for the original cost function by alternating between two steps, namely, the \emph{e-step} for optimizing the excitation and the \emph{h-step} for optimizing the filter. The individual steps consider a convex relaxation of the original cost function. We also proved that, with iterations, the reduction in the actual cost is upper-bounded by the reduction in the $\epsilon$-regularized surrogate cost, which in turn is non-increasing. This error reduction property ensures that, in practice, the iterations converge to a reasonable solution, a behavior that was also verified experimentally. As far as initialization is concerned, we considered the suitability of the regularized pseudo-inverse solution and established probabilistic guarantees on its distance from the ground-truth, which depends on the noise level, the condition number of the system, error in initial estimate of the filter, and regularization parameter. For bounded noise, the probabilistic bounds were tightened using Hoeffding's inequality. We then demonstrated an application of ALPA for blind deconvolution of voiced speech signals, into the smooth vocal tract and sparse excitation components. A comparison of the excitations obtained using ALPA, SOOT, SLP, and SDMM showed that ALPA yields the sparsest excitation and is also the fastest computationally.

                                \appendix
                                \section{Proof of Lemma~\ref{Proposition1}}
                                \label{appendix_proposition1}
                                Consider the convolution matrix
                                \scriptsize
                                \begin{equation}
                                \mb{H} = 
                                \begin{bmatrix}
                                h(0) & 0 & \dots \dots & 0 \\
                                h(1) & h(0) & \dots  \dots & 0 \\
                                h(2) & h(1) &  \dots  \dots & 0 \\
                                \vdots & \vdots & \ddots & \vdots \\
                                h(L-1) & h(L-2) & \dots  \dots& \dots \\
                                0 & h(L-1) & \dots \dots & \dots \\
                                0 & 0 & \dots \dots &  \dots \\
                                \vdots & \vdots & \vdots & \vdots \\
                                0 & 0 & \dots \dots & h(L-1)
                                \end{bmatrix}.
                                \normalsize
                                \label{convmatref}
                                \end{equation}
                                \normalsize
                                We assume, without loss of generality, that $h(0) \neq 0.$ The first entry of the vector $ \sum^{M-1}_{s = 0} \gamma_s \h_s$ will be zero if and only if $\gamma_0 = 0$. Similarly, the second entry will be zero if and only if $\gamma_0 = \gamma_1 = 0$. By mathematical induction, $\displaystyle \sum^{M-1}_{s = 0} \gamma_s \h_s = \mb{0}$ if and only if $\gamma_s = 0,\, \forall s$. Hence, $\{\mb{h}_s\}_{s=0}^{M-1}$ are linearly independent. From the definition of Riesz bases in finite-dimensional vector spaces \cite{mallat}, the Riesz bounds are given by the infimum and supremum of the quotient $\frac{\|\mb{Hx}\|_2^2}{\|\mb{x}\|_2^2}, \forall \mb{x} \in \mathbb{R}^M-\{\mb{0}\},$ which is, in fact, the Rayleigh quotient, and hence \eqref{Riesz_bounds} follows.\endproof\\

\section{Proof of Proposition~\ref{Proposition2}}
\label{appendix_theorem1} 
\indent Denote $\mb{R_{hh}} = \mb{H}^\text{\sc T}\mb{H}$, with its entries $(s_1,s_2)$ containing the autocorrelation terms corresponding to
$\mb{h}$, that is, 
\begin{equation}
[\mb{R_{hh}}]_{s_1, s_2} = 
\begin{cases}
\mb{h}^\text{\sc T}_{s_1} \mb{h}_{s_2} = \mb{h}^\text{\sc T}_{s_2} \mb{h}_{s_1}& \text{for } s_1 \ne s_2,\\
1 & \text{for } s_1 = s_2. \\
\end{cases}
\end{equation}
The entries also satisfy $\mb{R_{hh}}{(s_1,s_2)} = r_\mb{{hh}}(|s_1 - s_2|)$. In order to find the range of values $\sigma^2_{\text{min}}(\mb{H})$ can take, we
use the Gerschgorin disc theorem \cite{horn_johnson}. Due to the unit-norm constraint on $\mb{h}$, all the Gerschgorin discs pertaining to
$\mb{H}^\text{\sc T} \mb{H}$ will be centered at $1+0i$. Since $\mb{H}^\text{\sc T}\mb{H}$ is symmetric, all eigenvalues will be real.
According to Gerschgorin disc theorem, we have $\displaystyle \sigma^2_{\text{min}}(\mb{H}) \ge 1 - \underset{s_1}{\text{max}} \left(\sum_{s_2 = 0, s_2\ne s_1}^{M-1}
\left|[\mb{R_{hh}}]_{s_1,s_2}\right|\right).$ If $M$ is odd, then the $\left((M-1)/2\right)^{\text{th}}$ row of $\mb{R_{hh}}$  contains
autocorrelation terms
corresponding to all the lags to both left and right of $\mb{R_{hh}}(\frac{M-1}{2},\frac{M-1}{2})$ and hence the absolute sum of the elements of the
row excluding the diagonal element will be the largest. Thus, $\displaystyle \sigma^2_{\text{min}}(\mb{H}) \ge 1 - 2\left(\sum_{\ell = 1}^{(M-1)/2} \left|
r_{\mb{{hh}}}(\ell) \right| \right).$ In order to ensure that $ 0 < \eta \le \sigma^2_{\text{min}}(\mb{H}) \le 1$, we require that $\displaystyle 0 \le \sum_{\ell =
1}^{(M-1)/2} \left| r_\mb{hh}(\ell) \right| \le \frac{1-\eta}{2}.$ Similarly, $\sigma^2_{\text{max}}$ will also be contained within a circle of
radius $\displaystyle \sigma^2_{\text{max}}(\mb{H})\le 1 + 2 \sum_{\ell = 1}^{(M-1)/2} \left| r_{\mb{{hh}}}(\ell) \right|.$\endproof
\\

\section{Proof of Lemma~\ref{lemma1}}
\label{appendix1}
The difference
\begin{align}
F_{\epsilon}\left(\mb{h}^{(k)},\mb{e}^{(k)}\right) \!-\! F_{\epsilon}\left(\mb{h}^{(k)},\mb{e}^{(k+1)}\right) \! = \!
\delta& \sum_{j=0}^{M-1} \left(\left(\left(e_j^{(k)}\right)^2+\epsilon\right)^{p/2}  \!-\! 
\left(\left(e_j^{(k+1)}\right)^2+\epsilon\right)^{p/2}\right)\nonumber \\
&+ \left( \left\|\y-\mb{H}^{(k)}\mb{e}^{(k)}\right\|_{2}^{2}-\left\|\y-\mb{H}^{(k)}\mb{e}^{(k+1)}\right\|_{2}^{2}\right),\nonumber
\end{align}
can be rearranged as
\begin{align}
F_{\epsilon}&\left(\mb{h}^{(k)},\mb{e}^{(k)}\right) \!-\! F_{\epsilon}\left(\mb{h}^{(k)},\mb{e}^{(k+1)}\right) \!=\! \delta \sum_{j=0}^{M-1}
\left(\left(\left(e_j^{(k)}\right)^2+\epsilon\right)^{p/2}\! -\! \left(\left(e_j^{(k+1)}\right)^2+\epsilon\right)^{p/2}\right) \nonumber
\\ 
&+ \left\|\mb{H}^{(k)}\mb{e}^{(k)}-\mb{H}^{(k)}\mb{e}^{(k+1)}\right\|^2 +
\left(\y-\mb{H}^{(k)}\mb{e}^{(k+1)}\right)^{ \text{\sc T}}\left(\mb{H}^{(k)}\mb{e}^{(k+1)}-\mb{H}^{(k)}\mb{e}^{(k)}\right).\nonumber 
\end{align}
The last term is simplified by using \eqref{fixed-point} as,
\begin{align}
&F_{\epsilon}\left(\mb{h}^{(k)},\mb{e}^{(k)}\right)\! - \!F_{\epsilon}\left(\mb{h}^{(k)},\mb{e}^{(k+1)}\right)
\!=\! \left\|\mb{H}^{(k)}\mb{e}^{(k)}-\mb{H}^{(k)}\mb{e}^{(k+1)}\right\|_2^2\nonumber\\
&+\delta\sum_{j=0}^{M-1} \underbrace{\left(\left(\left(e_j^{(k)}\right)^2+\epsilon\right)^{p/2}  - 
\left(\left(e_j^{(k+1)}\right)^2+\epsilon\right)^{p/2} - p\frac{e_j^{(k+1)}\left(e_j^{(k)}-e_j^{(k+1)}\right)}{{\left(\left(e_j^{(k)}\right)^2+\epsilon\right)}^{(1-p/2)}}\right)}_{T_j}. 
 \nonumber
\end{align}
Consider the term $T_j$ inside the summation:
\small
\begin{align*}
&\frac{\left(\left(e_j^{(k)}\right)^2+\epsilon\right)-
\boxed{\left(\left(e_j^{(k+1)}\right)^2+\epsilon\right)^{(p/2)}\left(\left(e_j^{(k)}\right)^2+\epsilon\right)^{(1-p/2)}}
-pe_j^{(k+1)}\left(e_j^{(k)}-e_j^{(k+1)}\right)}{\left(\left(e_j^{(k)}\right)^2+\epsilon\right)^{(1-p/2)}}.
\end{align*}
Applying the inequality: arithmetic mean $\geq$ geometric mean, to the term inside the box, we get $$\left(\left(e_j^{(k+1)}\right)^2+\epsilon\right)^{(p/2)}\left(\left(e_j^{(k)}\right)^2+\epsilon\right)^{(1-p/2)} \leq \frac{p}{2}\left(\left(e_j^{(k+1)}\right)^2+\epsilon\right)+\left(1-\frac{p}{2}
\right)\left(\left(e_j^{(k)}\right)^2+\epsilon\right).$$
As a result, 
\begin{align*}
T_j&\ge\frac{ \left(\left(e_j^{(k)}\right)^2+\epsilon\right)\!-\!\frac{p}{2}\left(\left(e_j^{(k+1)}\right)^2+\epsilon\right)\!-\!\left(1-\frac{p}{2}
\right)\left(\left(e_j^{(k)}
\right)^2+\epsilon\right)
-p e_j^{(k+1)}\left(e_j^{(k)}-e_j^{(k+1)}\right) } {\left(\left(e_j^{(k)}\right)^2+\epsilon\right)^{(1-p/2)}} \nonumber\\
&=
\frac{p\left(e_j^{(k)}-e_j^{(k+1)}\right)^2}{\left(\left(e_j^{(k)}\right)^2+\epsilon\right)^{(1-p/2)}} \ge 0 \nonumber.
\end{align*}
\normalsize
\begin{eqnarray}
\text{Consequently,}\quad
F_{\epsilon}\left(\mb{h}^{(k)},\mb{e}^{(k)}\right) - F_{\epsilon}\left(\mb{h}^{(k)},\mb{e}^{(k+1)}\right) &\ge&
\left\|\mb{H}^{(k)}\left(\mb{e}^{(k)}-\mb{e}^{(k+1)}\right)\right\|_2^2 \geq 0,\nonumber\\
\Rightarrow F_{\epsilon}\left(\mb{h}^{(k)},\mb{e}^{(k+1)}\right) &\le& F_{\epsilon}\left(\mb{h}^{(k)},\mb{e}^{(k)}\right).\nonumber
\end{eqnarray} 
\endproof\\

\section{Upper Bound on MAE}
\label{Appendix_BRLS_expand}
Expanding \eqref{dbrls_expand} and rearranging terms, we get
\begin{eqnarray}
\left(\Hst \Hs + \delta \mb{I}\right) \dbrls = &\left(\Hs + \dH \right)^\tr \w -\left(\Hst \dH + \dH^\tr \dH + \delta \right)\es \nonumber\\ &-\left(\Hst \dH + \dH^\tr \Hs + \dH^\tr \dH\right)\dbrls.
\end{eqnarray}
Using the triangle inequality, $\|\mb{A}+\mb{B}\|_2\le \|\mb{A}\|_2+\|\mb{B}\|_2$ and compatibility of induced norms, $\|\mb{Ax}\|_2\le \|\mb{A}\|_2\|\mb{x}\|_2$ gives
\begin{align}
 &\|\dbrls\|_2 \le \|\left(\Hst \Hs + \delta \mb{I}\right)^{-1}\|_2\Big\{\|\Hs + \dH \|_2 \|\w\|_2 + \nonumber\\ &\left(\|\Hs\|_2 \|\dH\|_2 + \|\dH\|_2^2 + \delta \right)\|\es\|_2 +\left(2\|\Hs\|_2 \|\dH\|_2 + \|\dH\|_2^2\right)\|\dbrls\|_2 \Big\}.\nonumber
\end{align}
Now, $\|\dH\|_2 \le \sqrt{M}\|\dsh\|_2$ and hence, $\|\Hs + \dH\|_2 \le \|\Hs\|_2+\|\dH\|_2 \le \|\Hs\|_2+ \sqrt{M}\|\dsh\|_2$. Since $\|\left(\Hst \Hs + \delta \mb{I}\right)^{-1}\|_2 \|\Hs\|_2 = \kappa_q(\Hs)$ and $\|\Hs\|_2 \ge 1$, the term $\sqrt{M}\|\left(\Hst \Hs + \delta \mb{I}\right)^{-1}\|_2 \|\dsh\|_2 \|\w\|_2 \le \sqrt{M} \|\left(\Hst \Hs + \delta \mb{I}\right)^{-1}\|_2 \|\Hs\|_2 \|\dsh\|_2 \|\w\|_2 \\= \sqrt{M} \kappa_q{\Hs} \|\w\|_2$. Therefore,
\begin{align}
 \|\dbrls\|_2 &\le \left(\kappa_q(\Hs) + \sqrt{M} \kappa_q(\Hs) \|\dsh\|_2\right)  \|\w\|_2 + \left(\sqrt{M} \kappa_q(\Hs) \|\dsh\|_2 + \delta \right)\|\es\|_2 \nonumber\\ &+2\sqrt{M}\kappa_q(\Hs)\|\dsh\|_2\|\dbrls\|_2 + \mathcal{O}(\|\dsh\|^2_2).
\end{align}
Neglecting the $\mathcal{O}(\|\dsh\|^2_2)$ terms,
\begin{align}
 &\left(1-2\sqrt{M}\kappa_q(\Hs)\|\dsh\|_2\right)\|\dbrls\|_2 \le\nonumber\\ & \left(\kappa_q(\Hs) + \sqrt{M} \kappa_q(\Hs) \|\dsh\|_2\right)  \|\w\|_2 + \left(\sqrt{M} \kappa_q(\Hs) \|\dsh\|_2 + \delta \right)\|\es\|_2.
\end{align} 

Denoting $\sqrt{M} \kappa_q(\Hs) \|\dsh\|_2 = C_{\Delta\h}$. If $C_{\Delta\h}<\frac{1}{2}$, then
\begin{align}
\|\dbrls\|_2 \le& \frac{1}{\left(1-2C_{\Delta\h}\right)}\Bigg\{\left(\kappa_q(\Hs) + C_{\Delta\h}\right)  \|\w\|_2 + \left(C_{\Delta\h} + \delta \right)\|\es\|_2\Bigg\}.
\label{normdele}
\end{align}
Using equivalence of norms $\frac{1}{\sqrt{M}}\|\dbrls\|_1 \le \|\dbrls\|_2$, we get \eqref{dbrls_ineq}.
\endproof
\\
\section{Proof of Proposition~\ref{BRLS_prop}}
\label{appendix_BRLS_prop}
In view of the inequality in \eqref{dbrls_ineq}, 
\begin{align}
&\mathcal{P}\left(\frac{1}{M}\|\dbrls\|_1 > \xi\right) <\nonumber\\ &\mathcal{P}\Bigg(\frac{1}{\sqrt{M}\left(1-2C_{\Delta\h}\right)}\Bigg\{\left(\kappa_q(\Hs) + C_{\Delta\h}\right)  \|\w\|_2 + \left(\delta + C_{\Delta\h} \right)\|\es\|_2 \Bigg\} > \xi \Bigg).
\label{Markov_for_w1}
\end{align}
Rearranging the right-hand side terms and using Markov inequality, we get
\begin{align}
\mathcal{P}\left(\|\w\|^2_2 > \Bigg(\frac{\sqrt{M}(1-2C_{\Delta\h}) \xi - \left(\delta + C_{\Delta \h}\right)\|\es\|_2}{\kappa_q(\Hs)+ C_{\Delta \h}}\Bigg)^2\right) \nonumber\\< \frac{\mathcal{E}(\|\w\|^2_2) (\kappa_q(\Hs)+C_{\Delta \h})^2}{\left(\sqrt{M} (1-2C_{\Delta\h})\xi - \left(\delta + C_{\Delta \h}\right)\|\es\|_2\right)^2}.
\label{Markov_for_w}
\end{align}
Since the entries of $\w$ are i.i.d., $\mathcal{E}(\|\w\|^2_2) = M\sigma^2$, which results in the inequality \eqref{Markov_upperbound_bd}.$\blacksquare$

\section*{Acknowledgments}
The authors would like to thank Subhadip Mukherjee and Basty Ajay Shenoy for fruitful technical discussions.


\bibliographystyle{siamplain}
\bibliography{biblio}
\end{document}